\documentclass[longauth]{aa} 

\usepackage{graphicx}
\usepackage{txfonts}
\usepackage[dvipsnames]{xcolor}
\usepackage{comment}
\usepackage{amsmath}
\usepackage{nccmath}
\usepackage{caption}
\usepackage{subcaption}
\usepackage[colorlinks=true, citecolor=blue]{hyperref}
\usepackage[autostyle]{csquotes}
\bibpunct{(}{)}{;}{a}{}{,} 
\usepackage{longtable}
\usepackage{multirow}
\usepackage{lscape}
\usepackage{color, colortbl}
\usepackage{placeins}
\usepackage[normalem]{ulem}

\begin{document} 

   \title{$z$-GAL -- A NOEMA spectroscopic redshift survey of bright {\it Herschel} galaxies: [II] Dust properties}


\author{D. Ismail \inst{1} \and 
        A. Beelen \inst{1}  \and
        V. Buat \inst{1} \and 
        S. Berta\inst{2} \and 
        P. Cox\inst{3}   \and
        F. Stanley\inst{2}        \and
        A. Young\inst{4}          \and
        S. Jin\inst{5}             \and
        R. Neri\inst{2}          \and
        T. Bakx\inst{6,7,8}        \and  
        H. Dannerbauer\inst{9,10}  \and 
        K. Butler \inst{2}      \and
        A. Cooray\inst{11}      \and
        A. Nanni \inst{12, 13}   \and
        A. Omont\inst{3}        \and
        S. Serjeant\inst{14}    \and
        P. van der Werf\inst{15} \and
        C. Vlahakis\inst{16}    \and
        A. Wei{\ss}\inst{17}    \and
        C. Yang\inst{6}         \and
        A.~J. Baker \inst{4, 18}    \and
        G. Bendo \inst{19}      \and
        E. Borsato \inst{20}    \and
            N. Chartab \inst{11}        \and
            S. Dye \inst{21}            \and
        S. Eales \inst{22}      \and
        R. Gavazzi \inst{1}     \and
        D. Hughes \inst{23}     \and
            R. Ivison \inst{24,25,26,27}        \and
            B.~M. Jones \inst{28} \and
        M. Krips \inst{2}       \and
        M. Lehnert \inst{29}    \and 
        L. Marchetti \inst{30,31} \and  
        H. Messias \inst{32, 33}  \and  
        M. Negrello \inst{22}   \and  
        I. Perez-Fournon \inst{9,10} \and  
        D.~A. Riechers \inst{28} \and
        S. Urquhart\inst{14}
          }

   \institute{Aix-Marseille Université, CNRS and CNES, Laboratoire d’Astrophysique de Marseille, 38, rue Frédéric Joliot-Curie 13013 Marseille, France
              \email{diana.ismail@lam.fr} 
        \and 
             Institut de Radioastronomie Millim\'etrique (IRAM), 300 rue de la Piscine, 38400 Saint-Martin-d'H{\`e}res, France 
        \and     
             Sorbonne Universit{\'e}, UPMC Universit{\'e} Paris 6 and CNRS, UMR 7095, 
             Institut d'Astrophysique de Paris, 98bis boulevard Arago, 75014 Paris, France  
        \and 
             Department of Physics and Astronomy, Rutgers, The State University of New Jersey, 
             136 Frelinghuysen Road, Piscataway, NJ 08854-8019, USA 
        \and
             Cosmic DAWN Centre Radmandsgade 62, 2200 Copenhagen N, Denmark 
        \and 
            Departement of Space, Earth and Environment, Chalmers University of Technology, Onsala Space Observatory, 439 92 Onsala, Sweden 
        \and
            Division of Particle and Astrophysical Science, Graduate School of Science, 
            Nagoya University, Aichi 464-8602, Japan 
        \and 
            National Astronomical Observatory of Japan, 2-21-1, Osawa, Mitaka, Tokyo 181-8588, Japan 
        \and    
             Instituto Astrof{\'i}sica de Canarias (IAC), E-38205 La Laguna, Tenerife, Spain  
        \and
             Universidad de La Laguna, Dpto. Astrofísica, E-38206 La Laguna, Tenerife, Spain 
        \and
             University of California Irvine, Department of Physics \& Astronomy, FRH 2174, Irvine CA 92697, USA  
        \and
             National Centre for Nuclear Research, ul. Pasteura 7, 02-093 Warsaw, Poland 
        \and
             INAF - Osservatorio astronomico d'Abruzzo, Via Maggini, SNC 64100 Teramo, Italy 
        \and    
             Department of Physical Sciences, The Open University, Milton Keynes MK7 6AA, UK 
        \and 
            Leiden University, Leiden Observatory, PO Box 9513, 2300 RA Leiden, The Netherlands 
        \and 
            National Radio Astronomy Observatory, 520 Edgemont Road, Charlottesville VA 22903, USA 
        \and 
            Max-Planck-Institut f{\"u}r Radioastronomie, Auf dem H{\"u}gel 69, 53121 Bonn, Germany. 
        \and 
            Department of Physics and Astronomy, University of the Western Cape, Robert Sobukwe Road, Bellville 7535, South Africa 
        \and 
             UK ALMA Regional Center Node, Jodrell Bank Center for Astrophysics, Department of Physics and Astronomy, The University of Manchester, Oxford Road, Manchester M13 9PL, United Kingdom 
        \and 
             Dipartimento di Fisica \& Astronomia "G. Galilei", Universit\`a di Padova, vicolo dell'Osservatorio 3, Padova I-35122, Italy 
        \and 
             School of Physics and Astronomy, University of Nottingham, University Park, 
             Notthingham NG7 2RD, UK 
        \and 
             School of Physics and Astronomy, Cardiff University, Queens Building, The Parade, Cardiff, CF24 3AA, UK 
        \and
             Instituto Nacional de Astrofísica, \'Optica y Electr\'onica, Astrophysics Department, Apdo 51 y 216, Tonantzintla, Puebla 72000 Mexico 
        \and 
             European Southern Observatory, Karl-Schwarzschild-Strasse 2, D-85748 Garching, Germany 
        \and 
             Department of Physics and Astronomy, Macquarie University, North Ryde, New South Wales, Australia 
        \and
             School of Cosmic Physics, Dublin Institute for Advanced Studies, 31 Fitzwilliam Place, Dublin D02 XF86, Ireland 
        \and
             Institut for Astronomy, University of Edinburgh, Blackford Hill, Edinburgh EH9 3HJ, UK 
        \and 
             I. Physikalisches Institut, Iniversit\"at zu K\"oln, Z\"ulpicher Strasse 77, D-50937 K\"oln, Germany 
        \and
             Centre de Recherche Astrophysique de Lyon - CRAL, CNRS UMR 5574, UCBL1, ENS Lyon, 9 avenue Charles Andr\'e, F-69230 Saint-Genis-Laval, France 
        \and 
             Department of Physics and Astronomy, University of the Western Cape, Private Bag X17, Bellville 7535, Cape Town, South Africa 
        \and  
             Istituto Nazionale di Astrofisica, Istituto di Radioastronomia, via Gobetti 101, 40129 Bologna, Italy 
        \and
             Joint ALMA Observatory, Alonso de C\'ordova 3107, Vitacura 763-0355, Santiago de Chile, Chile 
        \and
             European Southern Observatory, Alonso de C\'ordova 3107, Vitacura, Casilla 19001, Santiago de Chile, Chile 
             }

   \date{Received dd mm, year; accepted dd mm, year}

 
  \abstract
    {We present the dust properties of 125 bright \textit{Herschel} galaxies selected from the \textit{z}-GAL NOEMA spectroscopic redshift survey. All the galaxies have precise spectroscopic redshifts in the range 1.3 < \textit{z} < 5.4. The large instantaneous bandwidth of NOEMA provides an exquisite sampling of the underlying dust continuum emission at 2 and 3 mm in the observed frame, with flux densities in at least four sidebands for each source. Together with the available \textit{Herschel} 250, 350, and 500~$\rm \rm \mu m$ and SCUBA-2 850~$\rm \mu m$ flux densities, the spectral energy distribution (SED) of each source can be analyzed from the far-infrared to the millimeter, with a fine sampling of the Rayleigh-Jeans tail. This wealth of data provides a solid basis to derive robust dust properties, in particular the dust emissivity index ($\beta$) and the dust temperature ($\rm T_{\rm dust}$). In order to demonstrate our ability to constrain the dust properties, we used a flux-generated mock catalog and analyzed the results under the assumption of an optically thin and optically thick modified black body emission. The robustness of the SED sampling for the \textit{z}-GAL sources is highlighted by the mock analysis that showed high accuracy in estimating the continuum dust properties. These findings provided the basis for our detailed analysis of the \textit{z}-GAL continuum data. We report a range of dust emissivities with $\beta \sim 1.5 - 3$ estimated up to high precision with relative uncertainties that vary in the range $7\%\ - 15$\%, and an average of $2.2 \pm 0.3$. We find dust temperatures varying from 20 to 50 K with an average of $\rm T_{\rm dust} \sim 30 $ K for the optically thin case and  $\rm T_{\rm dust} \sim 38$ K in the optically thick case. For all the sources, we estimate the dust masses and apparent infrared luminosities (based on the optically thin approach). An inverse correlation is found between $\rm T_{\rm dust}$ and $\beta$ with $\beta \propto T^{-0.69}_{\rm dust}$, which is similar to what is seen in the local Universe. Finally, we report an increasing trend in the dust temperature as a function of redshift at a rate of $6.5 \pm 0.5$ K/\textit{z} for this 500 $\mu$m-selected sample. Based on this study, future prospects are outlined to further explore the evolution of dust temperature across cosmic time.}

   \keywords{Dusty Star-Forming Galaxies - 
            galaxies: high redshift -
            galaxies: evolution - 
            infrared: galaxies -
            submillimeter: galaxies
            }

   \maketitle
%
\section{Introduction}
Following the detection of luminous infrared galaxies at high redshift with the IRAS satellite \citep[e.g.,][]{rowan-robinson1991iras,rowan-robinson1993iras}, the first observations at 850 $\rm \mu m$ with the Submillimeter Common-User Bolometer Array \citep[SCUBA;][]{holland1999scuba}, and later at 1.2 mm with the Max-Planck Millimeter Bolometer Array \citep[MAMBO;][]{kreysa1999mambo}, uncovered a population of very luminous dusty star-forming galaxies (DSFGs; with infrared luminosities in excess of $\geq 10^{12} L_\odot$) at high redshifts (\textit{z} $\geq$ 1) \citep[see reviews by][]{blain2002smg,carilli2013walter, casey2014dusty, hodge2020dacunha}. Subsequent extragalactic imaging surveys carried out with the \textit{Herschel} Space Observatory \citep{pilbratt2010herschel} increased the number of known DSFGs from hundreds to several hundreds of thousands \citep[e.g.,][]{eales2010herschel, oliver2012hermes, ward2022herschel}. These dust-enshrouded luminous galaxies absorb $\sim 99\%$ of the light emitted by young, forming stars \citep[e.g.,][]{clements1996irasgalaxies, lagache2005, buat2010sed}, and thermally radiate the reprocessed ultraviolet--optical light in the far-infrared (FIR) and submillimeter (submm) regimes. The discovery of this population of DSFGs has changed our understanding of galaxy evolution in the early Universe, and the advent of facilities such as the Atacama Large Millimeter/submillimeter Array (ALMA), the IRAM Northern Extended Millimeter Array (NOEMA), and the Karl G. Jansky Very Large Array (VLA) have enabled follow-up observations of large samples of DSFGs to study their physical properties by probing the molecular and atomic gas \citep[][and references therein]{hodge2020dacunha}, and tracing the FIR/submm spectral energy distribution (SED) of the dust emission up to \textit{z} $\sim 7$ \citep[e.g.,][]{riechers2013lambda, reuter2020complete, bouwens2021properties, sommovigo2022new}. 

Bright DSFGs are known to have very large dust reservoirs, with typical dust masses $M_{\rm dust} > 10^{8} M_\odot$ \citep[e.g.,][]{swinbank2014aless,reuter2020complete, dudzevivciute2021tracing, da2021measurements}. Dust grains, which are abundant in the interstellar medium (ISM) of galaxies in the local and distant Universe \citep[e.g.,][]{smail1997galaxyevolution, blain1999starformation}, play a vital role in many astrophysical processes, including the onset of the star-formation process. However, the origin of these large dust masses in DSFGs, particularly at high redshift, has been a matter of debate over the years. The mass buildup occurs mainly in the ejecta of supernovae (SNe) and in the envelopes of asymptotic giant branch (AGB) stars, but also via grain growth by accretion in the ISM \citep[e.g., see the review by][]{galliano2018review}. However, these mechanism cannot fully explain the amount of dust available in these high-\textit{z} galaxies given the short timescales involved \citep{michalowksi2015dust, nanni2020dustproduction}. It is therefore crucial to derive dust properties in DSFGs across cosmic time to search for any sign of evolution that could provide constraints on how the dust grains were formed in the early Universe.

Dust masses are commonly estimated by fitting a modified blackbody that best describes the thermal emission of dust in the FIR/submm domain. However, one of the caveats of this fitting method is the dependence of dust mass on the dust temperature estimates. For instance, \citet{casey2012sed} pointed out that a 4 K difference in temperature could result in a 150\% increase in dust mass. Therefore, it is important to quantify as well as constrain the dust properties in DSFGs from which we can derive the star formation rates \citep[SFRs;][]{sanders1996mirabel, kennicutt1998} and estimate the gas masses \citep[e.g.,][]{eales2010herschel, scoville2013gdr}.

To derive precise dust parameters, it is crucial to have a good coverage of the SED. With facilities like ALMA and NOEMA, follow-up observations of previously FIR/submm detected DSFGs have been carried out to measure the continuum emission along the Rayleigh-Jeans (RJ) tail from submm wavelengths down to 3~mm, which enabled a precise fitting of the FIR/submm SEDs \citep[e.g.,][]{reuter2020complete, berta2021close, da2021measurements, bendo2023bright}. However, uncertainties in measuring dust temperatures still exist, and arise from (i) the opacity (in other words, on the assumptions of optical depth), and (ii) the degeneracy with the dust emissivity index $\beta$. The wavelength at which the optical depth ($\tau$) becomes unity is often assumed to be $\lambda_{thick} \sim 100 \rm \mu m$ \citep{draine2006submillimeter}, although some galaxies have been found to have higher values, with $\lambda_{thick} \sim 200 \rm \mu m$ \citep{conley2011lambda0, riechers2013lambda, riechers2017rise}. For simplicity, some works assume that the dust emission in DSFGs can be approximated with an optically thin modified blackbody ($\tau \ll 1$), which can also lead to an underestimation of the dust temperature \citep[e.g., up to $\Delta T_{\rm dust} \sim 20$ K as shown by][]{da2021measurements}. To overcome these assumptions, high-resolution imaging is essential in order to derive robust values of dust opacities by recovering the galaxy size \cite[e.g.,][]{spilker2016alma}. On the other hand, a degeneracy has been found for $\beta$ and $T_{\rm dust}$  \citep{dupac2003degeneracy,shetty2009effect, paradis2010, planck2011planck, juvela2011galactic}, which is further amplified by a lack of photometric data at longer wavelengths. To avoid the effects of this degeneracy, the emissivity index $\beta$ is generally fixed based on measurements of the Milky Way \citep[$\beta = 1.5 - 2$, e.g.,][]{magnelli2012herschel}. However, recent studies show that dust emissivities of  DSFGs diverge from the classical values \citep[e.g.,][]{da2021measurements, cooper2022searching}, which would in turn affect the dust temperature estimate. \citet{bendo2023bright} found a difference of up to $\sim 9$ K when fixing $\beta$ for a sample of 37 \textit{Herschel}-selected galaxies observed with ALMA at 3 and 2~mm. With a good sampling of photometric data at longer wavelengths, the dust temperature estimate becomes isolated from the effect of fixing (or deriving) an inadequate value for $\beta$, thus breaking the degeneracy. 

Nonetheless, the growing number of surveys with FIR/submm observations has resulted in an abundance of dust temperature measurements in the literature that span a wide range of redshifts, reaching \textit{z} $\sim 8$ \citep[e.g.,][]{reuter2020complete, bakx2021, sommovigo2022new}. However, contradictory claims of dust temperature trends across cosmic time have been reported. Some studies argue that dust becomes hotter at higher redshifts as compared to the local Universe, which is attributed to higher specific star formation rates (sSFR) \citep[e.g.,][]{magnelli2014evolution, magdis2012evolving, bethermin2015evolution, schreiber2018dust, zavala2018s2cls, liang2019dust, faisst2020evolution, sommovigo2020warmdust, sommovigo2022new}. Conversely, \citet{dudzeviciute2020noevolution} argue that there is no evidence of temperature evolution, and the observed correlation is solely due to selection biases that limit observations to the brightest galaxies, especially at higher redshifts. 

Recently, \citet{neri2020noema} demonstrated the capability of the broadband receivers and the Polyfix correlator at NOEMA to measure up to ten continuum flux densities in the millimeter bands per source by observing 13 bright high-\textit{z} \textit{Herschel}-selected galaxies. Reliable spectroscopic redshifts were obtained for 11 of the selected sources, and by combining the available continuum flux densities from the \textit{Herschel}-Spire, SCUBA-2, and the 2 and 3~mm NOEMA data, well-sampled SEDs for each source were obtained from which the dust properties were derived \citep[infrared luminosities, dust masses, and temperatures;][]{neri2020noema, berta2021close}. Based on the successful NOEMA Pilot Program, the \textit{z}-GAL Large Program has extended this work by observing 126 high-redshift \textit{Herschel}-selected galaxies in the 2 and 3~mm NOEMA wavebands \citep{cox2023} with the goal being to measure precise spectroscopic redshifts. In addition to the molecular and atomic emission lines that were detected in all the sources, the continuum emission flux densities were extracted for each source in at least four, and up to ten, sidebands. The combined Pilot Program and $z$-GAL Large Program constitutes the final $z$-GAL sample of 137 sources, for which robust redshifts were derived for 135 sources, spanning the range 0.8<z<6.5.  The present study of the dust properties of high-$z$ galaxies is based on this sample, the largest with precise redshifts to date. 

This paper (Paper II) is part of a series of three papers reporting the results of the \textit{z}-GAL project. Paper I \citep{cox2023} gives an overview of the \textit{Herschel}-selected sources of this program and presents the overall properties of the molecular and atomic emission lines, as well as the derived spectroscopic redshifts. In Paper III, \citet{berta2023} analyze and describe the physical properties of the molecular and atomic gas in the \textit{z}-GAL sources using both the continuum and emission lines. Here, we study the continuum data of the \textit{z}-GAL sources. Our main goal is to derive the dust properties, taking advantage of the exquisite coverage in frequency of the SED. In order to demonstrate our ability to constrain the essential dust properties, such as dust temperature and emissivity index, we used a flux-generated mock catalog and analyzed the results under the assumptions of an optically thick and optically thin modified blackbody emission. We then performed a detailed analysis of the \textit{z}-GAL continuum data informed by the findings of the mock simulation. The main results focus on the optically thin approximation, which will serve as a basis for our analysis (specifically the dust masses and inferred FIR luminosities) in Paper III \citep{berta2023}. 

The paper is organized as follows. In Section \ref{section2}, we give a brief overview of the \textit{z}-GAL sample selected from the \textit{Herschel} catalog and the NOEMA observations in the 2 and 3~mm wavebands, and provide a detailed explanation of the continuum data reduction as well as an overview of the data that form the basis of this paper. In Section \ref{section:opt-thin-model}, we describe the optically thin approximation of the modified blackbody used to determine the continuum dust properties and analyze the model effects on each term. In Section \ref{section:zgal_opt_thin}, we present the optically thin dust properties of the \textit{z}-GAL sample of sources. In Section \ref{section5:gmbb}, we present the general modified blackbody (GMBB) model, and our derivation of parameters and \textit{z}-GAL dust properties. In Section \ref{section6:discussion}, we discuss the derived results and their evolution with redshift. Finally, in Section \ref{section7:conclusions}, we summarize our main conclusions and outline future studies. Throughout the paper, we adopt a spatially flat $\Lambda$CDM cosmology with
$H_0$ = 67.4 km s$^{-1}$ Mpc$^{-1}$ and $\Omega_M$ = 0.315 \citep{planck2018}.

\begin{figure}[ht!]
    \includegraphics[width=0.49\textwidth]{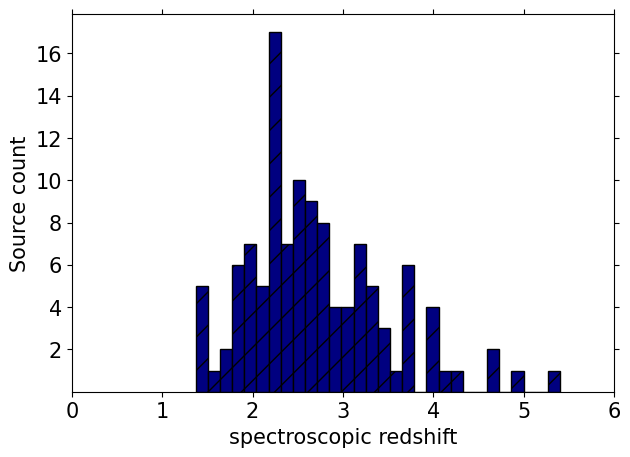}
    \caption{Redshift distribution of the galaxies used in this study, which includes 125 out of the 137 sources from the \textit{z}-GAL sample. In our analysis, we exclude multiples at different redshifts in the field of view observed by NOEMA, which are unresolved in the \textit{Herschel} field (see details in Section \ref{section:flux-sample}).}
    \label{fig:redshift-distribution-zgal}
\end{figure}

\section{\textit{z}-GAL sample}\label{section2}
The aim of the \textit{z}-GAL project is to measure robust spectroscopic redshifts of a sample of 137 Herschel-selected galaxies. The details of the sample selection are provided in Paper I \citep{cox2023}. The 126 sources for the \textit{z}-GAL Large Program (carried out under projects M18AB and D20AB - PI: P. Cox, T. Bakx \& H. Dannerbauer) were selected from the \textit{Herschel} Bright Sources (HerBS) sample, the HerMES Large Mode Survey (HeLMS), and the \textit{Herschel} Stripe 82 (HerS) Survey \citep{nayyeri2016candidate, bakx2018}, and also include the 11 sources of the Pilot Program \citep{neri2020noema} from the HerBS sample. The HeLMS and HerS fields cover 372 $\rm deg^2$, and HerBS sources were selected from the NGP and GAMA fields that cover 170.1 and 161.6 $\rm deg^2$, respectively. The selection was based on the 500 $\rm \mu m$ flux limit with $S_{500 \rm \mu m}$ > 80 mJy for the HerBS sources and $S_{500 \rm \mu m}$ > 100 mJy for both HeLMS and HerS sources, as well as photometric redshifts $z_{phot}$ > 2 \citep[see][for the sample selection and references therein]{cox2023}. The \textit{z}-GAL sources were observed using NOEMA by scanning the 2 and 3~mm wavebands, which successfully resulted in deriving spectroscopic redshifts based on at least two emission lines for all but two of the selected sources \citep{neri2020noema,cox2023}. The frequency setup for the 3~mm observations covers the range from 75.874 to 106.868~GHz and the 2~mm setup covers the range from 127.377 to 158.380~GHz. The fields of view at  2 and 3~mm are $\sim 33\arcsec$ and  $\sim 50\arcsec$, respectively, and the angular resolutions were in between $1\farcs2$ and $3\farcs5$ at 2~mm and between $1\farcs7$ and $\sim 6\arcsec$ at 3~mm. Alongside the line emission measurements, all sources were detected in the continuum in at least four sidebands thanks to the broad band receiver of NOEMA. The \textit{z}-GAL sources span the redshift range 1 < \textit{z} < 6 (see Fig. \ref{fig:redshift-distribution-zgal}), peaking at the peak of cosmic evolution at \textit{z} $\sim 2-3$, making it an ideal large sample with which to explore eventual changes of dust properties with redshift. 

\begin{figure}[ht!]
    \centering
    \includegraphics[width=0.5\textwidth]{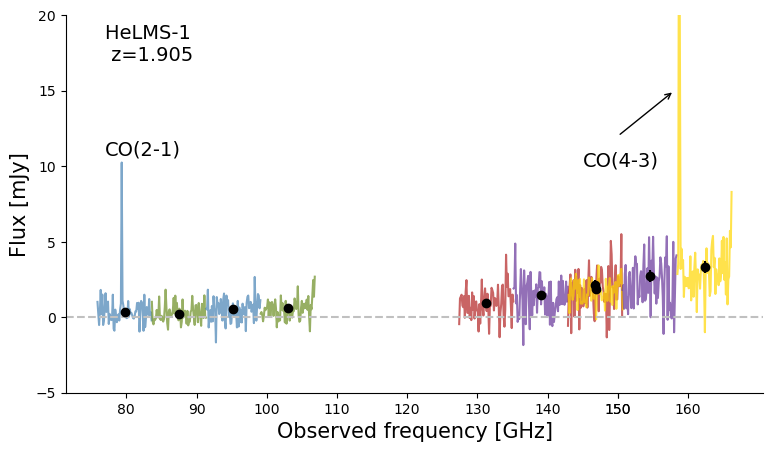}
    \caption{NOEMA spectral coverage and bandwidth of the 2 and 3~mm bands for the source HeLMS-1 at \textit{z} = 1.905. The ten alternating colors are the lower and upper sidebands for each setup. The black points show the flux density extracted in each band along with their uncertainties (very small error bars). The $^{12}$CO(4-3) and (2-1) emission lines detected in this source are identified in the figure.}
    \label{fig:noema-coverage-setup}
\end{figure}

\subsection{Data reduction}\label{section:data-reduction}
The NOEMA data were reduced and calibrated with the GILDAS\footnote{https://www.iram.fr/IRAMFR/GILDAS/} package and {\it uv}-tables of each sideband were produced in the standard way. The main calibrators adopted were MWC349 and LkH$\alpha$101. More details are provided in Paper-I. The absolute flux uncertainty is 10\% and the positional error is 0.2 arcsec.

For each $z$-GAL target, stacked maps were produced combining all channels of all available sidebands, including both continuum and spectral lines (hereafter called all-channels maps). Such stacked maps were then used as reference for the detection of sources in the NOEMA fields centered on the \textit{Herschel} positions. Continuum maps of each sideband were produced by combining all channels of the given sideband, but excluding the spectral ranges including the emission lines. As an example, Fig. \ref{fig:noema-coverage-setup} shows the ten sidebands observed for one of the sources, HeLMS-1, at 2 and 3~mm, each one with a bandwidth of approximately 8 GHz.

Source extraction and continuum measurements were conducted in the following way. For each target, the all-channels maps were used to identify all sources in the field of view of NOEMA using 3$\sigma$ threshold contours as guidance to define the aperture. The maps were cleaned using natural weights and support masks including all detected sources. The individual continuum maps were also cleaned with natural weight, adopting support masks re-adapted because of different beam size, orientation, and shape. In cases where sources were not detected in individual sidebands, the support masks follow the size and orientation of the beam. 

The measurement of continuum fluxes was then performed on the cleaned continuum maps of each individual sideband for the sources identified in the all-channels map. Aperture fluxes were measured using ad hoc extraction polygons and corrected for primary beam losses. Flux statistical uncertainties were computed by rescaling the map noise to the effective extraction aperture size. Source positions were computed as signal barycenters within the same extraction apertures.

\begin{figure}[ht!]
    \centering
    \includegraphics[width=0.45\textwidth]{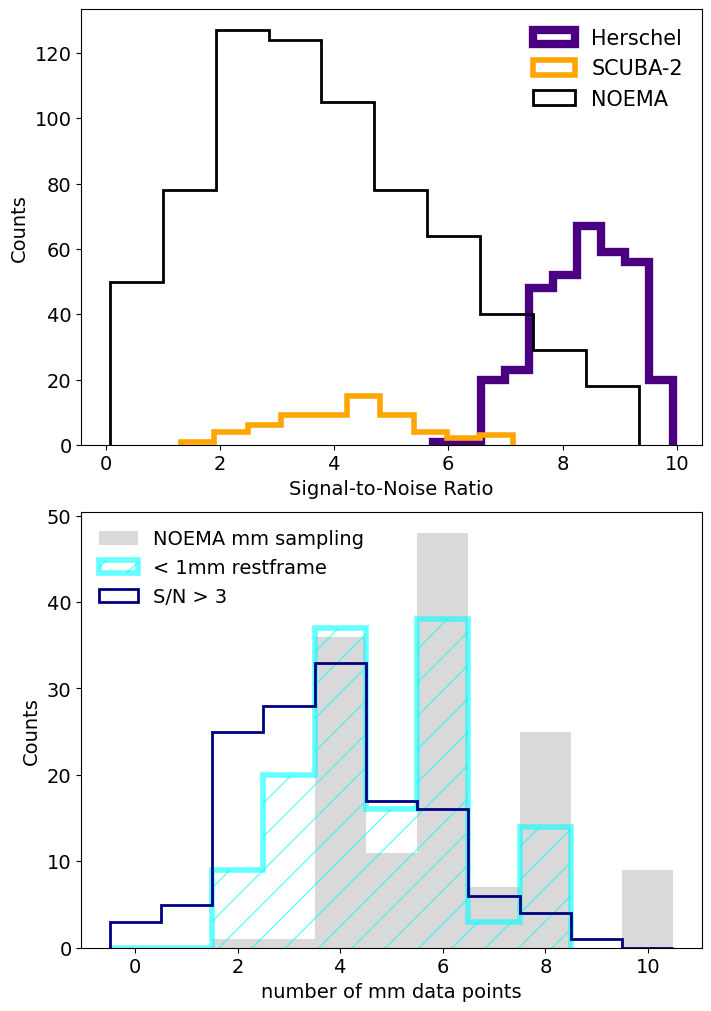}
    \caption{Sampling and S/N of the flux densities of the \textit{z}-GAL galaxies. \textit{Top panel}: Signal-to-noise ratio of the \textit{z}-GAL data split into \textit{Herschel} 250, 350, and 500 $\rm \mu m$ flux densities shown in indigo, which varies between 6 and 10; SCUBA-2 850 $\rm \mu m$ flux density, in orange, which varies between 1.5 and 7; and NOEMA 2 and 3~mm flux densities, which varies between 1 and 9, shown in black. \textit{Bottom panel}: Distribution of the number of millimeter flux data points available for each \textit{z}-GAL source shown in gray; all sources have at least 2, and up to 10, measurements. The cyan histogram shows the number of millimeter measurements used in the fitting procedure (below 1~mm in the rest-frame; see Section \ref{section:opt-thin-model}) with a median of 5 data points sampling the Rayleigh-Jeans of the \textit{z}-GAL sources. The indigo-lined histogram shows the number of data points with signal-to-noise ratio > 3.}
    \label{fig:zgal-mm-sampling}
\end{figure}

\subsection{Continuum flux densities of the \textit{z}-GAL sample}\label{section:flux-sample}

\begin{figure*}[ht!]
    \centering
    \includegraphics[width=0.8\textwidth]{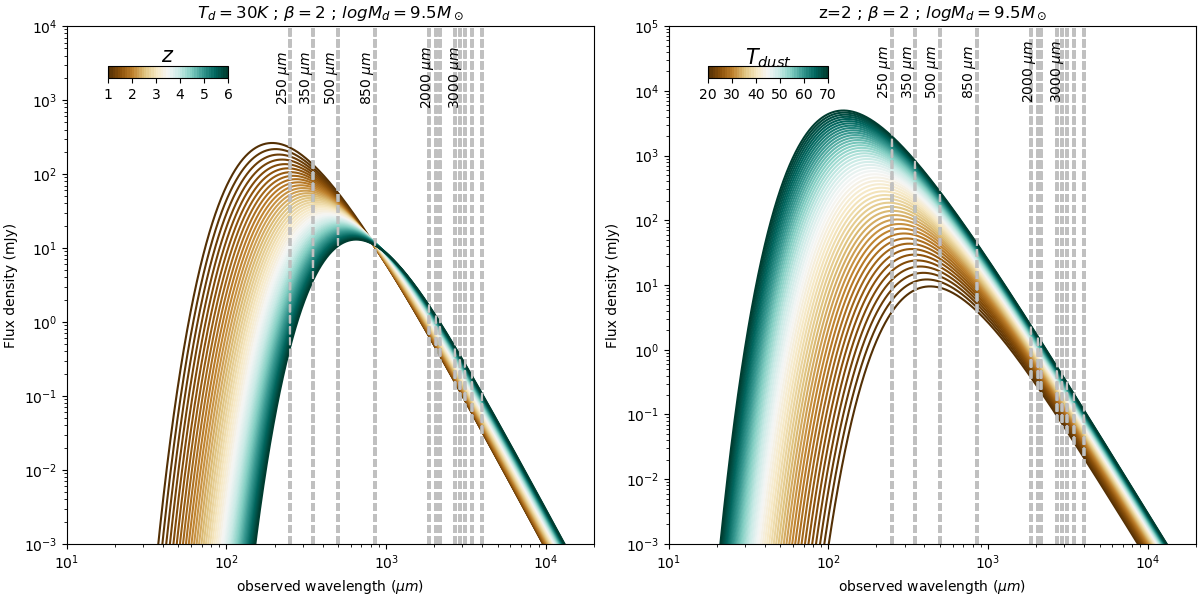}
    \caption{Coverage of the \textit{z}-GAL sample flux densities as a function of the physical properties of a galaxy. Left: When varying the redshift, we lose peak sensitivity at lower redshifts, which limits the recovery of the true $T_{\rm dust}$ of the SED. Right: With an intrinsically higher $T_{\rm dust}$, we lose peak sensitivity as well and we no longer recover the true temperature of the SED.}
    \label{fig:recover_parameter_issues}
\end{figure*}

The continuum flux densities are listed in Table \ref{tab:noema-fluxes} along with their uncertainties. The uncertainties listed are the ones estimated in the data reduction process, but in our subsequent analysis (Sections \ref{section:opt-thin-model} and \ref{section5:gmbb}), we add a 10\% error in quadrature to each flux density uncertainty to account for calibration uncertainty. Flux densities with S/N $\geq$ 3 are considered detections, and non-detections (S/N < 3) are listed with a \textquote{<} sign followed by the measured flux and uncertainty. The flux densities of the multiple sources detected in the NOEMA field that are unresolved by \textit{Herschel} and SCUBA-2 are listed individually and, in the case where sources are at the same \textit{z} (a total of 15 sources), as the sum of all components (see Notes of Table \ref{tab:noema-fluxes}). A total of 12 sources from 137 were discarded from our further analysis, 9 of which are multiple sources in the field of view at different redshifts (including 3 sources from the Pilot Program, i.e., HerBS-43, 70, 95), another two sources have tentative redshifts (HerS-19 and HerBS-82), and HerS-9 is a foreground galaxy at \textit{z} = 0.853. This brings the total number of sources retained in order to study their continuum and to extract dust-emission  properties   to 125. The selected sources span the redshift range 1.3 < \textit{z} < 5.4  (Fig. \ref{fig:redshift-distribution-zgal}).

In the top panel of Fig. \ref{fig:zgal-mm-sampling}, we summarize the signal-to-noise ratios of each of the wavebands, where the continuum coverage of the \textit{z}-GAL sample ranges from the FIR to the millimeter, with all sources having \textit{Herschel}-SPIRE 250, 350, and 500 $\rm \mu m$ flux densities with 7 < S/N < 10. The 850 $\rm \mu m$ SCUBA-2 flux density is only available for the HerBS sources \citep{bakx2018, bakx2020b} with 1.5 < S/N < 7 \citep[see][for a table of the \textit{Herschel} and SCUBA-2 flux densities]{cox2023}. The NOEMA 2 and 3~mm flux densities for all sources have been measured with 1 < S/N < 9. For two sources in the sample, additional data in the 1~mm band are available, namely for HeLMS-17 (see paper I) and HerBS-89 \citep{berta2021close}. In the bottom panel of Fig. \ref{fig:zgal-mm-sampling}, we summarize the \textit{z}-GAL millimeter sampling, illustrating the uniqueness of the $z$-GAL sources that have up to ten measured continuum flux densities in the 2 and 3~mm bands. In our further analysis, we use data below 1~mm in the rest-frame (see Section \ref{section:opt-thin-model-fitting-method}) so that the effective number of flux density measurements varies between 2 and 8 data points in the millimeter, where most of the sources have two or more detections (S/N > 3). 

\section{Modeling the optically thin dust emission}\label{section:opt-thin-model}
The main goal of this study is to estimate the properties of the dust continuum emission of the \textit{z}-GAL sample that covers the observed wavelength range from 250 $\rm \mu m$ to 3~mm. In this section, we present the dust emission model, the modified blackbody (MBB), used to fit the cold dust component between rest-frame wavelengths of 50 and 1000 $\rm \mu m$, excluding the warm dust and radio (free-free and synchrotron) emission, respectively. Our main focus is on the optically thin approximation of the MBB, which is presented in Section \ref{section:opt-thin-model-fitting-method}, which serves as a basis for the \textit{z}-GAL sample results presented in Section \ref{section:zgal_opt_thin}. We also study the robustness of the sampling and discuss parameter estimation in detail in Section \ref{section:mock-data-thin}.

\subsection{Optically thin modified blackbody}\label{section:opt-thin-model-fitting-method}
The thermal dust emission is most simplistically described by a single-temperature MBB. This is the solution to the radiative transfer equation where dust grains are in thermal equilibrium with the radiation field with an emergent flux density:

\begin{ceqn}
\begin{equation}\label{eq:general_mbb}
        S_{\nu_o} = \mu \frac{\Omega}{(1+z)^3} \epsilon_{\nu_r} B_{\nu_r}(T_{\rm dust}) 
,\end{equation}
\end{ceqn}

\noindent where $B_{\nu}(T_{\rm dust})$ is Planck's blackbody radiation, $\Omega$ is the solid angle of emission: $\Omega = (1+z)^4 A/D_L^2$, A, and D$_L$ are the physical area of the galaxy and luminosity distance, respectively, $\mu$ is the magnification factor for the sources that are gravitationally lensed, $\epsilon_{\nu}$ is the emissivity coefficient at rest-frame frequency $\nu$: $\epsilon_{\nu} = (1-e^{-\tau_{\nu}})$, and $\tau_\nu$ is the frequency-dependent optical depth that is written in terms of the dust mass surface density $\Sigma_{\rm dust}$ and the mass absorption coefficient \citep[e.g.,][]{beelen2006}: 

\begin{ceqn}
\begin{equation}\label{eq:tau_kappa}
    \tau_\nu = \kappa_\nu \Sigma_{\rm dust}  
,\end{equation}
\end{ceqn}

\noindent where $\kappa_\nu$ can be approximated by a power law at our wavelength of interest: $\kappa_\nu = \kappa_0 (\nu / \nu_0)^{\beta}$ with $\kappa_0$ being the mass absorption coefficient at frequency $\nu_0$ and $\beta$  the dust emissivity index. $\Sigma_{\rm dust} (= M_{\rm dust}/A$) is the dust mass density, with $ M_{\rm dust}$ being the dust mass. We adopt the values computed by \cite{draine2014kappa} for the mass absorption coefficient: $\nu_0 = 353$GHz (i.e., $\lambda_0 = 850 \rm \mu m$), $\kappa_0 = 0.047$ $m^2/kg$. \citet{berta2016} and \citet{bianchi2013} pointed out that a correction is needed when using the tabulated $\kappa_\nu$ values and fitting with $\beta \neq 2.08$ due to a dependence on the normalization ($M_{\rm dust}$ in this case), and so we correct $M_{\rm dust}$ by a factor of $(850\rm \mu m/500\rm \mu m)^{\beta - 2.08}$. 

\begin{figure*}[ht!]
    \centering
    \includegraphics[width=\textwidth]{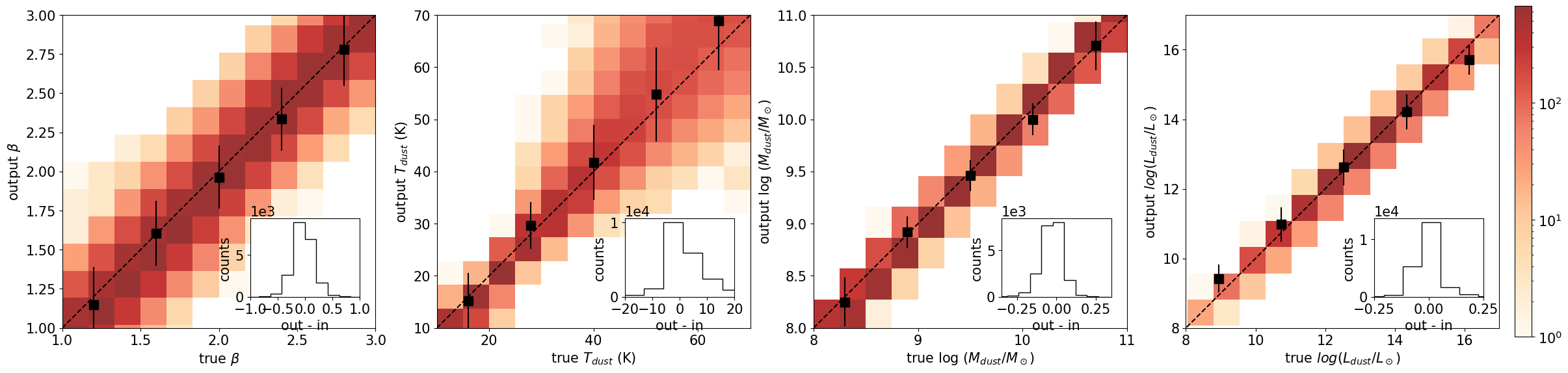}
    \caption{Input (\textit{true}) vs. output dust properties of the mock-generated catalog of sources. The black line is the identity line and the black dots show the median within the bins and the error bars represent the standard deviation in these bins. The boxes in the lower right corner describe the accuracy of the fitting method when using the optically thin MBB for this mock sample.}
    \label{fig:mock-thin-hist2d}
\end{figure*}
 
When assuming an optically thin medium in the observed frequencies, the optical depth $\tau_\nu \ll 1$, and therefore the emissivity coefficient will simplify to: $\epsilon_\nu = (1-e^{-\tau_\nu}) \approx \tau_\nu$ using Taylor's expansion. The observed flux can therefore be written:

\begin{ceqn}
\begin{equation}\label{eq:optically-thin-mbb}
    S_{\nu_o} = \mu \frac{(1+z)}{D_L^2} M_{\rm dust} \kappa_{\nu_r} B_{\nu_r}(T_{\rm dust}) 
.\end{equation}
\end{ceqn}

\noindent Hereafter, we refer to this form as MBB, which we correct for the cosmic microwave background (CMB) effects throughout this study, as described by \citet{da2013effect}.

The FIR luminosity of the cold dust component is estimated by integrating the fitted modified blackbody between rest-frame 50 and 1000 $\rm \mu m$: 

\begin{ceqn}
\begin{equation}\label{eq:luminosity-integration}
        L_{\rm FIR} = 4\pi D_L^2 \int_\nu S_{\nu} d\nu
.\end{equation}
\end{ceqn}

\subsection{SED fitting and parameter estimation}\label{section:mock-data-thin}
With the frequency coverage of the \textit{z}-GAL sample, especially along the RJ tail, we cover the millimeter wavebands with a varying number of flux measurements along with the FIR \textit{Herschel} and, when available, SCUBA-2 flux densities. This confers an advantage in that it allows us to measure the dust parameters with greater accuracy, especially $\beta$, which is governed by the RJ tail, without the need to fix them to an average standard (e.g.,  $\beta$ is usually fixed to 1.5 or 2, e.g., \citealt{magnelli2012herschel}). Here, we create a mock catalog using the MBB of simulated flux densities that mimic those in our sample in order to explore how well we can retrieve the dust parameters $\beta$, $T_{\rm dust}$, and $M_{\rm dust}$, which  are recovered from the SED fit of Eq. \ref{eq:optically-thin-mbb}, as well as $L_{\rm FIR}$, which is the integrated area under the curve using Eq. \ref{eq:luminosity-integration}.

To generate a mock catalog of flux densities using the MBB, we choose a range of initial parameters, with $\beta$ varying between 1 and 3, $T_{\rm dust}$ between 15 and 70 K, $M_{\rm dust}$ between $10^{8}-10^{11} M_{\odot}$, and a redshift range chosen between 1.3 and 6 that is representative of the \textit{z}-GAL sample (see Fig. \ref{fig:redshift-distribution-zgal}). Given the dust parameter ranges mentioned, we generate a grid of $12^4$ mock sources, without any prior correlation between the dust parameters and uniformly spread over the parameter space. Using Eq. \ref{eq:optically-thin-mbb}, the mock flux densities are then computed ---accounting for the CMB effect--- with these parameters and we add a random error bar (equivalent to 1$\sigma$) for each waveband, which was chosen to be similar to the real data signal-to-noise ratio (S/N), where the  S/N varies between 6 and 10  for the \textit{Herschel} flux densities, between 1.5 and 7 for the SCUBA-2 flux density, and between 1 and 9 for the NOEMA flux densities. To mimic real observations, we perturb the flux densities on a normal distribution with $1\sigma$, which is equivalent to the error bar of each flux point. The final mock catalog covers the 250, 350, and 500 $\rm \mu m$ \textit{Herschel} flux densities, the 850 $\rm \mu m$ SCUBA-2 flux, and five millimeter data points that vary between 79.76 and 162.32 GHz (or 3.7 and 1.8~mm). 

We then fit the SEDs to estimate the output dust parameters of the mock galaxies using the MBB. Throughout this paper, we use the EMCEE package \citep{emcee}, which is a Markov Chain Monte Carlo sampler, to fit the SEDs using the MBB corrected for CMB effects. We introduce the MCMC sampler with flat priors for each dust parameter:  (i) 0 < $T_{\rm dust}$ < 100 K, (ii) 0 < $\beta$ < 4, and (iii) $10^5$ < $M_{\rm dust}$ ($M_{\odot}$) < $10^{15}$. Given the flux densities with their uncertainties, we run the code with 150 walkers and 2000 steps with a 1000 burn-in phase and verify the chains converged (N$_{steps}$ > 10 $\times$ auto-correlation time). In the following subsections, we present a detailed study of how well the dust parameters are recovered (focusing mainly on the dust emissivity index in Section \ref{section:beta-recovery} and dust temperatures in Section \ref{section:temp-recovery}) and the impact of different aspects (redshift, intrinsic dust temperature, and S/N) on the accuracy of their estimates.
 
\subsubsection{Dust emissivity index \texorpdfstring{$\beta$}{}}\label{section:beta-recovery}
We find good agreement between \textit{true} and output $\beta$ values as shown in Fig. \ref{fig:mock-thin-hist2d} (first panel on the left) where the mean of each bin, represented by the black points, lies on the one-to-one relation, and is recovered with a standard deviation of between (output - input) $\pm 0.18$. We also find good agreement in recovering $\beta$ irrespective of the redshift or the recovery of the dust temperatures. Within different redshift bins, as shown in Fig. \ref{fig:mock-thin-hist2d-redshift-binned}, we see changes in the accuracy and scatter of the temperature, but we do not see any direct impact on the $\beta$ estimation.

\begin{figure}[ht!]
    \includegraphics[width=0.49\textwidth]{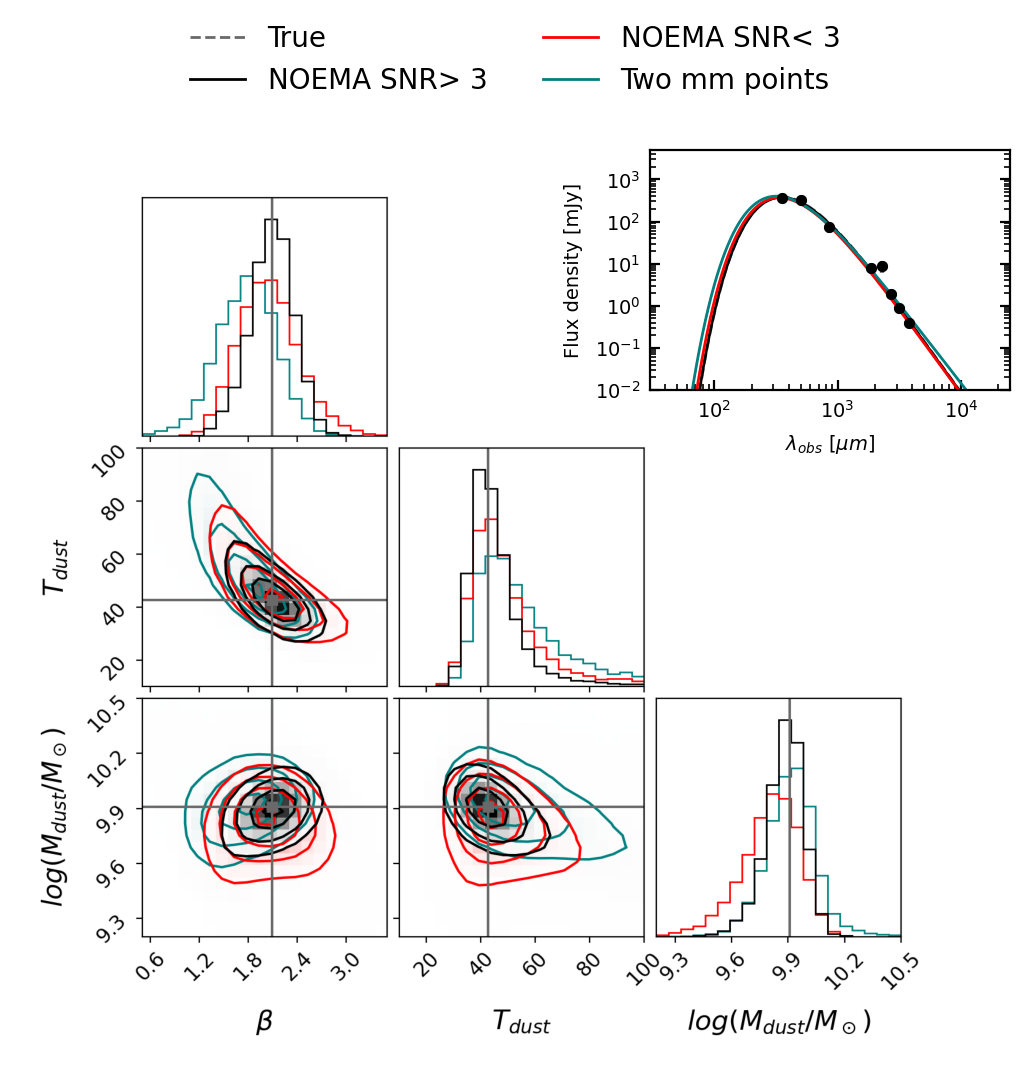}
    \caption{Corner plot of the posterior distribution and SED fit of a mock galaxy at redshift $z$ = 4.29. The posterior distribution shows dust parameter estimates: (i) with five flux points in the millimeter domain with S/N > 3 in black, (ii) five flux points in the millimeter domain with S/N < 3 in red, and (iii) two flux points in the millimeter domain with S/N > 3 in teal; over-plotted in gray are the \textit{true} values.}
\label{fig:corner-ot-mbb-mock-analysis}
\end{figure}
 
To look in more detail at the estimation of $\beta$, we show the posterior likelihood distributions of estimated parameters of a mock source chosen at \textit{z} = 4.29 in Fig. \ref{fig:corner-ot-mbb-mock-analysis}. We still find a slight degeneracy between dust temperature and $\beta,$ even with a large number of data points sampling the RJ tail. In Sect. \ref{section:discussion-beta-temp-relation}, we argue that this degeneracy points to an intrinsic origin between these two dust parameters. However, Fig. \ref{fig:corner-ot-mbb-mock-analysis} clearly shows that the dust emissivity is well constrained and independent of the peak sampling (or in other words, independent of the temperature estimate). Nevertheless, it is dependent on the quality of the data used to minimize the width of the posterior distribution (shown in blue for S/N< 3 and black for S/N> 3 in Fig. \ref{fig:corner-ot-mbb-mock-analysis}), an effect that was also described by \citet{shetty2009effect}. Unlike \textit{z}-GAL sources, many previous studies had a sampling of no more than two data points along the RJ tail, which results in a larger degeneracy between dust temperature and $\beta$ (illustrated in red contours in Fig. \ref{fig:corner-ot-mbb-mock-analysis}) especially when the peak is less well sampled. This result further underlines the robustness of the \textit{z}-GAL coverage for constraining the dust emissivity index $\beta$.

\subsubsection{Dust temperature \texorpdfstring{$T_{\rm dust}$}{}}\label{section:temp-recovery}
The dust temperature output is, on average, in good agreement with the true temperature and we recover it to within $\pm 6.8$ K, as shown in the second panel of Fig. \ref{fig:mock-thin-hist2d}. The significant dispersion in the estimates of $T_{\rm dust}$ is primarily related to the intrinsic temperature; that is, the higher the temperature, the lower the accuracy as the peak shifts towards shorter wavelengths, as shown in the right panel of Fig. \ref{fig:recover_parameter_issues}. Another important factor is the redshift, where the peak becomes poorly constrained at lower redshifts; given the wavelength coverage of  our sample, we lose sensitivity and accuracy in recovering $T_{\rm dust}$ as illustrated in the left panel of Fig. \ref{fig:recover_parameter_issues} and in the redshift-binned histogram in Fig. \ref{fig:mock-thin-hist2d-redshift-binned}. 

\begin{figure}[ht!]
    \centering
    \includegraphics[width=0.5\textwidth]{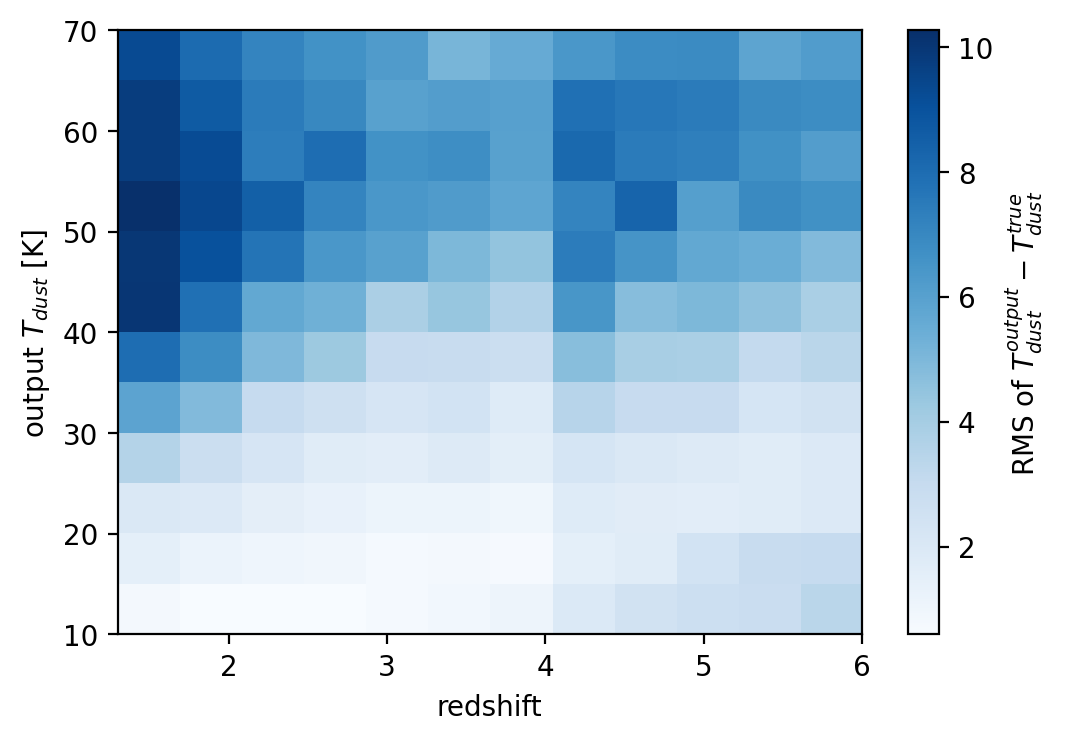}
    \caption{Root mean square of (output - \textit{true}) $T_{\rm dust}$ distribution as a function of the estimated dust temperatures and redshift.}
    \label{fig:rms-tdust-z-otmbb}
\end{figure}

Figure \ref{fig:rms-tdust-z-otmbb} displays the root mean square (RMS) of the $\Delta T_{\rm dust}$(= $T_{\rm dust}^{output} - T_{\rm dust}^{true}$) distribution as a function of the estimated $T_{\rm dust}$ and redshift. Coupling both redshift and temperature, $T_{\rm dust}$, is best recovered between redshifts 2 and 4, and with more precision toward higher temperatures. However, this precision slightly drops at \textit{z} > 4 because of the effect of our fitting method, where we remove any contribution from warm dust ($\lambda_{rest}<50 \rm \mu m$); that is, the \textit{Herschel} 250 $\rm \mu m$ flux point falls below that limit starting at \textit{z} $\geq$ 4, resulting in a lower accuracy in recovering $T_{\rm dust}$. 

Finally, the dust mass estimates are found to be in good agreement with the mock values and are recovered to within $\pm 0.066$ dex. In the MBB, this parameter acts as a normalization in the equation, but is also dependent on the dust temperature estimates. However, with good constraints on the dust temperature, we are able to obtain robust dust mass estimates. Furthermore, the dust luminosity that we recover is in very good agreement with mock values (to within $\pm 0.056$ dex on average), as shown in the right panel of Fig. \ref{fig:mock-thin-hist2d}, and is determined by the data rather than the fitting method. 

We performed a final check to measure the effect of the lack of a flux density measurement at 850$\rm \mu m$. Only the HerBS sources in the \textit{z}-GAL sample were observed at this wavelength with SCUBA-2 \citep{bakx2018}, whereas the HeLMS and HerS sources lack an 850 $\rm \mu m$ measurement (representing a total of 56 sources). To this effect, we fit the SEDs of the mock catalog generated using the same strategy but removing the flux density at 850 $\rm \mu m$; however, we do not find any significant change in the output. The standard deviation of the (output $-$ input) parameters is only slightly higher, with $\pm 0.2$ for $\beta$, $\rm \pm 7.07 \,K$ for $T_{\rm dust}$, $\pm 0.069$ dex for $M_{\rm dust}$, and $\pm 0.059$ dex for $L_{\rm FIR}$. Therefore, we do not distinguish between sources with and without this measurement because NOEMA data cover observations along the RJ tail, which allows us to constrain the dust parameters well.

\section{\textit{z}-GAL optically thin dust properties}\label{section:zgal_opt_thin}
In this section, we present the continuum dust properties of the \textit{z}-GAL sources using the MBB approach, which will provide the basis for the analysis presented in the series of \textit{z}-GAL papers, in particular, their infrared luminosities that are used to derive the physical properties of these galaxies \citep{berta2023}. The dust parameters are derived by fitting the SED of each source following the method described in Section \ref{section:mock-data-thin}.

\subsection{Results and quality of data}
\begin{figure}[ht!]
    \centering
    \includegraphics[width=0.5\textwidth]{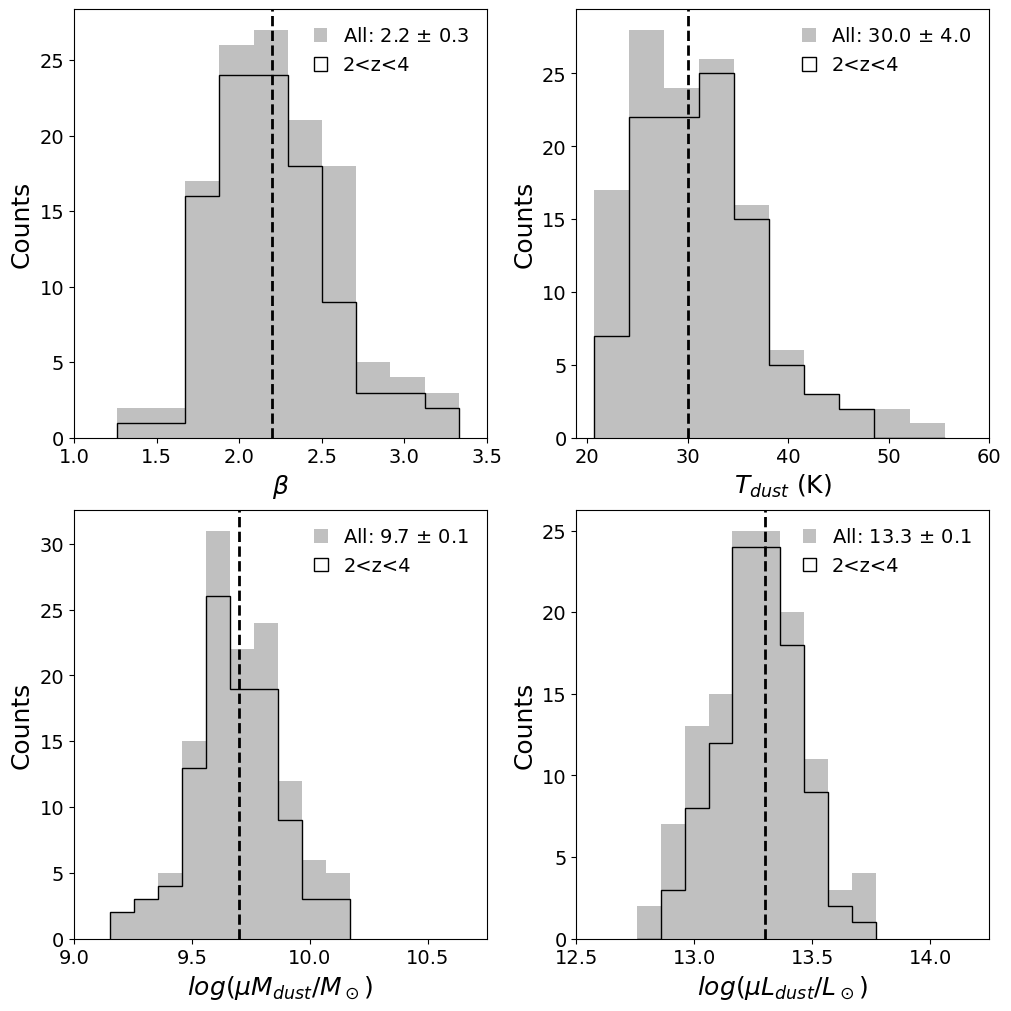}
    \caption{Distribution of the dust properties  of the \textit{z}-GAL sources derived from the MBB shown in gray. The median value of each parameter is plotted with a dashed line whose value is given in the upper right corner of each plot along with the median absolute deviation. The black contour histogram shows the distribution of each parameter within the redshift subset 2 < \textit{z} < 4.}
    \label{fig:opt-thin-properties-histogram}
\end{figure}

In Fig. \ref{fig:opt-thin-properties-histogram}, we summarize the distribution of the dust parameters for all the sources of the \textit{z}-GAL sample selected for this study (gray histogram). The results show a wide range of values around the median values of the dust emissivity index $\beta$ (2.16 $\pm 0.27$), the dust temperatures (30 $\pm$ 4.37 K), the apparent dust masses (0.58 $\pm$ 0.16) $\times\ 10^{10} M_\odot$, and the inferred apparent dust luminosities (1.8  $\pm$ 0.52) $\times\ 10^{13} L_\odot$. With the given \textit{z}-GAL sampling at the peak and the RJ tail of the SED, we find that on average, we recover robust dust parameters, as shown by the mock results, especially the dust masses and the infrared luminosities across all redshift bins. However, the accuracy of $T_{\rm dust}$ changes within each bin, as does the accuracy of the dust emissivity index $\beta$. 

\begin{figure}[ht!]
    \centering
    \includegraphics[width=0.5\textwidth]{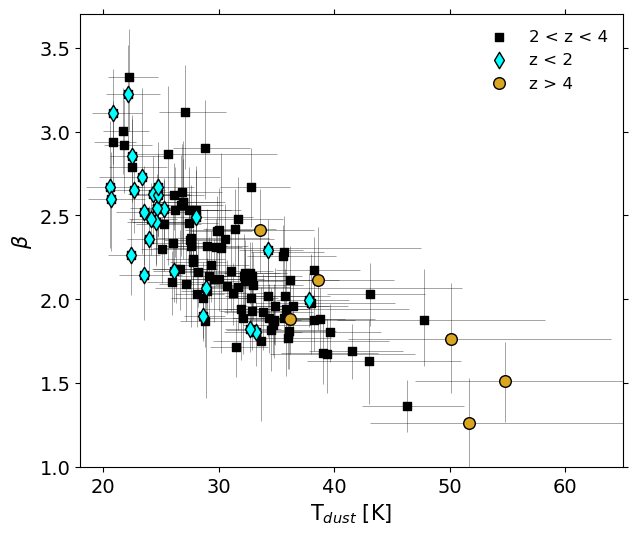}
    \caption{ $\beta - T_{\rm dust}$ relation and precision in estimating these parameters. The black squares, blue diamonds, and
yellow circles show the output for sources in the redshift ranges of 2 < \textit{z} < 4,   \textit{z} < 2,  and \textit{z} > 4, respectively.}
    \label{fig:zgal-beta-temp-uncertainties}
\end{figure}

One of the great advantages of the \textit{z}-GAL sample is the excellent coverage in the millimeter domain, which allows us to constrain the dust emissivity index with high accuracy. Indeed, we recover $\beta$ to within 10\% for most of the sources (the relative uncertainty varies in the range of approximately $7\%-15$\%); however, we still find some sources whose $\beta$ is less well constrained, with a relative uncertainty of $\Delta \beta / \beta$ > 15\%. In Section \ref{section:beta-recovery}, we show that the main reasons for a low precision in estimating $\beta$ come from a low S/N or poor-quality measurements, which also increases the degeneracy with dust temperatures. We show the distribution of estimated dust emissivity indices versus dust temperatures for our sample in Fig. \ref{fig:zgal-beta-temp-uncertainties}, where a total of 14 sources show a poorly constrained $\beta$. 

\begin{figure*}[ht!]
    \includegraphics[width=\textwidth]{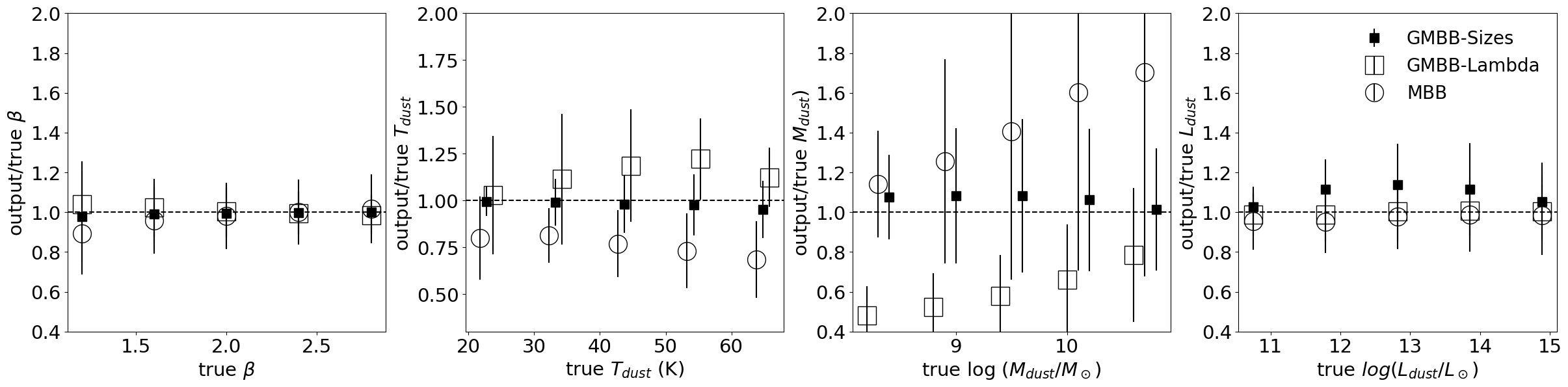}
        \caption{Ratio of output and {true} parameters to the {true} dust parameters of the GMBB mock-generated catalog of sources. The median within each bin is shown as a black filled square, an empty square, and an empty circle for the results fitted with GMBB-Sizes, GMBB-Lambda, and the MBB, respectively, with the error bars representing the standard deviation within each bin. With the \textit{z}-GAL sampling, $\beta$ is recovered very well with all methods. Fitting with GMBB-Sizes can overestimate the inferred luminosity by $\sim$ 10\%\ - 15\% due to an unrecovered source size (see Section \ref{section:gmbb-mock}). Fitting with GMBB-Lambda can overestimate the dust temperatures by $10\%-20$\% and consequently the dust masses become underestimated by up to 40\%. When fitting with MBB, the dust temperatures can be underestimated by $20\%-30$\% and consequently the dust masses become overestimated by up to 75\%.}
\label{fig:density_hist2d_gmbb}
\end{figure*}

\begin{figure*}[ht!]
    \centering
    \includegraphics[width=0.8\textwidth]{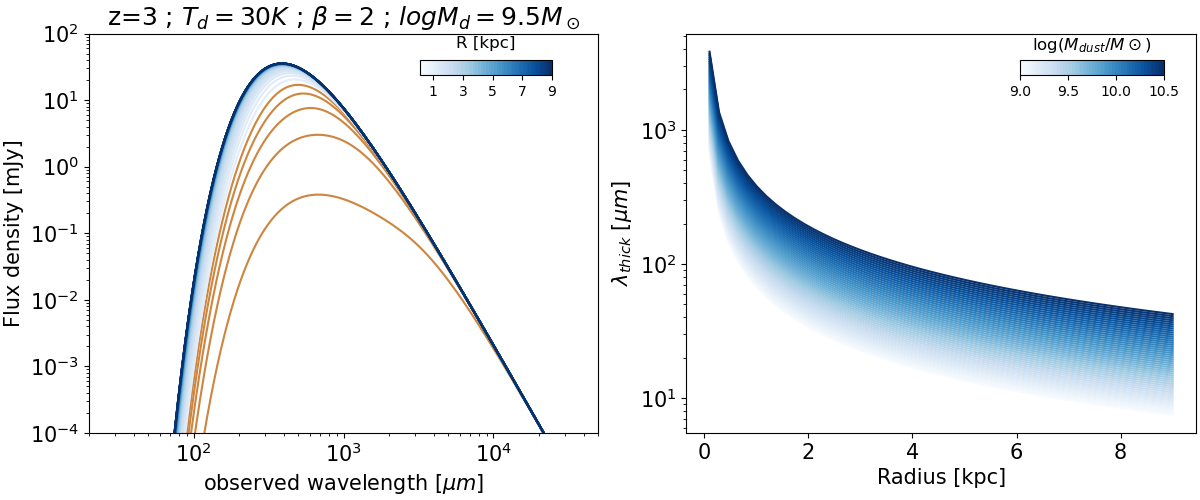}
    \caption{Effect of the source-size variation of SED fitting. Left: Effect of source size variation on the SED of a galaxy. Orange indicates the variation when the size R < 1 kpc, and  a gradient
of blues indicates the variation when R > 1 kpc. On the right side, we plot $\lambda_{thick}$ (wavelength at which optical depth is unity) versus source size (R), color coded according to dust mass. These two figures show that for source sizes R < 1 kpc, there is a large effect on the peak of the SED, where the medium is optically thick, reaching higher values of $\lambda_{thick}$.}
    \label{fig:effect-of-radius-variation}
\end{figure*}

The dust emissivity indices and dust temperatures clearly show an intrinsic relation between these parameters. In order to fully understand whether or not the uncertainties on $\beta$ are related to $T_{\rm dust}$, we looked at the dust temperature estimate and its accuracy, which we categorize into three subsamples. The first subsample is the most dominant in our sample, and covers the redshift range 2 < \textit{z} < 4, amounting to 101 of the total sample. This redshift subset gives the most robust constraints on parameter estimates, especially $T_{\rm dust}$ up to intrinsic temperatures of 40 K. Our results show that only five sources have temperatures above this limit, and that they vary between 42 and 47 K, as shown in Fig. \ref{fig:zgal-beta-temp-uncertainties}. After inspecting their posterior distribution in Fig. \ref{fig:corner-bad-sources} (particularly, HerBS-78, and HerBS-204), we find that the degeneracy is small and we find $\beta$ to be well constrained for those sources. We also checked whether the use of upper limits (as measured flux densities) biases the high $\beta$ values ($\beta \geq$ 3); however, Fig. \ref{fig:beta-corner-plots} shows that excluding them does not have a significant effect on the estimation of $\beta$ where the posterior distributions are overlapping.

The second subset is for sources at \textit{z} < 2, which turned out to be more challenging in constraining $T_{\rm dust}$ beyond 30 K. In our sample, 18 galaxies (at \textit{z} < 2) have estimated temperatures that reach $\sim$ 30 K and are recovered to within 15\% (denoted by blue diamonds in Fig. \ref{fig:zgal-beta-temp-uncertainties}), placing them in the unbiased region. Three sources in this subset show a poor constraint on $\beta$, while two of these (HeLMS-35 and HeLMS-44) have only two millimeter data points sampling the RJ tail, resulting in a lower precision, and HerBS-199 has poor millimeter data quality (low S/N). 

The third subset is for sources at \textit{z} > 4. In our sample, there are six sources within this range with estimated temperatures that vary between 33 and 56 K (shown as yellow filled circles in Fig. \ref{fig:zgal-beta-temp-uncertainties}). Similarly to the first subset, sources with $T_{\rm dust}$ > 40 K show lower precision in estimating dust temperatures. The relative uncertainties on these sources (HeLMS-19, HeLMS-24, and HeLMS-45) are due to a larger degeneracy with $T_{\rm dust}$, which is caused by a poorly sampled peak.

\section{Exploring the general modified blackbody model}\label{section5:gmbb}
Given its simplicity, the most commonly used model is the optically thin approximation, but it is nevertheless important to explore the general form of the modified blackbody (GMBB). Although the GMBB is also a simplification, where a single temperature is assumed, we need to take into consideration the fact that some sources could be optically thick up to $\sim$200 $\rm \mu m$ \citep[e.g.,][]{riechers2017rise,casey2019physical,dudzevivciute2021tracing}. It also has been shown that  $T_{\rm dust}$  tends to be underestimated  by up to $\sim$20 K when using the optically thin model \citep[e.g.,][]{da2021measurements, cortzen2020deceptively}. In this section, we explore the GMBB SED fitting and show the accuracy and the different biases in estimating dust properties. 

We start by modeling the GMBB using Eq. \ref{eq:general_mbb} and the expression of the optical depth in Eq. \ref{eq:tau_kappa}. The observed flux in its expanded form can now be written as:

\begin{ceqn}
\begin{equation}\label{eq:general_mbb_size_expanded}
        S_{\nu_o} = \mu \frac{A}{D_L^2} (1+z) \left( 1 - \exp\left[ - (\nu_r/\nu_0)^\beta \kappa_0 \frac{M_{\rm dust}}{A}\right]\right) B_{\nu_r}(T_{\rm dust}) 
,\end{equation}
\end{ceqn}

\noindent and we refer to this equation as GMBB-Sizes hereafter. The GMBB-Sizes contains an extra parameter, namely the source size, which we set as a free parameter along with $\beta$, $T_{\rm dust}$, and $M_{\rm dust}$.

It is also common to represent the optical depth as $\tau_\nu = (\nu / \nu_{thick})^{\beta}$, where $\nu_{thick}$ is the frequency at which the optical depth is unity \citep{spilker2016alma,draine2006submillimeter}. Substituting this equation of the optical depth into Eq. \ref{eq:general_mbb}, the observed flux can be written as:

\begin{ceqn}
\begin{equation}\label{eq:general_mbb_lambda_thick}
        S_{\nu_o} = \mu \frac{A}{D_L^2} (1+z) \left( 1 - \exp\left[ - (\nu_r/\nu_{thick})^\beta \right]\right) B_{\nu_r}(T_{\rm dust}) 
,\end{equation}
\end{ceqn}

 \noindent and we refer to this equation as GMBB-Lambda hereafter. $\lambda_{thick}$ (= c/$\nu_{thick}$) is usually set to 100 $\rm \mu m$, but estimates go as low as 55 $\rm \mu m$ \citep[][derived a range between 55 and 90 $\rm \mu m$]{simpson2017scuba}, and could reach as high as 200 $\rm \mu m$ \citep{riechers2017rise, riechers2021gadot}. \citet{spilker2016alma} also find a median of 140 $\rm \mu m$ with a range between $\sim$50 and 250 $\rm \mu m$. $\lambda_{thick}$ is dependent on intrinsic dust properties (dust mass absorption coefficient $\kappa_0$ and dust mass surface density $\Sigma_{\rm dust}$) and by equating the two forms of optical depth ($\tau_\nu = (\nu / \nu_{thick})^{\beta}$ and Eq. \ref{eq:tau_kappa}), it can be expressed as

\begin{ceqn}
\begin{equation}\label{eq:lambda-thick}
    \lambda_{thick} = \lambda_0 \left(\frac{\kappa_0 M_{\rm dust}}{A}\right)^{1/\beta}
.\end{equation}
\end{ceqn}

This means that for a given dust mass, $\lambda_{thick}$  decreases as a function of source size (R ; A = 4$\pi R^2$) and the medium becomes optically thin at shorter wavelengths. Taking the median apparent dust mass derived for \textit{z}-GAL sources, namely $M_{\rm dust} = 10^{9.7} M_\odot$, and a median $\beta=2$, the medium becomes optically thin at 33 $\rm \mu m$ for an apparent source size of 5 kpc and at 165 $\rm \mu m$ for a size of 1 kpc. 

\subsection{SED fitting and parameter estimation with GMBB}\label{section:gmbb-mock}
We perform similar tests for GMBB as done for the optically thin mock catalog and explore how well we can constrain the dust parameters when using the general opacity model in both its forms (Eq. \ref{eq:general_mbb_size_expanded} and Eq. \ref{eq:general_mbb_lambda_thick}) to re-estimate the GMBB-generated mock catalog. We also show the degeneracies that arise when using a general model (GMBB-Sizes and GMBB-Lambda) as well as the biases when assuming solely an optically thin approximation by fitting MBB to the GMBB-generated mock catalog. 

\begin{figure}[ht!]
    \centering
    \includegraphics[width=0.5\textwidth]{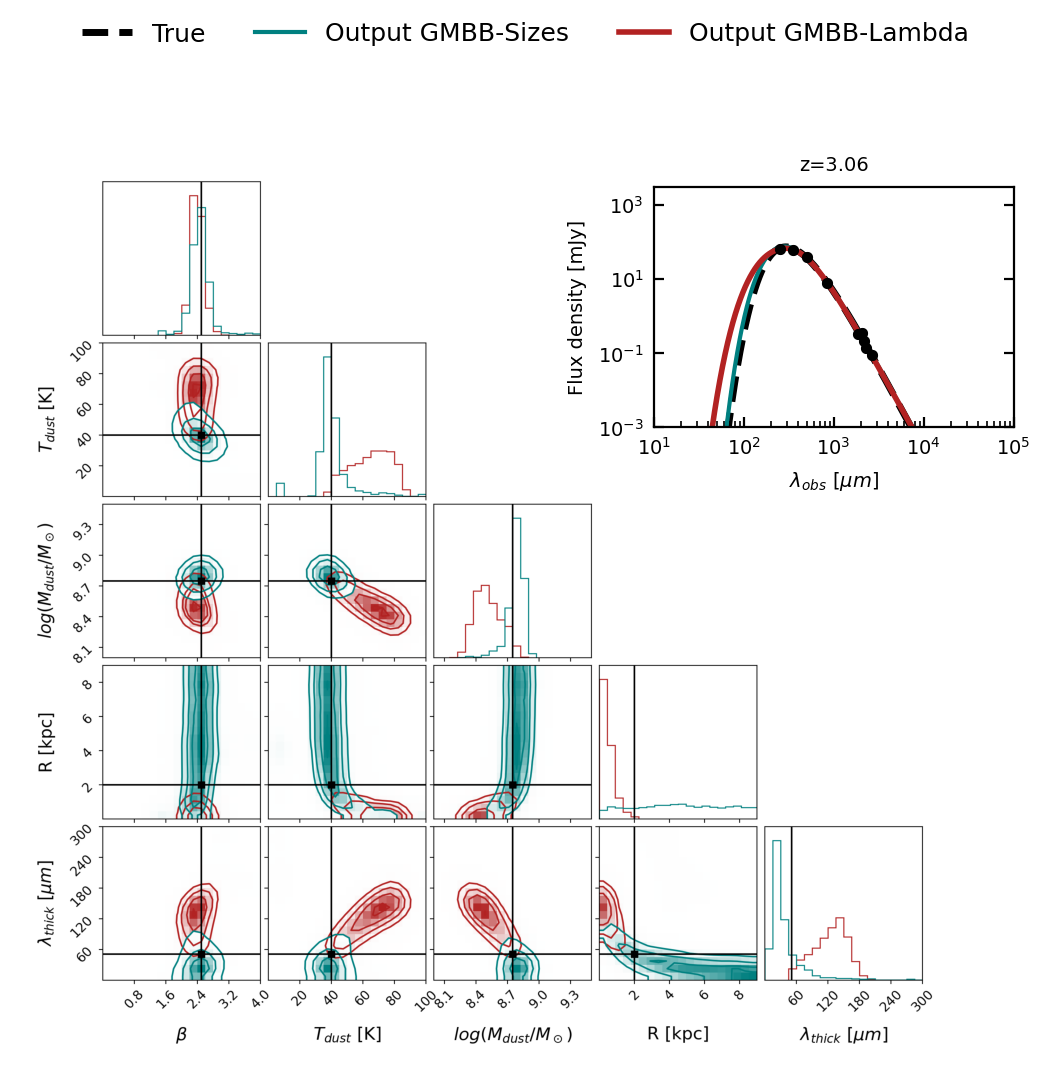}
    \caption{Corner plot of the posterior distribution and SED fit of a mock source at redshift \textit{z} = 3.06. The posterior distribution shows dust parameter estimation using (i) GMBB-Sizes (green) and (ii) GMBB-Lambda (red), which are over-plotted with the true values in black. In this fit, $\lambda_{thick}$ for GMBB-Sizes is estimated using Eq. \ref{eq:lambda-thick} given the posterior distribution of the output parameters ($\beta, M_{\rm dust}$ and A), and $M_{\rm dust}$ distribution for GMBB-Lambda is also estimated with Eq. \ref{eq:lambda-thick} using the posterior distribution of the output parameters ($\beta$, A = 4$\pi R^2$, and $\lambda_{thick}$)}
    \label{fig:corner-gmbb-mock}
\end{figure}

To generate the mock catalog of flux densities, we follow the same procedure as in Section \ref{section:mock-data-thin} and choose the same range of initial parameters. In addition to \textit{z}, $\beta$, $T_{\rm dust}$, and $M_{\rm dust}$, we need to choose a range of the intrinsic size (R) of the emission region. ALMA observations show that the typical FIR/submm dusty galaxies have R = 1 - 5 kpc \citep[e.g.,][]{hodge2015sizes, hodge2016sizes, hodge2019sizes, simpson2015scuba, rujopakarn2016vla, fujimoto2017demonstrating}, which will be the basis of the range chosen here. We then generate a grid of $9^{5}$ mock sources and estimate the corresponding $\lambda_{thick}$ values of the parameter space, which varies between 0.02 and 731 $\rm \mu m$, but we narrow it down to sources within the range of 40 < $\lambda_{thick}$ < 300 $\rm \mu m,$ keeping only the physical values, giving a total of 27,054 mock sources in this catalog. Finally, we compute the flux densities using Eq. \ref{eq:general_mbb_size_expanded}, and we follow the procedure in Section \ref{section:mock-data-thin} to derive their respective uncertainties and scatter to mimic real observations. 

\begin{figure}[ht!]
    \centering
    \includegraphics[width=0.5\textwidth]{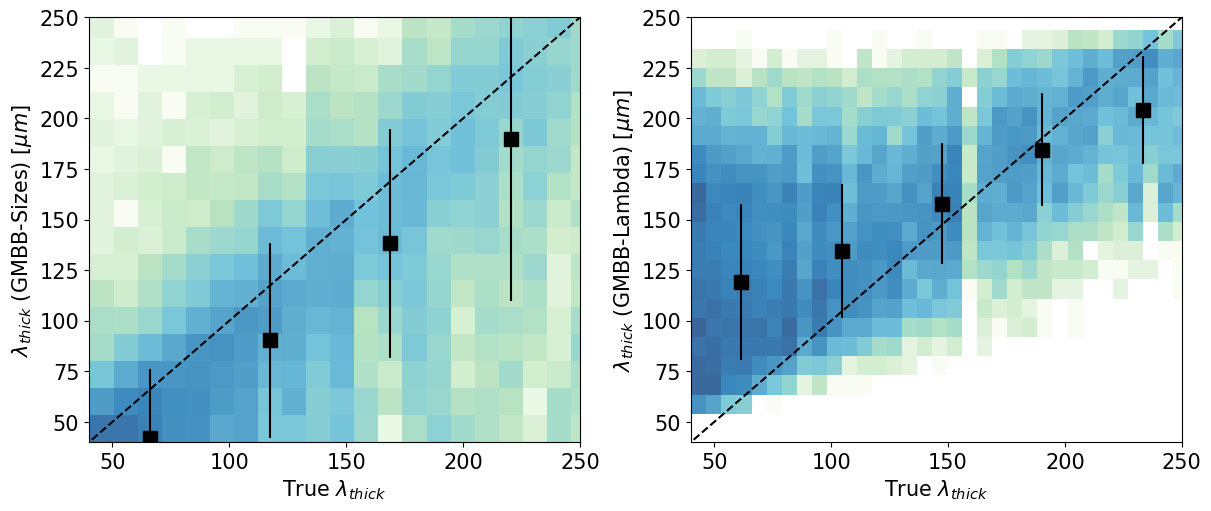}
    \caption{Output versus {true} $\lambda_{thick}$ estimation. The left figure shows the accuracy of estimating $\lambda_{thick}$ when using GMBB-Sizes to fit the SED, {in which it is derived }using Eq. \ref{eq:lambda-thick}. The right figure shows the accuracy of estimating $\lambda_{thick}$ when using GMBB-Lambda to fit the SED, in which it is a free parameter in the equation. The black dashed line is the identity line and the black squares show the median within each bin.}
    \label{fig:lambda-thick-estimated-gmbb}
\end{figure}

Figure \ref{fig:density_hist2d_gmbb} summarizes the output versus true parameters of the SED fits using GMBB-Sizes by setting the size as a free parameter with a prior between 0.1 and 9 kpc in the MCMC sampler. On average, we find good agreement in recovering the dust parameters using GMBB-Sizes where $\beta$ is recovered to within $\pm 0.23$, dust temperatures within $\pm 6.9$K, dust masses within $\pm 0.1$ dex, and dust luminosity within $\pm 0.067$ dex. However, the ratio of output/input $L_{\rm FIR}$ is on average overestimated by $\sim 10\% - 15$\% (the reason for this is explained in the following paragraph). We also looked at the generated catalog split into two categories: where the medium becomes optically thin (i) at lower wavelengths ($\lambda_{thick} < 100$ $\rm \mu m$), and (ii) at higher wavelengths ($\lambda_{thick} > 100$ $\rm \mu m$). GMBB-Sizes results in estimated parameters that agree with the true ones in both scenarios, as shown in the density histogram of figures \ref{fig:mock-thick-hist2d-forced-thin} and \ref{fig:mock-thick-hist2d-forced-thick} (first row). 

\begin{figure*}[ht!]
    \centering
    \includegraphics[width=\textwidth]{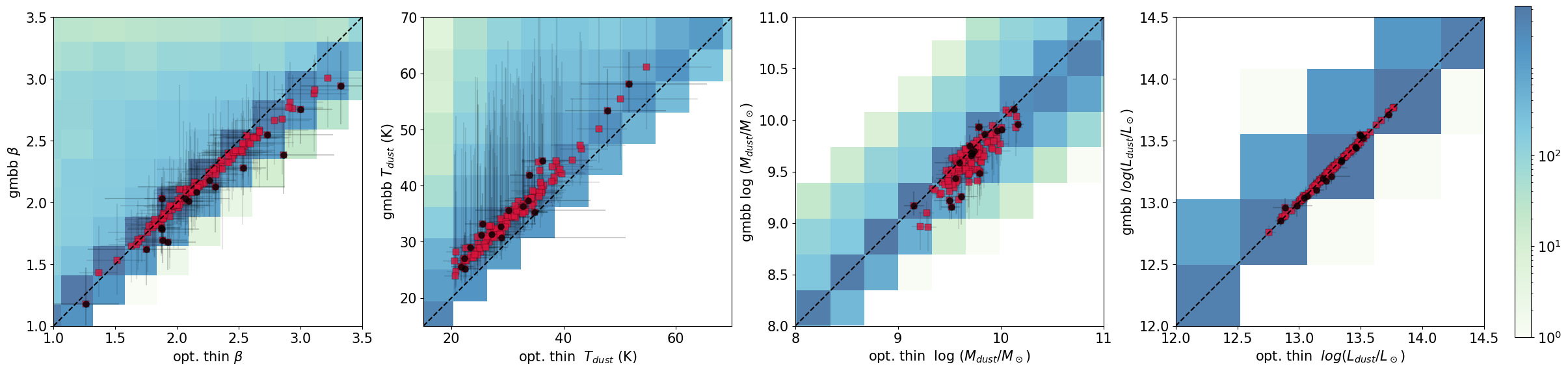}
    \caption{Dust properties of the \textit{z}-GAL sample of sources derived using GMBB-Sizes versus MBB shown in red squares with the derived error bars. Sources denoted in black are some of the problematic sources that we treated with an extra step (discussed in Section \ref{section:zgal-gmbb-results}). For comparison, in the background, we show the output results of the mock catalog when computing the dust parameters using MBB versus GMBB-Sizes. The black dotted line is the identity line. }
    \label{fig:gmbb-vs-thin-free-sizes}
\end{figure*}

We note that having the source size as an extra free parameter gives rise to a new type of covariance between parameters. In Fig. \ref{fig:corner-gmbb-mock}, we demonstrate the posterior distribution of the output parameters of a mock source when using GMBB-Sizes, which shows that we recover the dust emissivity index, the dust temperature, and dust masses  with high accuracy, but not the {true} size where the likelihood shows a flat distribution. Moreover, the output sizes show a flat distribution for almost all cases where R (output) $\sim$ 4.5 kpc (midpoint of the range set for the prior).  This, in turn, affects the FIR luminosity overestimation by $\sim$10\% on average. To explain the flat distribution, we illustrate the variation of the SED fit as a function of source size (R) in Fig. \ref{fig:effect-of-radius-variation}, which shows that for a source size at R > 3 kpc (for a given $M_{\rm dust}$), the medium becomes optically thin at $\lambda_{thick} \sim 50$ $\rm \mu m$. But for R < 1 kpc, the variation at the peak is very large, and the medium is optically thick up to longer wavelengths for all $M_{\rm dust}$, which creates a degeneracy between the source size and dust temperature, as well as between the source size and dust mass. Nevertheless, the derived $\lambda_{thick}$ remains close to the true value with a systematic shift of $\sim 30$ $\rm \mu m,$ as shown in the left panel of Fig. \ref{fig:lambda-thick-estimated-gmbb}, although with a large scatter.

We also fit the SEDs using GMBB-Lambda ---which is a more common way to use the GMBB \citep[e.g.,][]{cortzen2020deceptively, da2021measurements}--- by setting $\lambda_{thick}$ as a free parameter that varies between 50 and 250 $\rm \mu m$. Results show (Fig. \ref{fig:density_hist2d_gmbb}) that dust temperatures are (on average) always overestimated, which is a result of the large degeneracy between $\lambda_{thick}$ and $T_{\rm dust}$, where $\lambda_{thick}$ is poorly constrained (as illustrated in the corner plot of Fig. \ref{fig:corner-gmbb-mock}). With the given sampling of the peak, the likelihood shows a preference for the longer wavelengths, with a consequent rise in dust temperatures by up to 20\%. Especially in the region where the medium becomes optically thin at lower wavelengths ($\lambda_{thick}$ < 100$\rm \mu m$), this method does not prove to be a reliable way to constrain dust parameters, and the output $\lambda_{thick}$ is overestimated by a factor of $\sim 1.5-2$, as shown in Fig. \ref{fig:lambda-thick-estimated-gmbb} (right panel). However, it holds true for an optically thick medium for all output parameters and especially $T_{\rm dust}$, where the output $\lambda_{thick}$ appears to be slightly better constrained between 125 and 200 $\rm \mu m,$ as also shown in Fig. \ref{fig:lambda-thick-estimated-gmbb} (see also density histograms of Appendix \ref{Appendix:forced-thick-thin}). In addition, we see the effect on the dust mass estimate, which is underestimated by 1.5 times\footnote{In the case of GMBB-Lambda, $M_{\rm dust}$ is not a parameter used in the fit, but is estimated afterwards. We use Eq. \ref{eq:lambda-thick} to calculate the dust masses using the derived parameters ($\lambda_{thick}$, $\beta$, $M_{\rm dust}$, and area (A)).}  on average.

Finally, we fit the GMBB-generated catalog using the MBB, which shows on average an underestimation of the dust temperatures, as expected, with an increasing offset towards higher intrinsic temperatures and consequently an overestimation of dust masses, which can reach 1.5 times the true value. These trends become even more evident when we check the split catalog, where the optically thin approximation holds true when $\lambda_{thick}$ < 100 $\rm \mu m$. Otherwise, the dust temperatures are underestimated with an offset of $\sim 10$ K for an intrinsic $T_{\rm dust} =$ 30 K, which increases to an offset of $\sim 25$ K for intrinsic $T_{\rm dust} =$ 60 K (see Figs. \ref{fig:mock-thick-hist2d-forced-thin} and \ref{fig:mock-thick-hist2d-forced-thick} for density histograms).

\subsection{\textit{z}-GAL dust properties using the GMBB}\label{section:zgal-gmbb-results}
Following the results of the GMBB mock analysis in Section \ref{section:gmbb-mock}, the GMBB-Sizes provides more accurate results, especially for $T_{\rm dust}$ (and consequently $M_{\rm dust}$), where we are also able to quantify the optical depth to a certain level of accuracy, unlike with GMBB-Lambda. Based on this result, we estimate the dust properties of the \textit{z}-GAL sample using GMBB-Sizes alone. In Fig. \ref{fig:helms48-corner-plot}, we demonstrate the posterior likelihood for one of the sources, HeLMS-48, with a corner plot. The results of GMBB-Sizes versus MBB are displayed in Fig. \ref{fig:gmbb-vs-thin-free-sizes}, where the red dots represent \textit{z}-GAL sources with their respective uncertainties; over-plotted are the output results of the mock sample, which clearly show the similarity in trends between real and synthetic data. Shown in black circles are the problematic sources with two probable solutions (a double-peaked posterior distribution) because of the degeneracy between the source size and dust temperature, as illustrated in Fig. \ref{fig:effect-of-radius-variation}. For these sources, we constrain the prior additionally by choosing the highest probability values between the dust temperature and dust masses from the first run, then rerun the MCMC sampler with an additional Gaussian prior on dust temperatures. The Gaussian is centered on the maximum probability from the previous chains with a width of 7K.


\begin{figure}[ht!]
    \centering
    \includegraphics[width=0.49\textwidth]{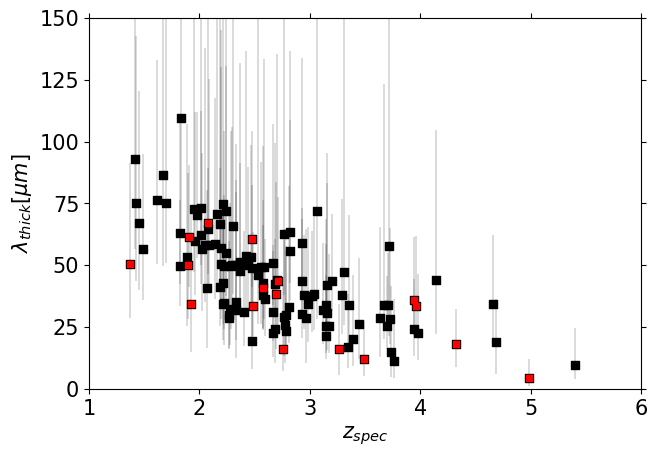}
    \caption{Distribution of derived $\lambda_{thick}$ values using Eq. \ref{eq:lambda-thick} and the output results of GMBB-Sizes as a function of redshift for the \textit{z}-GAL sample of galaxies. In red, we show the problematic sources (see Section \ref{section:zgal-gmbb-results}).}
    \label{fig:lambda-thick-zgal}
\end{figure}

We find $\beta$ to lie on the identity relation when using either method with a slight systematic shift when $\beta \sim 3$ and a median $\beta$ = 2.17 ($\beta^{MBB}$ = 2.16), which underlines once more the robustness of the \textit{z}-GAL sampling in constraining the value of $\beta$. As expected, the dust temperatures are higher when using the GMBB with $\Delta T_{\rm dust}$ = 5 - 15 K where the median $T_{\rm dust}^{GMBB}$ = 37.7 K (7.7 K higher than $T_{\rm dust}^{MBB}$).  With the data in hand, we cannot confirm that the derived temperatures using GMBB are correct until the source size is known. Consequently, the difference in the derived dust temperatures affects the estimated dust masses as shown in the third panel of Fig. \ref{fig:gmbb-vs-thin-free-sizes}, where the MBB-derived $M_{\rm dust}$ are lower than the GMBB-derived ones. Finally, we find that the inferred apparent dust luminosities are in very good agreement between MBB and GMBB-Sizes.

Similar to the mock output results, our sources show a flat distribution when setting the size as a free parameter with R(output) varying between 3 and 4.5 kpc for most sources. In Fig. \ref{fig:lambda-thick-zgal}, we plot the derived $\lambda_{thick}$ values using the output parameters we fit for our SEDs, and we find that it varies between 25 and 75 $\rm \mu m$ for most sources, but with very low precision resulting from a flat distribution in the derived source sizes. As we find similar trends between the mock and the \textit{z}-GAL sources, we cannot distinguish whether our sources are optically thick or thin until we have new high-angular resolution data that will constrain their sizes. 

A decreasing trend with redshift is also observed ---although statistically insignificant because of the large uncertainties--- which appears to be influenced by the trend in $\beta$ values. $\lambda_{thick}$ is dependent on three free parameters (see Eq. \ref{eq:lambda-thick}), namely the dust mass, which does not exhibit a clear evolution with redshift (slope $= -0.05 \pm 0.026$); the source size, which shows a flat distribution with a rather constant value; and the dust emissivity index ($\beta$) with a slight slope of -0.3 $\pm$ 0.014. However, this trend cannot be regarded as real because the source size is not actually constant for all sources. Additionally, the mock results show no clear evidence of the evolution of $\lambda_{thick}$ either.

\section{Discussion}\label{section6:discussion}
In the previous sections, we estimate the dust properties using the optically thin MBB of the \textit{z}-GAL sample of galaxies, which covers a wide range of redshifts from 1.3 to 6. We also explore the general opacity model that has shown that additional information is needed (i.e., the source size) in order to be certain about the dust parameters. In this section, we focus on the results of the MBB and compare them to other samples found in the literature. We first present a comparison of the dust masses and FIR luminosities of the \textit{z}-GAL sample with sources similarly selected from the literature in Sect. \ref{section:discussion-mass-lum}. In Sect. \ref{section:discussion-beta}, we discuss the dust emissivity indices derived in Section \ref{section:zgal_opt_thin}. In Section \ref{section:discussion-temp}, the dust temperature evolution derived for the \textit{z}-GAL sample is estimated. Finally, in Section \ref{section:discussion-beta-temp-relation}, we discuss the $\beta - T_{\rm dust}$ relation. 

\begin{figure}[ht!]
    \centering
    \includegraphics[width=0.49\textwidth]{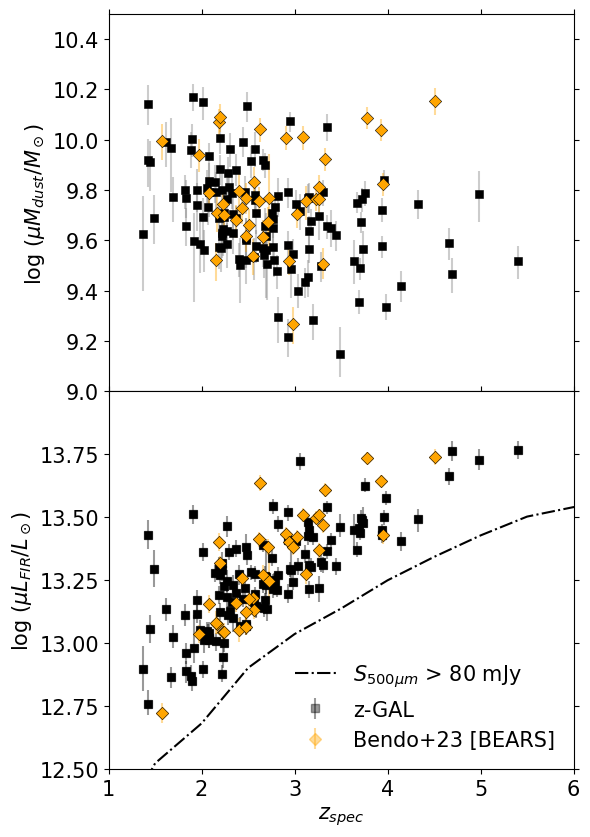}
    \caption{Evolution of \textit{z}-GAL apparent dust mass (upper panel) and apparent FIR luminosity (lower panel)  with redshift for the optically thin MBB, shown in black squares. The BEARS sample \citep{bendo2023bright} is over-plotted in orange diamonds. In the lower panel, the black dashed line shows the lower limit resulting from the 500 $\rm \mu m$ flux density selection.}
    \label{fig:lum-mdust-evolution-redshift}
\end{figure}

\subsection{Dust masses and far-infrared luminosities}\label{section:discussion-mass-lum}
We compare the dust parameters of the \textit{z}-GAL sample with those of 37 gravitationally lensed galaxies selected by \citet{bendo2023bright} from the BEARS sample \citep{urquhart2022bears}. BEARS sources, similarly to the \textit{z}-GAL HerBS sources, were selected from the same \textit{Herschel} catalog \citep{bakx2018}, with $S_{500 \rm \mu m}$ > 80 mJy, and have continuum flux density measurements in the FIR/submm wavelength range including 250, 350, and 500 $\rm \mu m$ \textit{Herschel}-SPIRE, SCUBA-2 850 $\rm \mu m$ data, and two millimeter measurements at 101 and 151 GHz ($\geq 5\sigma$ detections). The sources have spectroscopic redshifts in the range 1.5 < \textit{z} < 4.5. \citet{bendo2023bright} report dust emissivity indices and dust temperature values derived by fitting a MBB using a similar method to ours; for consistency, we refit the BEARS sources to include the dust masses and dust luminosities in our comparison. In Fig. \ref{fig:lum-mdust-evolution-redshift}, we show the distribution of the apparent dust masses and inferred FIR luminosities across different redshifts, which demonstrates the Malmquist bias for the selection at 500 $\rm \mu m$, which approximately samples the peak. The BEARS sources follow the same trend as the \textit{z}-GAL sources, as expected given that they were selected in a similar way to the \textit{z}-GAL sources. At this stage, we cannot say much about the intrinsic trends due to gravitational lensing effects, which will change the intrinsic relation depending on the magnification factors that are still unknown for both the \textit{z}-GAL and BEARS samples.

\subsection{Dust emissivity index \texorpdfstring{$\beta$}{}}\label{section:discussion-beta}
\begin{figure}[ht!]
    \centering
    \includegraphics[width=0.49\textwidth]{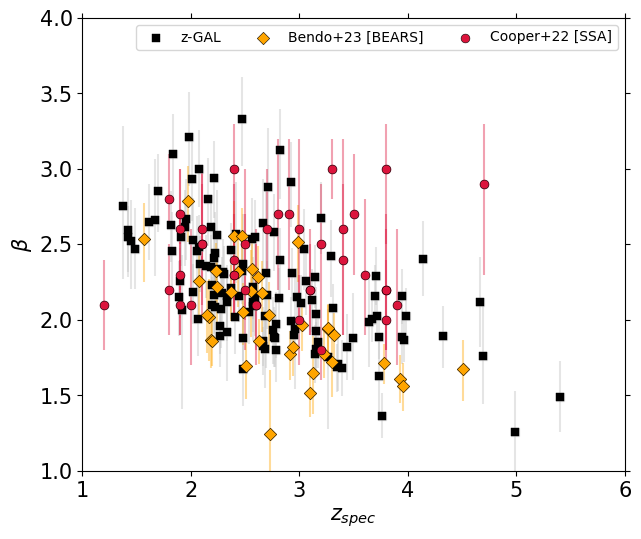}
    \caption{Dust emissivity index ($\beta$) derived using MBB as a function of redshift of the \textit{z}-GAL sample in black squares, \citet{bendo2023bright} BEARS sample in orange diamonds, and \citet{cooper2022searching} SSA sample in red circles.}
    \label{fig:beta-redshift-relation}
\end{figure}

In addition to the BEARS sample, we compare the results from \textit{z}-GAL to the dust emissivities derived by \citet{cooper2022searching} for the bright 850 $\mu$m-selected sample of 39 sources (SSA) for which AzTEC 1.1~mm and ALMA 2~mm flux densities are available with reported detections at S/N > 3 (and a few non-detections). It is important to mention that our comparison to the SSA sample (throughout the discussion) is restricted to the dust emissivity results, as only photometric redshifts (1.2 < $z_{\rm phot}$ < 4.7) are available.
 
Our findings show a distribution of $\beta$ of the \textit{z}-GAL sample that varies between roughly 1.5 and 3 for the majority of the sources, with an average uncertainty of 0.25 and 0.3 using the MBB and GMBB, respectively. However, we derived unusually high dust emissivities, $\beta \geq$ 3, for a total of five sources, which is not common in the literature. While degeneracies exist in the MBB fit in the presence of noise and lack of data, the millimeter coverage of NOEMA has demonstrated that we can reach high precision in estimating $\beta$ and thus lift this degeneracy; in particular, we find that there are only slight to no differences in the $\beta$ estimates when fitting with either MBB or GMBB (see Fig. \ref{fig:gmbb-vs-thin-free-sizes}). Intrinsic degeneracy could also be affecting $\beta$, depending on the grain type or composition and its environment (see our discussion on the  $\beta-T_{\rm dust}$ relation in Section \ref{section:discussion-beta-temp-relation}).

In Fig. \ref{fig:beta-redshift-relation}, we plot the dust emissivity indices as a function of redshift, which agree with the BEARS sample between redshifts 2 and 4. The SSA sample $\beta$ estimates follow a similar distribution to those of the \textit{z}-GAL sources, apart from a few sources with very large uncertainties. In general, we note that we do not find an evolution of $\beta$; for the sources in the populated redshift range (2 < \textit{z} < 3), the dust emissivities follow a similar variation to the total sample. With all three samples across a wide redshift range, dusty galaxies tend to show a wider distribution of $\beta$ values than the range generally assumed in the literature ( 1.5 < $\beta$ < 2, e.g., \citealp{magnelli2012herschel, scoville2016ism}) based on the Milky Way estimates.

The dust emissivity index could potentially provide information about the dust grain composition and/or size \citep[e.g.,][]{ysard2019grains}. However, at this stage, this may be an over-interpretation of the results, because the MBB model is a simplification of the true dust emission, which is spatially heterogeneous and more complex. Nevertheless, the findings reported suggest that we cannot limit $\beta$ to 2 because of the possible variety of grain properties. Moreover, steeper dust emissivities could be due to growth of icy mantles on dust grains. \citet{aannestad1975absorptive} found that silicate grains coated with icy mantles could have $\beta \sim$ 3 and could reach as high as 3.5. \citet{kuan1996three} also reported a high dust emissivity measured for Sgr B2 ($\beta = 3.7 \pm 0.7$) fitted to the millimeter data.

\subsection{Dust temperature \texorpdfstring{$T_{\rm dust}$}{}}\label{section:discussion-temp}
\begin{figure}[ht!]
    \centering
    \includegraphics[width=0.49\textwidth]{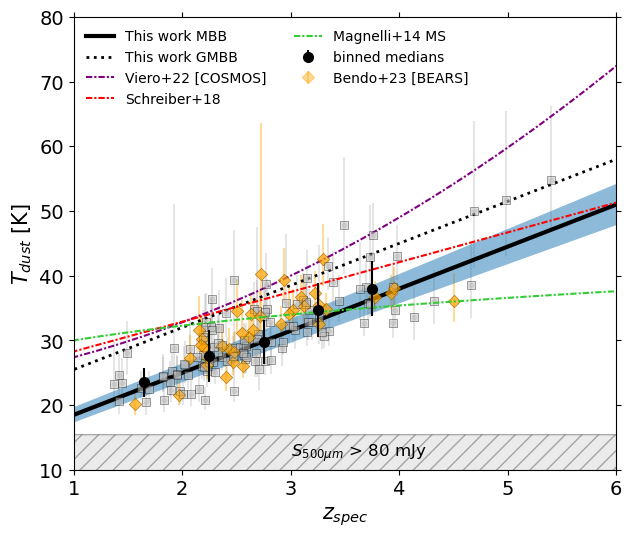}
    \caption{\textit{z}-GAL MBB dust temperatures as a function of redshift shown in gray squares with their respective uncertainties. Best-fit slope estimated for the MBB is plotted with a solid black line and for GMBB with a dotted black line with the confidence interval shaded in blue. The median binned temperatures are plotted in black circles up to redshift \textit{z} = 4 with a bin size of $\Delta$\textit{z} = 0.5. The selection effect is shown with a gray shaded region. For comparison, we over-plot the BEARS-derived MBB dust temperatures by \citet{bendo2023bright} shown as orange diamonds. Evolutionary trends from the literature are also plotted in purple \citep{viero2022evolution}, in red \citep[][extrapolated to $z=6$]{schreiber2018dust}, and in green \citep[][extrapolated to $z=6$]{magnelli2014evolution}.} 
    \label{fig:temp-evolution-with-redshift}
\end{figure}

The distribution of the \textit{z}-GAL dust temperatures ---derived using MBB and as a function of redshift--- is shown in Fig. \ref{fig:temp-evolution-with-redshift}, which exhibits an increasing trend with redshift. Fitting a linear model accounting for intrinsic scatter, $T_{\rm dust}$ = a\textit{z} + b + $\epsilon$, we find a = 6.5$\pm$0.5 K/\textit{z} and b = 12$^{+1.1}_{-2}$ K  with a minimal intrinsic scatter of $\epsilon = $ 1.1$^{+1.1}_{0.74}$ K/\textit{z}\footnote{The linear fit is found using a Bayesian tool (Linmix) for linear models that takes into account the intrinsic scatter.}. A similar trend is found for the temperatures derived with the GMBB, but shifted $\sim 7$ K higher. The shaded region in Fig. \ref{fig:temp-evolution-with-redshift}, representing the 500 $\rm \mu m$ selection effect on dust temperatures, shows no clear evidence of bias towards the trend. 

In general, our result agrees with the increasing $T_{\rm dust}$ versus $z$ evolutionary trends, which are attributed to higher specific star formation rates (sSFRs) at higher redshift \citep{liang2019dust, magdis2012evolving}. However, the rate we find for this evolution is steeper than that found by \citet{magnelli2014evolution} (for main sequence galaxies) and \citet{schreiber2018dust} (for \textit{Herschel}-like galaxy models), but is comparable to that found by \citet{viero2022evolution} (for the stacked COSMOS galaxies) between redshifts 2 and 4 (see Fig. \ref{fig:temp-evolution-with-redshift}). Moreover, all three trends show temperatures that are $\sim 7-10$ K higher than the \textit{z}-GAL ones at lower redshifts. Although more compatible with the GMBB intercept, all three samples used the optically thin approximation to estimate dust temperatures. \citet{schreiber2016lowSFR} demonstrated that more massive star-forming galaxies tend to have lower dust temperatures at lower redshifts as compared to main sequence galaxies, suggesting that they are undergoing a decline in star-formation activity.

\begin{figure}[ht!]
    \includegraphics[width=0.49\textwidth]{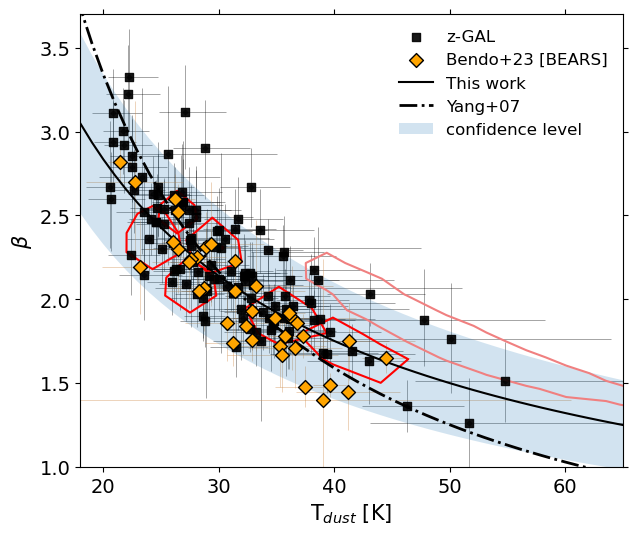}
    \caption{$\beta$ -- $T_{\rm dust}$ distribution for the \textit{z}-GAL sources  (black squares) compared to the BEARS sources \citep[orange diamonds;][]{bendo2023bright}. The solid black line is the fitted relation we find for the \textit{z}-GAL sources with $\alpha = 0.69 \pm 0.04$ (see Sect. \ref{section:discussion-beta-temp-relation}) and the shaded region is the confidence level. We over-plot the trend found by \citet{yang2007} for local LIRGs in black dashed line. The 1$\sigma$ level distribution of the MCMC chains is represented by the red contours for a few sources, that demonstrate a high level constraint. The lighter pink contour demonstrates one of the few cases where the 1$\sigma$ level shows large degeneracy.}
    \label{fig:beta-temp-relation-zgal} 
\end{figure}

On the other hand, there is contradictory evidence in the literature of a lack of evolution of dust temperatures over cosmic time. \citet{drew2022no} argue that dust temperatures do not evolve, based on samples from IRAS, H-ATLAS, and COSMOS surveys at \textit{z} = $0-2$. Similarly, \citet{reuter2020complete} do not provide any conclusive evidence about the evolution trend for the SPT sample of 81 gravitationally lensed galaxies selected in the millimeter ($S_{1.4mm}$ > 20 mJy and $S_{870 \rm \mu m}$ > 25 mJy) that cover the redshift range between 1.9 and 6.9. It is also argued that selection effects could induce an increasing trend that limits observations to the brightest galaxies, especially at higher redshifts \citep[e.g.,][]{dudzeviciute2020noevolution, riechers2020bias}. The strong cut on the luminosity induced by the 500 $\rm \mu m$ selection of the \textit{z}-GAL sources (as shown in Fig. \ref{fig:lum-mdust-evolution-redshift}) could explain the observed increasing trend in $T_{\rm dust}$ following the temperature--luminosity relation \citep[$T \propto L_{\rm FIR}^{0.28}$ and $T \propto L_{\rm FIR}^{0.14}$, as shown by][respectively]{chapman2005temp,casey2012temp}. However, this needs to be further checked after obtaining the magnification factors for the \textit{z}-GAL galaxies. 

Due to the various aspects that influence the dust temperature estimates, such as the fitting method (e.g., MBB, GMBB-Sizes, GMBB-Lambda) and observational limits, which make each sample significantly different, we intend to compare our results to other samples \citep[e.g.,][]{bethermin2015evolution, zavala2018s2cls, reuter2020complete, sommovigo2020warmdust} in a forthcoming paper. This will help us to identify the main cause, or causes, of the differences, which cannot be achieved using a homogeneous flux-generated catalog that does not represent the real galaxy emission.

\subsection{\texorpdfstring{$\beta - T_{\rm dust}$}{} relation}\label{section:discussion-beta-temp-relation}
In Fig. \ref{fig:beta-temp-relation-zgal}, we show the distribution of dust emissivity $\beta$ versus dust temperature for the \textit{z}-GAL sample derived from the optically thin MBB in Section \ref{section:zgal_opt_thin}. We see a clear anti-correlation between the two parameters. It has been argued that there is an intrinsic relation between $\beta$ and $T_{\rm dust}$ \citep[e.g.,][]{desert2008, paradis2010, smith2012herschel, juvela2013degeneracy, kirkpatrick2014untangling, cortese2014pacs}, which was also found in laboratory experiments using interstellar grain analogs \citep{agladze1996laboratory, mennella1998temperature}. We derive an empirical relation between the two parameters described by the following equation: 

\begin{ceqn}
\begin{equation}
    \beta = A T_{\rm dust}^{-\alpha}
.\end{equation}
\end{ceqn}

We find $\alpha = 0.69 \pm 0.04$ for the \textit{z}-GAL sources, which is shallower than the one found by \citet{yang2007} ($\alpha =1.07$) estimated for a sample of 18 local luminous infrared galaxies (LIRGs). We over-plot the results found for the BEARS sample, which also follow a very similar trend to ours. We note that our fitting method did not introduce any bias towards the trend \citep[a systematic study was performed by][who show that both the fitting method and the noise increase the degeneracy between $\beta$ and $T_{\rm dust}$]{juvela2013degeneracy}. Using the mock catalog, we found a uniform distribution in the resulting $\beta - T_{\rm dust}$ parameter space, resulting from the fact that our mock catalog is uniformly distributed over a grid with no prior relations between parameters. 

Additionally, we checked the $1\sigma$ contours of the \textit{z}-GAL sources (a few of the contours are shown in red in Fig. \ref{fig:beta-temp-relation-zgal}) and we find that for the bulk of them, the degeneracy is very minimal; it only becomes evident for sources with noisy (or poor-quality) continuum data or when the peak is a poorly sampled (e.g., HeLMS-19), but such cases do not dominate our sample. Fitting this relation to the well-constrained sample (a total of 27 sources), where the relative uncertainties are within 10\%, yields a result of $\alpha = 0.57 \pm 0.09,$ which falls within the confidence range. The underlying nature of the $\beta - T_{\rm dust}$ relation, whether it is physical or is a result of parameter degeneracy, remains an open question. Previous studies used a hierarchical Bayesian fitting method \citep[e.g.,][]{kelly2012bayeshier, juvela2013degeneracy, lamperti2019hierarbayes}, which resulted in a significantly reduced degeneracy. However, these studies focused on datasets with limited millimeter data that cover the RJ tail. In the case of \textit{z}-GAL, the rich sampling along the RJ tail provides better constraints and larger weights on the derived $\beta$ values, which reduces the necessity for more complex statistical approaches.

The $\beta - T_{\rm dust}$ relation has been studied in more detail in our Galaxy \citep[e.g.,][]{juvela2011galactic, planck2011planck, abergel2014planck}. The value of $\alpha$ could carry physical information about the cloud metallicity \citep{cortese2014pacs}, and could also explain the physical conditions that dust grains are exposed to in different parts of the galaxy \citep[such as heating in star-forming clouds in the outskirts of a galaxy;][]{smith2012herschel}. However, it is challenging to retrieve the properties of the environments where the dust grains are located at this stage in the \textit{z}-GAL sources, and more in-depth studies are needed to confirm the extreme dust emissivities. 

\section{Summary and conclusions}\label{section7:conclusions}

In this paper, we present the continuum dust properties of 125 galaxies selected from the \textit{z}-GAL sample that cover the redshift range 1.3 < \textit{z} < 5.4 and are centered around \textit{z} $\sim 2-3$ at the peak epoch of cosmic SFR density. To explore how the derivation of the dust parameters is influenced by the data that are available along the SED from the FIR to the millimeter wavelengths, we performed an in-depth mock-data analysis using both the general opacity and optically thin approximation modified blackbody models. Our principal findings and conclusions are as follows:

\begin{itemize}
    \item We demonstrate the unique wavelength sampling of the \textit{z}-GAL sample of galaxies in the FIR/submm regime through a detailed mock analysis. The \textit{Herschel}-SPIRE flux densities, together with the SCUBA-2 850 $\rm \mu m$ data, when available, which sample roughly the peak of the SED, and the NOEMA 2 and 3~mm fluxes, which sample the RJ tail, allowed us to constrain the dust properties $\beta$, $T_{\rm dust}$, and $M_{\rm dust}$ 
with great precision. 
    
    \item The dust emissivity index is derived with high confidence based on the sampling of an average of five NOEMA flux densities along the RJ tail. $\beta$ estimates show no change (within error bars) when using either the optically thin approximation or the general opacity modified blackbody, even though the dust temperatures vary from one model to the other. We report an average value of $\beta$ = 2.2 $\pm  0.3$ and a range of between  roughly 1.5 and 3 with relative uncertainties that vary in the range $7\%\ - 15$\%. This range is wider than previous Milky Way estimates ($\beta = 1.5 - 2$); however, it coincides with more recent DSFG studies \citep{cooper2022searching}. 
    
    \item We find an average $T^{MBB}_{\rm dust} \sim 30 \pm 4$ K and $T^{GMBB}_{\rm dust} \sim 38$ K when fitting with a MBB and a GMBB, respectively. Although the temperature estimate is decoupled from the uncertainties on $\beta$, this parameter is still dependent on the optical depth assumptions. Our mock analysis shows that accurate $T_{\rm dust}$ values can be estimated using GMBB with a measured source size. While many studies are based on the use of $\lambda_{thick}$, the wavelength at which the optical depth reaches unity, the results remain degenerate with $T_{\rm dust}$ and could overestimate the dust temperatures by $\sim 20\%$. Consequently, the change in $T_{\rm dust}$ affects the $M_{\rm dust}$ estimates, leading to an overestimation that varies between $10\%-75$\% depending on the opacity model assumed.

    \item We find an anti-correlation between $\beta$ and $T_{\rm dust}$ following the trend $\beta \propto T_{\rm dust}^{- (0.69 \pm 0.04)}$. The flux-generated mock catalog shows that this relation is induced neither by the sample selection nor the fitting method. This confirms that the relation is intrinsic, as shown previously by \citet{juvela2013degeneracy}. The relationship between $\beta$ and $T_{\rm dust}$ has direct implications on the nature of dust grains and the physical environment to which they are exposed.

    \item We find an evolution of the dust temperature with redshift given by $T_{\rm dust}$ = ($6.5 \pm 0.5$) \textit{z} + 12$^{+1.1}_{-2}$ K. However, this trend cannot represent the entire galaxy population, because the \textit{z}-GAL sample is flux limited. A deeper sample, covering a large redshift range, will be needed to complement the bright zGAL and BEARS \textit{Herschel}-selected sources in order to verify this evolutionary trend across cosmic time. This study will be presented in a following paper.

    \item Due to potential gravitational lensing, the dust masses and inferred luminosities are only apparent values. However, once the magnification factor is derived, we will be able to add the intrinsic properties for the lensed sources in the \textit{z}-GAL sample, and compare them with those of other samples for which this information is available in order to further explore the nature of the dust grains and their evolution across cosmic time.

\end{itemize}

This work has highlighted the robustness of the \textit{z}-GAL flux sampling from the FIR to the millimeter regime in determining dust properties, particularly the dust emissivity index $\beta$, whose value could vary significantly from the Milky Way range, leading to biased dust temperature estimates. The results from our mock analysis indicate that the accurate measurement of dust temperatures requires estimates of the source sizes with high-resolution imaging of the continuum, and that this will result in a well-constrained dust mass. Future work will focus on exploring the evolution of the dust temperature across cosmic time and comparing the \textit{z}-GAL sample with other similar samples of high-\textit{z} dusty star-forming galaxies in order to further explore the cause or causes of the differences, which cannot be well understood with a flux-generated mock catalog.

\begin{acknowledgements}
This work is based on observations carried out under project numbers M18AB and, subsequently D20AB, with the IRAM NOEMA Interferometer. IRAM is supported by INSU/CNRS (France), MPG (Germany) and IGN (Spain). 
The authors are grateful to IRAM for making this work possible and for the continuous support that we received over the past four years to make this large program a success. 
The authors are also grateful to the IRAM director for approving the DDT proposal that enabled to complete the survey. 
This work benefited from the support of the project Z-GAL ANR-AAPG2019 of the French National Research Agency (ANR). 
The anonymous referee is thanked for providing useful comments that helped to improve the contents of this paper. 
The authors would like to thank I. Cortzen and Cynthia Herrera for their contributions in the early stages of this project and recognise the essential work of the z-GAL Cat-Team and Tiger-Team, who performed the reduction, calibration and delivery of the z-GAL data. 
DI extends her heartfelt appreciation to Alfred Gulum for being a pillar of support.
A.J.B. and A.J.Y acknowledge support from the National Science Foundation grant AST-1716585.
A.N. acknowledges support from the Narodowe Centrum Nauki (UMO-2020/38/E/ST9/00077).
D.A.R. acknowledges support from the National Science Foundation under grant numbers AST-1614213 and AST-1910107 and from the Alexander von Humboldt Foundation through a Humboldt Research Fellowship for Experienced Researchers. 
SS was partly supported by the ESCAPE project. ESCAPE - The European Science Cluster of Astronomy \& Particle Physics ESFRI Research Infrastructures has received funding from the European Union’s Horizon 2020 research and innovation program under Grant Agreement no.~824064.  
CY acknowledges support from ERC Advanced Grant 789410.
HD acknowledges financial support from the Agencia Estatal de Investigación del Ministerio de Ciencia e Innovación (AEI-MCINN) under grant (La evolución de los cíumulos de galaxias desde el amanecer hasta el mediodía cósmico) with reference (PID2019-105776GB-I00/DOI:10.13039/501100011033) and acknowledge support from the ACIISI, Consejería de Economía, Conocimiento y Empleo del Gobierno de Canarias and the European Regional Development Fund (ERDF) under grant with reference PROID2020010107.
TB acknowledges support from NAOJ ALMA Scientific Research Grant Nos. 2018-09B and JSPS KAKENHI No.~17H06130, 22H04939, and 22J21948.
AN acknowledges support from INAF - Osservatorio astronomico d'Abruzzo Via maggini SNC 64100 Teramo.
RJI acknowledges funding by the Deutsche Forschungsgemeinschaft (DFG, German Research Foundation) under Germany's Excellence Strategy -- EXC-2094 -- 390783311.
\end{acknowledgements}

\bibliographystyle{aa} 
\bibliography{citation} 

\onecolumn

\begin{landscape}
\small
\newcolumntype{C}{ @{}>{${}}c<{{}$}@{} }
\begin{longtable}[c]{llll *8{r@{}c@{}l}}
\caption{Millimeter continuum flux densities of the \textit{z}-GAL sources.}
\label{tab:noema-fluxes} \\
\hline \hline
\multirow{2}{*}{Source} & RA & Dec & \multirow{2}{*}{$z_{spec}$} & \multicolumn{24}{c}{$S_\nu$ (mJy)} \\
 & hh:mm:ss & hh:mm:ss &  & 
 \multicolumn{3}{c}{154 GHz} &
 \multicolumn{3}{c}{146 GHz} & 
 \multicolumn{3}{c}{139 GHz} & 
 \multicolumn{3}{c}{131 GHz} & 
 \multicolumn{3}{c}{103 GHz} & 
 \multicolumn{3}{c}{95 GHz} &
 \multicolumn{3}{c}{87 GHz} &  
 \multicolumn{3}{c}{79 GHz} \\ \hline
\endfirsthead
\multicolumn{28}{c}%
{{\bfseries Table \thetable\: continued}} \\
\hline \hline
\multirow{2}{*}{Source} & RA & Dec & \multirow{2}{*}{$z_{spec}$} & \multicolumn{24}{c}{$S_\nu$ (mJy)} \\
 & hh:mm:ss & hh:mm:ss &  & 
 \multicolumn{3}{c}{154 GHz} &
 \multicolumn{3}{c}{146 GHz} & 
 \multicolumn{3}{c}{139 GHz} & 
 \multicolumn{3}{c}{131 GHz} & 
 \multicolumn{3}{c}{103 GHz} & 
 \multicolumn{3}{c}{95 GHz} &
 \multicolumn{3}{c}{87 GHz} &  
 \multicolumn{3}{c}{79 GHz} \\ \hline
\endhead
\hline
\endfoot
\endlastfoot
\multicolumn{28}{c}{HeLMS Sources} \\ \hline
HeLMS-1  & 23:34:40.97 & -06:52:20.2 & 1.9047 & 2.75 & $\pm$ & 0.28 & 2.14 & $\pm$ & 0.26 & 1.47 & $\pm$ & 0.19 & 0.93 & $\pm$ & 0.16 & 0.64 & $\pm$ & 0.11 & 0.52 & $\pm$ & 0.11 & 0.22 & $\pm$ & 0.07 & <0.35 & $\pm$ & 0.36 \\
HeLMS-3  & 00:02:15.96 & -01:28:30.5 & 1.4202 & 1.69 & $\pm$ & 0.22 &  & {---} &  & 1.3 & $\pm$ & 0.17 &  & {---} &  &  & {---} &  & 0.43 & $\pm$ & 0.09 &  & {---} &  & <0.23 & $\pm$ & 0.24 \\
HeLMS-11 & 00:39:29.45 & 00:24:25.9 & 2.4829 &  & {---} &  & 3.49 & $\pm$ & 0.21 &  & {---} &  & 2.4 & $\pm$ & 0.18 & 1.01 & $\pm$ & 0.11 & 0.56 & $\pm$ & 0.09 & 0.35 & $\pm$ & 0.07 & 0.35 & $\pm$ & 0.08 \\
HeLMS-12 & 23:56:01.47 & -07:11:43.1 & 2.3699 & 1.89 & $\pm$ & 0.26 &  & {---} &  & 1.28 & $\pm$ & 0.2 &  & {---} &  & 0.41 & $\pm$ & 0.11 & 0.44 & $\pm$ & 0.11 & <0.25 & $\pm$ & 0.27 & 0.36 & $\pm$ & 0.12 \\
HeLMS-14 & 00:36:19.77 & 00:24:17.7 & 1.6169 &  & {---} &  & 1.0 & $\pm$ & 0.18 &  & {---} &  & 0.55 & $\pm$ & 0.15 & 0.26 & $\pm$ & 0.07 & 0.41 & $\pm$ & 0.1 & <0.15 & $\pm$ & 0.18 & <0.11 & $\pm$ & 0.18 \\
HeLMS-16 & 23:18:57.18 & -05:30:34.7 & 2.8187 &  & {---} &  & 1.31 & $\pm$ & 0.28 &  & {---} &  & 1.61 & $\pm$ & 0.28 & <0.19 & $\pm$ & 0.33 & 0.48 & $\pm$ & 0.13 & <0.24 & $\pm$ & 0.33 & <0.1 & $\pm$ & 0.33 \\
HeLMS-17 \tablefootmark{\dag} &  \multicolumn{1}{c}{----} &  \multicolumn{1}{c}{----} & 2.2978 & 1.93 & $\pm$ & 0.28 & <1.14 & $\pm$ & 1.17 & 1.31 & $\pm$ & 0.21 & 1.65 & $\pm$ & 0.41 & <0.02 & $\pm$ & 0.6 & <0.36 & $\pm$ & 0.66 & 0.69 & $\pm$ & 0.2 & 0.8 & $\pm$ & 0.25 \\
HeLMS-17 E & 23:25:58.56 & -04:45:26.8 & 2.2983 & 0.97 & $\pm$ & 0.2 & <0.54 & $\pm$ & 0.93 & 0.53 & $\pm$ & 0.14 & 0.97 & $\pm$ & 0.3 & <-0.01 & $\pm$ & 0.42 & <0.08 & $\pm$ & 0.45 & <0.26 & $\pm$ & 0.36 & <0.28 & $\pm$ & 0.54 \\
HeLMS-17 W & 23:25:57.92 & -04:45:25.5 & 2.2972 & 0.96 & $\pm$ & 0.19 & <0.6 & $\pm$ & 0.69 & 0.78 & $\pm$ & 0.16 & <0.68 & $\pm$ & 0.84 & <0.03 & $\pm$ & 0.42 & <0.28 & $\pm$ & 0.48 & <0.43 & $\pm$ & 0.48 & <0.52 & $\pm$ & 0.54 \\
HeLMS-19 EW \tablefootmark{\dag} & 23:22:10.05 & -03:36:00.8 & 4.6885 & 4.14 & $\pm$ & 0.34 & 2.16 & $\pm$ & 0.37 & 3.12 & $\pm$ & 0.24 & 1.37 & $\pm$ & 0.32 & 0.86 & $\pm$ & 0.19 & 0.57 & $\pm$ & 0.17 & <0.4 & $\pm$ & 0.48 & 0.73 & $\pm$ & 0.22 \\
HeLMS-19 E & 23:22:10.20 & -03:35:59.8 & 4.6871 & 1.97 & $\pm$ & 0.24 & 1.17 & $\pm$ & 0.27 & 1.21 & $\pm$ & 0.16 & 0.83 & $\pm$ & 0.25 & <0.2 & $\pm$ & 0.36 & <0.04 & $\pm$ & 0.33 & <0.08 & $\pm$ & 0.27 & <0.16 & $\pm$ & 0.39 \\
HeLMS-19 W & 23:22:09.96 & -03:36:01.4 & 4.6882 & 2.17 & $\pm$ & 0.24 & 0.99 & $\pm$ & 0.26 & 1.91 & $\pm$ & 0.18 & <0.54 & $\pm$ & 0.6 & 0.66 & $\pm$ & 0.15 & 0.53 & $\pm$ & 0.13 & <0.32 & $\pm$ & 0.39 & 0.57 & $\pm$ & 0.18 \\
HeLMS-20 & 23:37:28.83 & -04:51:06.3 & 2.1947 & 1.54 & $\pm$ & 0.22 & 1.36 & $\pm$ & 0.21 & 1.15 & $\pm$ & 0.16 & 1.44 & $\pm$ & 0.18 & 0.39 & $\pm$ & 0.07 & <0.19 & $\pm$ & 0.21 & <0.15 & $\pm$ & 0.18 & <0.15 & $\pm$ & 0.27 \\
HeLMS-21 & 00:18:00.21 & -06:02:35.3 & 2.7710 & 1.93 & $\pm$ & 0.3 &  & {---} &  & 1.89 & $\pm$ & 0.22 &  & {---} &  &  & {---} &  & <0.26 & $\pm$ & 0.27 &  & {---} &  & 0.27 & $\pm$ & 0.09 \\
HeLMS-23 & 00:58:41.09 & -01:11:49.0 & 1.4888 & 0.81 & $\pm$ & 0.2 &  & {---} &  & 0.58 & $\pm$ & 0.14 &  & {---} &  &  & {---} &  & <0.22 & $\pm$ & 0.48 &  & {---} &  & <0.02 & $\pm$ & 0.36 \\
HeLMS-24 & 00:38:14.00 & -00:22:52.5 & 4.984 &  & {---} &  &  & {---} &  &  & {---} &  &  & {---} &  &  & {---} &  & 1.4 & $\pm$ & 0.12 &  & {---} &  & 0.96 & $\pm$ & 0.11 \\
HeLMS-25 & 00:41:24.12 & -01:03:07.8 & 2.1404 &  & {---} &  & 1.39 & $\pm$ & 0.19 &  & {---} &  & 0.72 & $\pm$ & 0.13 & 0.24 & $\pm$ & 0.07 & <0.16 & $\pm$ & 0.21 & <0.13 & $\pm$ & 0.18 & <0.12 & $\pm$ & 0.18 \\
HeLMS-26 \tablefootmark{\dag} &  \multicolumn{1}{c}{----} &  \multicolumn{1}{c}{----} & 2.6887 & <0.49 & $\pm$ & 0.54 & 1.03 & $\pm$ & 0.21 & 0.69 & $\pm$ & 0.16 & 0.74 & $\pm$ & 0.2 & 0.54 & $\pm$ & 0.14 & 0.29 & $\pm$ & 0.09 &  & {---} &  & <0.02 & $\pm$ & 0.27 \\
HeLMS-26 E & 00:47:47.56 & 06:14:39.4 & 2.6899 & <0.18 & $\pm$ & 0.36 & 0.42 & $\pm$ & 0.13 & 0.43 & $\pm$ & 0.11 & <0.38 & $\pm$ & 0.39 & <0.11 & $\pm$ & 0.27 & <0.14 & $\pm$ & 0.21 & <0.01 & $\pm$ & 0.21 & <0.07 & $\pm$ & 0.21 \\
HeLMS-26 W & 00:47:46.76 & 06:14:45.3 & 2.6875 & <0.31 & $\pm$ & 0.39 & 0.61 & $\pm$ & 0.17 & <0.26 & $\pm$ & 0.33 & <0.36 & $\pm$ & 0.45 & 0.43 & $\pm$ & 0.11 & <0.15 & $\pm$ & 0.18 & <-0.01 & $\pm$ & 0.18 & <-0.05 & $\pm$ & 0.18 \\
HeLMS-27 & 00:37:58.07 & -01:06:20.1 & 3.765 &  & {---} &  & 4.11 & $\pm$ & 0.29 &  & {---} &  & 2.98 & $\pm$ & 0.19 &  & {---} &  & 0.95 & $\pm$ & 0.11 &  & {---} &  & 0.54 & $\pm$ & 0.09 \\
HeLMS-28 & 00:30:09.28 & -02:06:25.1 & 2.5322 & 2.61 & $\pm$ & 0.29 &  & {---} &  & 1.52 & $\pm$ & 0.21 &  & {---} &  &  & {---} &  & 0.31 & $\pm$ & 0.1 &  & {---} &  & <0.24 & $\pm$ & 0.27 \\
HeLMS-30 & 00:10:27.16 & -02:46:26.4 & 1.8197 & 0.73 & $\pm$ & 0.16 &  & {---} &  & 0.75 & $\pm$ & 0.16 &  & {---} &  &  & {---} &  & <0.14 & $\pm$ & 0.42 &  & {---} &  & <0.01 & $\pm$ & 0.42 \\
HeLMS-31 & 00:13:53.53 & -06:02:00.2 & 1.9494 & 1.14 & $\pm$ & 0.2 &  & {---} &  & 0.63 & $\pm$ & 0.15 &  & {---} &  & <0.1 & $\pm$ & 0.36 & <0.18 & $\pm$ & 0.24 & <-0.08 & $\pm$ & 0.36 & <0.2 & $\pm$ & 0.27 \\
HeLMS-32 C \tablefootmark{\ddag} & 00:03:37.03 & 01:40:12.2 & 1.7149 & <0.4 & $\pm$ & 0.45 & <0.05 & $\pm$ & 0.39 & <0.06 & $\pm$ & 0.33 & <-0.01 & $\pm$ & 0.3 & <0.08 & $\pm$ & 0.18 & <0.07 & $\pm$ & 0.15 & <0.11 & $\pm$ & 0.15 & <-0.03 & $\pm$ & 0.21 \\
HeLMS-32 E \tablefootmark{\ddag} & 00:03:37.51 & 01:40:10.6 & \multicolumn{1}{c}{----} & <0.38 & $\pm$ & 0.42 & <0.04 & $\pm$ & 0.42 & <-0.03 & $\pm$ & 0.42 & <0.22 & $\pm$ & 0.3 & <0.14 & $\pm$ & 0.18 & <0.08 & $\pm$ & 0.18 & <0.1 & $\pm$ & 0.15 & <0.04 & $\pm$ & 0.24 \\
HeLMS-34 & 00:27:19.60 & 00:12:02.5 & 2.2714 & 1.59 & $\pm$ & 0.19 &  & {---} &  & 1.31 & $\pm$ & 0.16 &  & {---} &  & 0.41 & $\pm$ & 0.09 & <0.17 & $\pm$ & 0.21 & 0.23 & $\pm$ & 0.06 & <0.19 & $\pm$ & 0.24 \\
HeLMS-35 & 23:24:59.96 & -00:56:44.1 & 1.6684 &  & {---} &  & <0.81 & $\pm$ & 0.87 &  & {---} &  & <0.47 & $\pm$ & 0.75 & <0.06 & $\pm$ & 0.33 & <0.45 & $\pm$ & 0.69 & <0.18 & $\pm$ & 0.33 & <0.24 & $\pm$ & 0.6 \\
HeLMS-36 & 23:43:14.11 & 01:21:55.4 & 3.9802 & 1.81 & $\pm$ & 0.2 &  & {---} &  &  & {---} &  &  & {---} &  &  & {---} &  & 0.41 & $\pm$ & 0.1 &  & {---} &  & <0.03 & $\pm$ & 0.24 \\
HeLMS-37 & 01:08:01.83 & 05:32:01.2 & 2.7576 & 2.05 & $\pm$ & 0.26 &  & {---} &  & 1.42 & $\pm$ & 0.17 &  & {---} &  &  & {---} &  & <0.27 & $\pm$ & 0.3 &  & {---} &  & <0.08 & $\pm$ & 0.24 \\
HeLMS-38 & 00:22:08.10 & 03:40:42.0 & 2.1898 &  & {---} &  & 0.76 & $\pm$ & 0.16 &  & {---} &  & 0.41 & $\pm$ & 0.11 & 0.24 & $\pm$ & 0.08 & <0.19 & $\pm$ & 0.21 & <0.06 & $\pm$ & 0.15 & <0.07 & $\pm$ & 0.18 \\
HeLMS-39 & 00:29:36.02 & 02:07:13.1 & 2.7659 & 1.19 & $\pm$ & 0.17 &  & {---} &  & 0.73 & $\pm$ & 0.12 &  & {---} &  &  & {---} &  & 0.24 & $\pm$ & 0.07 &  & {---} &  & <0.14 & $\pm$ & 0.21 \\
HeLMS-40 \tablefootmark{\dag} & \multicolumn{1}{c}{----} & \multicolumn{1}{c}{----} & 3.14 & 2.71 & $\pm$ & 0.26 &  & {---} &  & 2.13 & $\pm$ & 0.2 &  & {---} &  &  & {---} &  & 0.45 & $\pm$ & 0.13 &  & {---} &  & <0.25 & $\pm$ & 0.36 \\
HeLMS-40 E & 23:53:31.89 & 03:17:19.9 & 3.1445 & 1.69 & $\pm$ & 0.19 &  & {---} &  & 1.42 & $\pm$ & 0.15 &  & {---} &  &  & {---} &  & <0.19 & $\pm$ & 0.24 &  & {---} &  & <0.12 & $\pm$ & 0.27 \\
HeLMS-40 W & 23:53:31.578 & 03:17:21.0 & 3.1395 & 1.02 & $\pm$ & 0.18 &  & {---} &  & 0.71 & $\pm$ & 0.13 &  & {---} &  &  & {---} &  & <0.26 & $\pm$ & 0.3 &  & {---} &  & <0.13 & $\pm$ & 0.24 \\
HeLMS-41 & 23:36:33.72 & -03:21:20.2 & 2.3353 & 1.31 & $\pm$ & 0.26 &  & {---} &  & 0.93 & $\pm$ & 0.21 &  & {---} &  & 0.75 & $\pm$ & 0.12 & 0.3 & $\pm$ & 0.09 & 0.28 & $\pm$ & 0.07 & <0.23 & $\pm$ & 0.33 \\
HeLMS-42 & 23:40:14.29 & -07:07:36.5 & 1.9558 & 1.08 & $\pm$ & 0.22 &  & {---} &  & <0.4 & $\pm$ & 0.45 &  & {---} &  & <0.17 & $\pm$ & 0.24 & <0.2 & $\pm$ & 0.24 & <0.13 & $\pm$ & 0.21 & <-0.01 & $\pm$ & 0.24 \\
HeLMS-43 E \tablefootmark{\ddag}& 23:34:20.55 & -00:34:56.8 & \multicolumn{1}{c}{----} & <0.35 & $\pm$ & 0.51 &  & {---} &  & <0.26 & $\pm$ & 0.42 &  & {---} &  & <-0.15 & $\pm$ & 0.81 & <0.16 & $\pm$ & 0.24 &  & {---} &  & <0.09 & $\pm$ & 0.3 \\
HeLMS-43 W \tablefootmark{\ddag}& 23:34:20.08 & -00:34:58.1 & 2.2912 & 0.61 & $\pm$ & 0.2 &  & {---} &  & <0.44 & $\pm$ & 0.48 &  & {---} &  & <0.03 & $\pm$ & 0.87 & <0.12 & $\pm$ & 0.24 & <0.38 & $\pm$ & 0.6 & <0.05 & $\pm$ & 0.24 \\
HeLMS-44 & 23:14:47.61 & -04:56:56.0 & 1.37 &  & {---} &  & <0.26 & $\pm$ & 0.54 &  & {---} &  & <0.31 & $\pm$ & 0.51 &  & {---} &  &  & {---} &  &  & {---} &  & <0.04 & $\pm$ & 0.3 \\
HeLMS-45 & 00:12:27.07 & 02:08:06.4 & 5.3994 &  & {---} &  & 4.72 & $\pm$ & 0.3 &  & {---} &  & 3.19 & $\pm$ & 0.21 & <0.92 & $\pm$ & 0.93 & 0.84 & $\pm$ & 0.14 & <0.4 & $\pm$ & 0.69 & 0.55 & $\pm$ & 0.11 \\
HeLMS-46 & 00:46:22.27 & 07:35:18.2 & 2.5765 &  & {---} &  & 2.18 & $\pm$ & 0.2 &  & {---} &  & 1.38 & $\pm$ & 0.17 &  & {---} &  & 0.28 & $\pm$ & 0.07 &  & {---} &  & 0.24 & $\pm$ & 0.07 \\
HeLMS-47 & 23:49:51.73 & -03:00:17.5 & 2.2232 & 1.53 & $\pm$ & 0.24 & 0.95 & $\pm$ & 0.19 & 1.02 & $\pm$ & 0.17 & 0.55 & $\pm$ & 0.13 & 0.41 & $\pm$ & 0.09 & <0.19 & $\pm$ & 0.24 & <0.06 & $\pm$ & 0.18 & <0.13 & $\pm$ & 0.24 \\
HeLMS-48 & 23:28:33.52 & -03:14:18.9 & 3.3514 &  & {---} &  & 3.9 & $\pm$ & 0.57 &  & {---} &  & 2.25 & $\pm$ & 0.53 & 1.09 & $\pm$ & 0.14 & 0.83 & $\pm$ & 0.18 & 0.65 & $\pm$ & 0.1 & <0.56 & $\pm$ & 0.63 \\
HeLMS-49 & 23:37:21.72 & -06:47:42.0 & 2.2154 & 0.94 & $\pm$ & 0.21 &  & {---} &  & 0.88 & $\pm$ & 0.17 &  & {---} &  &  & {---} &  & 0.33 & $\pm$ & 0.08 &  & {---} &  & <0.17 & $\pm$ & 0.27 \\
HeLMS-50 & 23:51:01.86 & -02:44:23.5 & 2.0531 & 1.04 & $\pm$ & 0.18 &  & {---} &  & 0.65 & $\pm$ & 0.14 &  & {---} &  & 0.35 & $\pm$ & 0.07 & <0.11 & $\pm$ & 0.18 & 0.29 & $\pm$ & 0.07 & <0.02 & $\pm$ & 0.18 \\
HeLMS-51 & 23:26:17.61 & -02:53:19.6 & 2.1567 & 0.62 & $\pm$ & 0.16 & <0.58 & $\pm$ & 0.6 & 0.75 & $\pm$ & 0.14 & <0.61 & $\pm$ & 0.63 & <0.38 & $\pm$ & 0.42 & <0.18 & $\pm$ & 0.45 & <0.15 & $\pm$ & 0.27 & <0.06 & $\pm$ & 0.36 \\
HeLMS-52 & 23:37:27.01 & -00:23:41.1 & 2.2092 &  & {---} &  & 1.19 & $\pm$ & 0.2 &  & {---} &  & <0.29 & $\pm$ & 0.36 & <0.07 & $\pm$ & 0.63 & <0.12 & $\pm$ & 0.18 & <-0.04 & $\pm$ & 0.39 & <0.08 & $\pm$ & 0.21 \\
HeLMS-54 & 00:27:17.79 & 02:39:46.5 & 2.707 & 0.74 & $\pm$ & 0.17 & <0.16 & $\pm$ & 0.54 & <0.36 & $\pm$ & 0.39 & 0.79 & $\pm$ & 0.18 & <0.16 & $\pm$ & 0.24 & <0.28 & $\pm$ & 0.33 & <0.09 & $\pm$ & 0.18 & <0.21 & $\pm$ & 0.24 \\
HeLMS-55 & 23:28:31.77 & -00:40:36.6 & 2.2834 & 1.65 & $\pm$ & 0.23 &  & {---} &  & 1.01 & $\pm$ & 0.17 &  & {---} &  & <-0.54 & $\pm$ & 1.02 & 0.33 & $\pm$ & 0.09 & <0.13 & $\pm$ & 0.54 & <0.16 & $\pm$ & 0.39 \\
HeLMS-56 & 00:13:25.82 & 04:25:07.3 & 3.3896 &  & {---} &  &  & {---} &  &  & {---} &  &  & {---} &  & 0.71 & $\pm$ & 0.1 & 0.37 & $\pm$ & 0.08 & 0.37 & $\pm$ & 0.07 & 0.31 & $\pm$ & 0.07 \\
HeLMS-57 & 00:35:19.56 & 07:28:03.7 & 1.9817 & <0.34 & $\pm$ & 0.39 & <0.25 & $\pm$ & 0.36 & 0.45 & $\pm$ & 0.12 & <0.22 & $\pm$ & 0.33 & <0.15 & $\pm$ & 0.21 & <0.09 & $\pm$ & 0.18 & <0.21 & $\pm$ & 0.21 & <0.26 & $\pm$ & 0.27 \\ 
\hline \multicolumn{28}{c}{HerS Sources} \\ \hline
HerS-2 & 01:20:41.65 & -00:27:05.8 & 2.0151 & 2.96 & $\pm$ & 0.37 &  & {---} &  & 1.8 & $\pm$ & 0.24 &  & {---} &  & 0.6 & $\pm$ & 0.1 & 0.38 & $\pm$ & 0.06 & 0.26 & $\pm$ & 0.07 & 0.4 & $\pm$ & 0.07 \\
HerS-3 NE & 01:27:54.06 & 00:49:38.9 & 3.0608 & 3.06 & $\pm$ & 0.47 &  & {---} &  & 2.18 & $\pm$ & 0.37 &  & {---} &  & 0.3 & $\pm$ & 0.07 & <0.28 & $\pm$ & 0.39 & 0.43 & $\pm$ & 0.08 & <0.2 & $\pm$ & 0.21 \\
HerS-5 & 01:26:20.60 & 01:29:49.9 & 1.4493 & 0.81 & $\pm$ & 0.22 &  & {---} &  & 0.9 & $\pm$ & 0.18 &  & {---} &  &  & {---} &  & 0.3 & $\pm$ & 0.06 &  & {---} &  & <0.02 & $\pm$ & 0.18 \\
HerS-7 C1 \tablefootmark{\ddag}& 01:01:33.92 & 00:31:54.6 &   & <0.26 & $\pm$ & 0.51 &  & {---} &  & 0.38 & $\pm$ & 0.12 &  & {---} &  &  & {---} &  & <0.02 & $\pm$ & 0.33 &  & {---} &  & <0.03 & $\pm$ & 0.27 \\
HerS-7 C2 \tablefootmark{\ddag}& 01:01:33.66 & 00:31:54.1 &   & 0.76 & $\pm$ & 0.21 &  & {---} &  & 0.88 & $\pm$ & 0.17 &  & {---} &  &  & {---} &  & <0.11 & $\pm$ & 0.39 &  & {---} &  & <0.08 & $\pm$ & 0.33 \\
HerS-7 N \tablefootmark{\ddag}& 01:01:33.83 & 00:32:04.6 &   & <0.45 & $\pm$ & 0.48 &  & {---} &  & 0.52 & $\pm$ & 0.13 &  & {---} &  &  & {---} &  & <0.22 & $\pm$ & 0.3 &  & {---} &  & <0.11 & $\pm$ & 0.27 \\
HerS-7 W \tablefootmark{\ddag}& 01:01:32.87 & 00:31:58.1 &   & <0.39 & $\pm$ & 0.54 &  & {---} &  & 0.32 & $\pm$ & 0.1 &  & {---} &  &  & {---} &  & <0.18 & $\pm$ & 0.33 &  & {---} &  & <0.05 & $\pm$ & 0.3 \\
HerS-8 & 01:09:39.20 & -01:48:24.4 & 2.2431 & 1.14 & $\pm$ & 0.24 &  & {---} &  & <0.33 & $\pm$ & 0.66 &  & {---} &  & <0.32 & $\pm$ & 0.33 & <0.24 & $\pm$ & 0.39 & <0.09 & $\pm$ & 0.21 & 0.5 & $\pm$ & 0.14 \\
HerS-9 \tablefootmark{c} & 01:09:11.86 &        -01:17:33.4 & 0.8530 & <0.47 & $\pm$ & 0.51 & 0.39 & $\pm$ & 0.07 & <0.41 & $\pm$ & 0.45 & 0.33 & $\pm$ & 0.06 & 0.27 & $\pm$ & 0.06 & 0.26 & $\pm$ & 0.08 & <0.02 & $\pm$ & 0.09 & <0.07 & $\pm$ & 0.21 \\
HerS-10 & 01:17:22.40 & 00:56:22.8 & 2.4688 & 1.07 & $\pm$ & 0.24 &  & {---} &  & 1.31 & $\pm$ & 0.24 &  & {---} &  & 0.38 & $\pm$ & 0.07 & 0.23 & $\pm$ & 0.06 & <0.12 & $\pm$ & 0.21 & <0.13 & $\pm$ & 0.15 \\
HerS-11 & 00:58:47.20 & -01:00:16.2 & 4.6618 & 3.42 & $\pm$ & 0.28 &  & {---} &  & 2.32 & $\pm$ & 0.22 &  & {---} &  &  & {---} &  & 0.41 & $\pm$ & 0.11 &  & {---} &  & 0.3 & $\pm$ & 0.09 \\
HerS-12 & 01:25:46.18 & -00:11:42.2 & 2.2706 & 1.88 & $\pm$ & 0.36 &  & {---} &  & 1.43 & $\pm$ & 0.28 &  & {---} &  & 0.48 & $\pm$ & 0.09 & 0.25 & $\pm$ & 0.06 & 0.24 & $\pm$ & 0.07 & 0.3 & $\pm$ & 0.08 \\
HerS-13 & 01:25:20.95 & 01:17:24.3 & 2.4759 & 1.94 & $\pm$ & 0.3 &  & {---} &  & 0.82 & $\pm$ & 0.18 &  & {---} &  & 0.63 & $\pm$ & 0.09 & 0.45 & $\pm$ & 0.07 & 0.27 & $\pm$ & 0.05 & 0.24 & $\pm$ & 0.05 \\
HerS-14 & 01:40:57.35 & -01:05:46.8 & 3.3441 & 2.72 & $\pm$ & 0.38 & 2.4 & $\pm$ & 0.12 & 2.2 & $\pm$ & 0.28 & 1.76 & $\pm$ & 0.1 & 0.59 & $\pm$ & 0.1 & 0.45 & $\pm$ & 0.1 & 0.47 & $\pm$ & 0.08 & <0.21 & $\pm$ & 0.24 \\
HerS-15 & 01:21:06.83 & 00:34:55.8 & 2.3018 & 1.81 & $\pm$ & 0.33 &  & {---} &  & 1.44 & $\pm$ & 0.25 &  & {---} &  & 0.33 & $\pm$ & 0.08 & 0.25 & $\pm$ & 0.06 & <0.18 & $\pm$ & 0.21 & 0.24 & $\pm$ & 0.07 \\
HerS-16 & 02:14:34.48 & 00:59:23.9 & 2.1981 & 2.32 & $\pm$ & 0.21 & 1.89 & $\pm$ & 0.13 & 1.53 & $\pm$ & 0.16 & 1.2 & $\pm$ & 0.09 & 0.36 & $\pm$ & 0.09 & <0.23 & $\pm$ & 0.27 & 0.24 & $\pm$ & 0.06 & <0.15 & $\pm$ & 0.21 \\
HerS-17 & 02:14:02.54 & -00:46:11.0 & 3.0182 &  & {---} &  & 2.17 & $\pm$ & 0.11 &  & {---} &  & 1.41 & $\pm$ & 0.09 & 0.35 & $\pm$ & 0.06 & 0.26 & $\pm$ & 0.08 & 0.19 & $\pm$ & 0.06 & <0.18 & $\pm$ & 0.24 \\
HerS-18 E & 01:32:12.31 & 00:17:55.6 & 1.6926 & 0.87 & $\pm$ & 0.29 & 0.44 & $\pm$ & 0.1 & <0.29 & $\pm$ & 0.72 & 0.43 & $\pm$ & 0.09 & <0.18 & $\pm$ & 0.24 & <0.19 & $\pm$ & 0.27 & <0.13 & $\pm$ & 0.21 & <0.17 & $\pm$ & 0.24 \\
HerS-18 W & 01:32:11.78 & 00:17:58.2 & 0.5636 & <0.58 & $\pm$ & 0.81 & 0.58 & $\pm$ & 0.1 & <0.16 & $\pm$ & 0.57 & 0.33 & $\pm$ & 0.07 & <0.09 & $\pm$ & 0.18 & <0.04 & $\pm$ & 0.21 & <0.02 & $\pm$ & 0.15 & <-0.05 & $\pm$ & 0.15 \\
HerS-19 SE\tablefootmark{a} & 02:05:29.26 & 00:04:57.0 & \textit{3.1678} &  & {---} &  & 1.17 & $\pm$ & 0.1 &  & {---} &  & 0.68 & $\pm$ & 0.08 &  & {---} &  & 0.22 & $\pm$ & 0.07 &  & {---} &  & <0.01 & $\pm$ & 0.18 \\
HerS-19 W\tablefootmark{a} & 02:05:28.82 & 00:05:01.2 & \textit{3.1394} &  & {---} &  & 0.56 & $\pm$ & 0.07 &  & {---} &  & 0.5 & $\pm$ & 0.06 &  & {---} &  & <0.15 & $\pm$ & 0.18 &  & {---} &  & <0.14 & $\pm$ & 0.18 \\
HerS-19 NE\tablefootmark{a} & 02:05:29.32 & 00:05:04.5 & \multicolumn{1}{c}{----}  &  & {---} &  & 0.44 & $\pm$ & 0.07 &  & {---} &  & 0.33 & $\pm$ & 0.06 &  & {---} &  & <0.15 & $\pm$ & 0.18 &  & {---} &  & <0.1 & $\pm$ & 0.18 \\
HerS-20 & 01:02:46.14 & 01:05:40.1 & 2.0792 & 0.93 & $\pm$ & 0.18 & <0.83 & $\pm$ & 2.64 & 0.59 & $\pm$ & 0.13 & <0.19 & $\pm$ & 0.99 & <0.06 & $\pm$ & 0.12 & <0.22 & $\pm$ & 0.3 & <0.09 & $\pm$ & 0.12 & <0.06 & $\pm$ & 0.21 \\
\hline \multicolumn{28}{c}{HerBS Sources} \\ \hline
HerBS-38 SE \tablefootmark{\ddag}& 14:46:09.08 & 02:19:19.4 & 2.4775 & 0.93 & $\pm$ & 0.14 &  & {---} &  & <0.37 & $\pm$ & 0.39 &  & {---} &  & 0.28 & $\pm$ & 0.08 & <0.27 & $\pm$ & 0.3 & 0.22 & $\pm$ & 0.06 & <0.14 & $\pm$ & 0.3 \\
HerBS-38 W \tablefootmark{\ddag}& 14:46:08.33 & 02:19:29.8 & 2.4158 & 0.76 & $\pm$ & 0.15 &  & {---} &  & <0.23 & $\pm$ & 0.36 &  & {---} &  & 0.3 & $\pm$ & 0.09 & <0.06 & $\pm$ & 0.27 & <0.04 & $\pm$ & 0.15 & <0.07 & $\pm$ & 0.27 \\
HerBS-38 NE \tablefootmark{\ddag}& 14:46:09.10 & 02:19:34.1 & 6.5678 & 0.59 & $\pm$ & 0.13 &  & {---} &  & 0.48 & $\pm$ & 0.13 &  & {---} &  & <0.15 & $\pm$ & 0.21 & <-0.02 & $\pm$ & 0.27 & <0.14 & $\pm$ & 0.18 & <0.12 & $\pm$ & 0.3 \\
HerBS-46 & 14:45:56.30 & -00:48:51.8 & 1.8349 & 0.64 & $\pm$ & 0.12 &  & {---} &  & 0.56 & $\pm$ & 0.11 &  & {---} &  & <0.3 & $\pm$ & 0.36 & <0.15 & $\pm$ & 0.27 & <0.02 & $\pm$ & 0.33 & <0.18 & $\pm$ & 0.33 \\
HerBS-48 & 12:13:01.53 & -00:49:23.2 & 3.1438 & 1.38 & $\pm$ & 0.26 &  & {---} &  & 0.95 & $\pm$ & 0.17 &  & {---} &  &  & {---} &  & <0.24 & $\pm$ & 0.3 &  & {---} &  & <0.03 & $\pm$ & 0.21 \\
HerBS-50 & 12:03:19.14 & -01:12:54.6 & 2.9283 &  & {---} &  & 1.88 & $\pm$ & 0.19 &  & {---} &  & 1.22 & $\pm$ & 0.16 & 0.55 & $\pm$ & 0.13 & <0.17 & $\pm$ & 0.27 & <0.14 & $\pm$ & 0.36 & <0.02 & $\pm$ & 0.18 \\
HerBS-51 & 12:07:09.12 & -01:47:01.8 & 2.1827 & 0.88 & $\pm$ & 0.22 &  & {---} &  & 0.52 & $\pm$ & 0.14 &  & {---} &  & 0.38 & $\pm$ & 0.09 & <0.12 & $\pm$ & 0.21 & <0.15 & $\pm$ & 0.15 & <-0.01 & $\pm$ & 0.15 \\
HerBS-53 \tablefootmark{\dag} & \multicolumn{1}{c}{----} & \multicolumn{1}{c}{----} &  & 0.73 & $\pm$ & 0.22 & 0.87 & $\pm$ & 0.17 & <0.46 & $\pm$ & 0.78 & <0.37 & $\pm$ & 0.48 & 0.51 & $\pm$ & 0.13 & <0.16 & $\pm$ & 0.36 & <0.26 & $\pm$ & 0.36 & <0.31 & $\pm$ & 0.33 \\
HerBS-53 E & 11:51:11.99 & -01:26:34.0 & 1.4219 & <0.26 & $\pm$ & 0.45 & 0.48 & $\pm$ & 0.13 & <0.03 & $\pm$ & 0.54 & <0.17 & $\pm$ & 0.3 & <0.21 & $\pm$ & 0.24 & <0.05 & $\pm$ & 0.27 & <0.19 & $\pm$ & 0.24 & <0.17 & $\pm$ & 0.24 \\
HerBS-53 W & 11:51:12.27 & -01:26:38.3 & 1.4236 & <0.47 & $\pm$ & 0.48 & 0.39 & $\pm$ & 0.11 & <0.43 & $\pm$ & 0.57 & <0.2 & $\pm$ & 0.36 & <0.3 & $\pm$ & 0.3 & <0.11 & $\pm$ & 0.24 & <0.07 & $\pm$ & 0.27 & <0.14 & $\pm$ & 0.24 \\
HerBS-61 & 12:01:27.59 & -01:40:45.8 & 3.7293 &  & {---} &  & 3.19 & $\pm$ & 0.24 &  & {---} &  & 1.75 & $\pm$ & 0.17 &  & {---} &  & 0.49 & $\pm$ & 0.11 &  & {---} &  & <0.13 & $\pm$ & 0.24 \\
HerBS-62 & 12:15:42.81 & -00:52:20.8 & 2.5738 &  & {---} &  & 1.62 & $\pm$ & 0.19 &  & {---} &  & 0.82 & $\pm$ & 0.15 &  & {---} &  & 0.34 & $\pm$ & 0.09 &  & {---} &  & <0.21 & $\pm$ & 0.24 \\
HerBS-65 & 13:44:22.58 & 23:19:50.1 & 2.6858 & 2.33 & $\pm$ & 0.23 &  & {---} &  & 1.07 & $\pm$ & 0.17 &  & {---} &  &  & {---} &  & <0.14 & $\pm$ & 0.63 &  & {---} &  & <0.16 & $\pm$ & 0.27 \\
HerBS-72 & 14:45:12.16 & -00:15:10.5 & 3.638 & 2.64 & $\pm$ & 0.22 & 0.95 & $\pm$ & 0.09 & 1.96 & $\pm$ & 0.15 & 0.72 & $\pm$ & 0.07 & 0.49 & $\pm$ & 0.09 & 0.29 & $\pm$ & 0.07 & 0.26 & $\pm$ & 0.07 & <0.26 & $\pm$ & 0.3 \\
HerBS-74 & 12:06:00.46 & 00:34:01.1 & 2.5596 &  & {---} &  & 0.78 & $\pm$ & 0.13 &  & {---} &  & 0.84 & $\pm$ & 0.13 & 1.34 & $\pm$ & 0.12 & 1.38 & $\pm$ & 0.13 & 1.54 & $\pm$ & 0.1 & 1.59 & $\pm$ & 0.12 \\
HerBS-76 EW & 13:35:34.09 & 34:18:34.8 & 2.3302 & 1.56 & $\pm$ & 0.21 &  & {---} &  & 1.11 & $\pm$ & 0.18 &  & {---} &  & 0.29 & $\pm$ & 0.08 & <0.23 & $\pm$ & 0.39 & <0.19 & $\pm$ & 0.21 & <0.13 & $\pm$ & 0.45 \\
HerBS-78 & 14:33:52.56 & 02:04:17.2 & 3.7344 & 3.41 & $\pm$ & 0.13 & 1.52 & $\pm$ & 0.08 & 2.35 & $\pm$ & 0.09 & 1.14 & $\pm$ & 0.06 &  & {---} &  & 0.65 & $\pm$ & 0.1 &  & {---} &  & <0.16 & $\pm$ & 0.21 \\
HerBS-82\tablefootmark{a} & 12:11:44.88 & 01:06:38.1 & \textit{2.0583} & 0.99 & $\pm$ & 0.21 & 0.88 & $\pm$ & 0.17 & 0.3 & $\pm$ & 0.1 & 0.95 & $\pm$ & 0.16 &  & {---} &  & <-0.1 & $\pm$ & 0.75 &  & {---} &  & <0.31 & $\pm$ & 0.39 \\
HerBS-83 & 12:18:13.06 & 01:18:43.2 & 3.9438 & 3.05 & $\pm$ & 0.32 & 2.48 & $\pm$ & 0.24 & 1.64 & $\pm$ & 0.24 & 1.44 & $\pm$ & 0.19 &  & {---} &  & 0.4 & $\pm$ & 0.12 &  & {---} &  & <0.16 & $\pm$ & 0.24 \\
HerBS-85 & 11:47:52.85 & -00:58:32.0 & 2.8169 & <0.32 & $\pm$ & 0.42 & 0.5 & $\pm$ & 0.11 & 0.38 & $\pm$ & 0.11 & <0.23 & $\pm$ & 0.27 & <0.21 & $\pm$ & 0.24 & <0.12 & $\pm$ & 0.27 & <-0.16 & $\pm$ & 0.42 & <0.01 & $\pm$ & 0.33 \\
HerBS-91 \tablefootmark{\dag} & \multicolumn{1}{c}{----} & \multicolumn{1}{c}{----} & 2.4048 & 1.01 & $\pm$ & 0.14 & 1.08 & $\pm$ & 0.21 & 0.85 & $\pm$ & 0.15 & 0.61 & $\pm$ & 0.2 & <0.33 & $\pm$ & 0.36 & 0.33 & $\pm$ & 0.08 & <0.08 & $\pm$ & 0.33 & <-0.01 & $\pm$ & 0.27 \\
HerBS-91 E & 09:21:35.82 & 00:01:30.9 & 2.4048 & 0.58 & $\pm$ & 0.1 & 0.55 & $\pm$ & 0.14 & 0.49 & $\pm$ & 0.11 & <0.22 & $\pm$ & 0.33 & 0.22 & $\pm$ & 0.07 & <0.15 & $\pm$ & 0.15 & <-0.02 & $\pm$ & 0.21 & <0.02 & $\pm$ & 0.21 \\
HerBS-91 C & 09:21:35.55 & 00:01:30.4 & 2.4047 & 0.31 & $\pm$ & 0.08 & <0.24 & $\pm$ & 0.3 & <0.16 & $\pm$ & 0.21 & <0.21 & $\pm$ & 0.33 & <0.04 & $\pm$ & 0.18 & <0.06 & $\pm$ & 0.12 & <0.03 & $\pm$ & 0.18 & <-0.06 & $\pm$ & 0.15 \\
HerBS-91 W & 09:21:35.22 & 00:01:29.0 &   & <0.12 & $\pm$ & 0.15 & <0.29 & $\pm$ & 0.36 & <0.2 & $\pm$ & 0.24 & <0.18 & $\pm$ & 0.39 & <0.07 & $\pm$ & 0.21 & 0.12 & $\pm$ & 0.04 & <0.07 & $\pm$ & 0.18 & <0.03 & $\pm$ & 0.12 \\
HerBS-92 EW \tablefootmark{\dag} & 13:38:09.15 & 25:51:55.4 & 3.264 & 2.43 & $\pm$ & 0.32 & 1.38 & $\pm$ & 0.17 & 2.39 & $\pm$ & 0.28 & 1.25 & $\pm$ & 0.17 & 0.35 & $\pm$ & 0.09 & <1.32 & $\pm$ & 1.41 & 0.54 & $\pm$ & 0.11 & <0.23 & $\pm$ & 0.6 \\
HerBS-92 E & 13:38:09.15 & 25:51:55.4 & 3.264 & 1.56 & $\pm$ & 0.26 & 0.84 & $\pm$ & 0.14 & 1.78 & $\pm$ & 0.24 & 0.82 & $\pm$ & 0.14 & <0.17 & $\pm$ & 0.21 & <0.78 & $\pm$ & 1.14 & 0.31 & $\pm$ & 0.08 & <0.23 & $\pm$ & 0.51 \\
HerBS-92 W & 13:38:08.27 & 25:51:48.9 & 3.264 & 0.87 & $\pm$ & 0.19 & 0.54 & $\pm$ & 0.1 & 0.61 & $\pm$ & 0.15 & 0.43 & $\pm$ & 0.1 & 0.18 & $\pm$ & 0.06 & <0.54 & $\pm$ & 0.81 & 0.23 & $\pm$ & 0.07 &  & {---} & \\
HerBS-105 SW\tablefootmark{b} & 08:39:31.97 & -01:18:00.1 & 2.6684  & 1.01 & $\pm$ & 0.17 &  & {---} &  & 0.43 & $\pm$ & 0.1 &  & {---} &  &  & {---} &  & 0.19 & $\pm$ & 0.06 &  & {---} &  & <0.05 & $\pm$ & 0.12 \\
HerBS-105 NE 1 & 08:39:32.36 & -01:17:56.5 &   & <0.19 & $\pm$ & 0.3 &  & {---} &  & <0.14 & $\pm$ & 0.21 &  & {---} &  &  & {---} &  & <0.03 & $\pm$ & 0.09 &  & {---} &  & <0.04 & $\pm$ & 0.12 \\
HerBS-105 NE 2 & 08:39:33.15 & -01:17:54.1 &   & <0.27 & $\pm$ & 0.3 &  & {---} &  & <0.15 & $\pm$ & 0.21 &  & {---} &  &  & {---} &  & <0.11 & $\pm$ & 0.12 &  & {---} &  & <0.04 & $\pm$ & 0.12 \\
HerBS-108 & 08:38:17.42 & -00:41:34.4 & 3.7168 & 2.9 & $\pm$ & 0.27 & 2.67 & $\pm$ & 0.13 & 2.11 & $\pm$ & 0.2 & 1.74 & $\pm$ & 0.12 &  & {---} &  & 0.31 & $\pm$ & 0.07 &  & {---} &  & 0.27 & $\pm$ & 0.07 \\
HerBS-109 NW \tablefootmark{\ddag}& 13:29:00.35 & 28:19:18.5 & 1.585 &  & {---} &  & 0.33 & $\pm$ & 0.09 &  & {---} &  & <0.16 & $\pm$ & 0.21 & <0.07 & $\pm$ & 0.12 & <0.07 & $\pm$ & 0.15 & <-0.01 & $\pm$ & 0.12 & <0.05 & $\pm$ & 0.15 \\
HerBS-109 S \tablefootmark{\ddag}& 13:29:00.31 & 28:19:07.5 & 1.5843 &  & {---} &  & 0.6 & $\pm$ & 0.1 &  & {---} &  & 0.32 & $\pm$ & 0.09 & <0.05 & $\pm$ & 0.15 & <0.03 & $\pm$ & 0.15 &  & {---} &  & <0.02 & $\pm$ & 0.15 \\
HerBS-109 NE \tablefootmark{\ddag}& 13:29:00.77 & 28:19:16.3 & 2.8385 &  & {---} &  & 0.57 & $\pm$ & 0.11 &  & {---} &  & 0.43 & $\pm$ & 0.09 & <0.13 & $\pm$ & 0.15 & <0.1 & $\pm$ & 0.18 & <0.02 & $\pm$ & 0.12 & <0.1 & $\pm$ & 0.15 \\
HerBS-110 & 14:18:33.24 & 01:02:10.9 & 2.681 & 2.39 & $\pm$ & 0.18 &  & {---} &  & 1.46 & $\pm$ & 0.12 &  & {---} &  &  & {---} &  & 0.36 & $\pm$ & 0.1 &  & {---} &  & 0.33 & $\pm$ & 0.08 \\
HerBS-115 & 13:35:38.20 & 26:57:40.1 & 2.3706 &  & {---} &  & 0.79 & $\pm$ & 0.18 &  & {---} &  & 0.43 & $\pm$ & 0.09 & 0.22 & $\pm$ & 0.06 & <0.15 & $\pm$ & 0.15 & 0.12 & $\pm$ & 0.04 & <0.12 & $\pm$ & 0.18 \\
HerBS-116 EW \tablefootmark{\dag} & 12:13:48.05 & 01:08:11.0 & 3.1547 & 2.2 & $\pm$ & 0.37 &  & {---} &  & 1.26 & $\pm$ & 0.25 &  & {---} &  & 0.63 & $\pm$ & 0.13 & 0.44 & $\pm$ & 0.14 & 0.36 & $\pm$ & 0.11 & <0.11 & $\pm$ & 0.24 \\
HerBS-116 E & 12:13:48.13 & 01:08:10.5 & 3.1547 & 1.05 & $\pm$ & 0.26 &  & {---} &  & 0.51 & $\pm$ & 0.17 &  & {---} &  & 0.33 & $\pm$ & 0.1 & <0.32 & $\pm$ & 0.33 & <0.13 & $\pm$ & 0.21 & <0.04 & $\pm$ & 0.18 \\
HerBS-116 W & 12:13:47.93 & 01:08:10.3 & 3.1547 & 1.15 & $\pm$ & 0.26 &  & {---} &  & 0.75 & $\pm$ & 0.19 &  & {---} &  & 0.3 & $\pm$ & 0.09 & <0.12 & $\pm$ & 0.27 & <0.23 & $\pm$ & 0.27 & <0.07 & $\pm$ & 0.18 \\
HerBS-124 EW \tablefootmark{\dag} & \multicolumn{1}{c}{----} & \multicolumn{1}{c}{----} &  & 1.46 & $\pm$ & 0.27 &  & {---} &  & <0.54 & $\pm$ & 0.63 &  & {---} &  & 0.4 & $\pm$ & 0.11 & <0.08 & $\pm$ & 0.3 & <0.19 & $\pm$ & 0.21 & <0.09 & $\pm$ & 0.27 \\
HerBS-124 W & 12:21:58.48 & 00:33:24.9 & 2.2772 & 1.0 & $\pm$ & 0.22 &  & {---} &  & <0.38 & $\pm$ & 0.48 &  & {---} &  & 0.38 & $\pm$ & 0.09 & <0.05 & $\pm$ & 0.21 & <0.1 & $\pm$ & 0.15 & <0.06 & $\pm$ & 0.21 \\
HerBS-124 E & 12:21:58.65 & 00:33:25.0 & 2.2781 & <0.46 & $\pm$ & 0.48 &  & {---} &  & <0.16 & $\pm$ & 0.39 &  & {---} &  & <0.02 & $\pm$ & 0.18 & <0.03 & $\pm$ & 0.21 & <0.09 & $\pm$ & 0.15 & <0.03 & $\pm$ & 0.18 \\
HerBS-125 & 13:04:32.18 & 29:53:39.3 & 2.5739 &  & {---} &  & 1.39 & $\pm$ & 0.19 &  & {---} &  & 0.88 & $\pm$ & 0.15 & <0.1 & $\pm$ & 0.24 & 0.31 & $\pm$ & 0.09 & 0.4 & $\pm$ & 0.1 & <0.07 & $\pm$ & 0.3 \\
HerBS-126 & 14:51:35.27 & -01:14:17.3 & 2.5875 &  & {---} &  & 0.71 & $\pm$ & 0.08 &  & {---} &  & 0.49 & $\pm$ & 0.07 &  & {---} &  & <0.29 & $\pm$ & 0.33 &  & {---} &  & <0.03 & $\pm$ & 0.27 \\
HerBS-127 & 13:21:28.75 & 28:20:23.8 &   & 0.86 & $\pm$ & 0.25 & 0.76 & $\pm$ & 0.12 & <0.19 & $\pm$ & 0.6 & 0.31 & $\pm$ & 0.09 &  & {---} &  & 0.24 & $\pm$ & 0.06 &  & {---} &  & <0.09 & $\pm$ & 0.15 \\
HerBS-128 & 13:04:14.47 & 30:35:38.3 & 2.0681 & <0.79 & $\pm$ & 0.81 & 0.97 & $\pm$ & 0.13 & 0.79 & $\pm$ & 0.24 & 0.74 & $\pm$ & 0.11 & 0.34 & $\pm$ & 0.1 & 0.26 & $\pm$ & 0.08 & <0.19 & $\pm$ & 0.24 & <0.28 & $\pm$ & 0.36 \\
HerBS-129 & 13:00:53.80 & 26:03:00.1 & 3.3074 &  & {---} &  & 2.47 & $\pm$ & 0.2 &  & {---} &  & 1.5 & $\pm$ & 0.16 & 0.45 & $\pm$ & 0.1 & 0.26 & $\pm$ & 0.07 & 0.25 & $\pm$ & 0.08 & <0.23 & $\pm$ & 0.33 \\
HerBS-134 & 13:34:40.42 & 35:31:39.1 & 3.1725 & 2.46 & $\pm$ & 0.22 &  & {---} &  & 1.75 & $\pm$ & 0.19 &  & {---} &  & 0.46 & $\pm$ & 0.07 & <0.39 & $\pm$ & 0.45 & 0.21 & $\pm$ & 0.06 & <0.18 & $\pm$ & 0.69 \\
HerBS-136 & 08:53:08.46 & -00:57:28.9 & 3.2884 & 1.43 & $\pm$ & 0.23 & 1.27 & $\pm$ & 0.11 & 0.75 & $\pm$ & 0.17 & 0.41 & $\pm$ & 0.1 &  & {---} &  & 0.36 & $\pm$ & 0.08 &  & {---} &  & <0.22 & $\pm$ & 0.3 \\
HerBS-137 & 14:53:37.15 & 00:04:10.2 & 3.0408 & 0.73 & $\pm$ & 0.12 &  & {---} &  & 0.82 & $\pm$ & 0.12 &  & {---} &  & 0.25 & $\pm$ & 0.07 & <0.11 & $\pm$ & 0.27 & <0.11 & $\pm$ & 0.18 & <0.23 & $\pm$ & 0.27 \\
HerBS-140 & 14:21:40.42 & 00:04:46.3 & 2.7799 & 1.2 & $\pm$ & 0.15 &  & {---} &  & 1.01 & $\pm$ & 0.11 &  & {---} &  &  & {---} &  & 0.28 & $\pm$ & 0.09 &  & {---} &  & <0.13 & $\pm$ & 0.24 \\
HerBS-143 & 14:18:10.10 & -00:37:45.5 & 2.2406 & 1.05 & $\pm$ & 0.18 &  & {---} &  & 0.69 & $\pm$ & 0.13 &  & {---} &  & <0.19 & $\pm$ & 0.21 & 0.28 & $\pm$ & 0.09 & <0.16 & $\pm$ & 0.21 & <0.09 & $\pm$ & 0.21 \\
HerBS-147 & 14:34:03.66 & 00:02:30.1 & 3.115 & 1.08 & $\pm$ & 0.16 &  & {---} &  & 1.14 & $\pm$ & 0.13 &  & {---} &  & <0.21 & $\pm$ & 0.21 & <0.17 & $\pm$ & 0.24 & 0.18 & $\pm$ & 0.06 & <0.16 & $\pm$ & 0.24 \\
HerBS-149 & 13:38:27.61 & 31:39:55.6 & 2.665 & 1.63 & $\pm$ & 0.22 &  & {---} &  & 1.08 & $\pm$ & 0.16 &  & {---} &  &  & {---} &  & <0.13 & $\pm$ & 0.42 &  & {---} &  & <0.04 & $\pm$ & 0.42 \\
HerBS-150 \tablefootmark{\dag} & \multicolumn{1}{c}{----} & \multicolumn{1}{c}{----} &  & 3.64 & $\pm$ & 0.38 & 2.13 & $\pm$ & 0.33 & 1.67 & $\pm$ & 0.27 & 1.58 & $\pm$ & 0.22 & <0.37 & $\pm$ & 0.39 & 0.41 & $\pm$ & 0.11 & 0.32 & $\pm$ & 0.1 & <0.14 & $\pm$ & 0.3 \\
HerBS-150 E & 12:24:59.27 & -00:56:51.9 & 3.6732 & 0.93 & $\pm$ & 0.2 & <0.28 & $\pm$ & 0.45 & 0.5 & $\pm$ & 0.14 & <0.26 & $\pm$ & 0.3 & <0.01 & $\pm$ & 0.21 & <0.03 & $\pm$ & 0.15 & <0.14 & $\pm$ & 0.18 & <0.02 & $\pm$ & 0.12 \\
HerBS-150 C & 12:24:58.98 & -00:56:47.5 & 3.6682 & 1.17 & $\pm$ & 0.21 & 0.77 & $\pm$ & 0.21 & 0.49 & $\pm$ & 0.15 & 0.52 & $\pm$ & 0.13 & <0.14 & $\pm$ & 0.24 & <0.04 & $\pm$ & 0.12 & <0.11 & $\pm$ & 0.18 &  & {---} & \\
HerBS-150 W & 12:24:58.70 & -00:56:48.7 & 3.6787 & 1.54 & $\pm$ & 0.25 & 1.08 & $\pm$ & 0.2 & 0.68 & $\pm$ & 0.17 & 0.8 & $\pm$ & 0.14 & <0.22 & $\pm$ & 0.24 & 0.34 & $\pm$ & 0.09 & <0.07 & $\pm$ & 0.15 & <0.12 & $\pm$ & 0.24 \\
HerBS-153 & 14:42:43.41 & 01:55:04.3 & 3.1501 & 2.09 & $\pm$ & 0.12 &  & {---} &  & 1.65 & $\pm$ & 0.13 &  & {---} &  &  & {---} &  & 0.31 & $\pm$ & 0.07 &  & {---} &  & <0.12 & $\pm$ & 0.18 \\
HerBS-157 & 08:49:57.77 & 01:07:10.8 & 1.8964 & 1.51 & $\pm$ & 0.14 & 1.15 & $\pm$ & 0.2 & 0.94 & $\pm$ & 0.11 & 0.58 & $\pm$ & 0.13 & 0.26 & $\pm$ & 0.07 & 0.16 & $\pm$ & 0.04 & <0.14 & $\pm$ & 0.21 & <0.11 & $\pm$ & 0.15 \\
HerBS-162 SW & 14:43:34.35 & -00:30:35.7 & 2.474 &  & {---} &  & 0.36 & $\pm$ & 0.07 &  & {---} &  & <0.17 & $\pm$ & 0.21 & <0.2 & $\pm$ & 0.3 & 0.38 & $\pm$ & 0.11 & <0.01 & $\pm$ & 0.18 & <-0.07 & $\pm$ & 0.24 \\
HerBS-164 & 12:14:16.30 & -01:37:03.7 & 2.0126 & <0.55 & $\pm$ & 0.57 & 0.56 & $\pm$ & 0.13 & 0.45 & $\pm$ & 0.12 & <0.17 & $\pm$ & 0.3 &  & {---} &  & <0.04 & $\pm$ & 0.36 &  & {---} &  & <0.09 & $\pm$ & 0.24 \\
HerBS-165 & 09:06:13.91 & -01:00:41.1 & 2.2251 &  & {---} &  & 0.67 & $\pm$ & 0.2 &  & {---} &  & <0.41 & $\pm$ & 0.42 & <0.17 & $\pm$ & 0.24 & <0.09 & $\pm$ & 0.18 & <-0.01 & $\pm$ & 0.18 & <0.03 & $\pm$ & 0.15 \\
HerBS-167 & 13:03:41.80 & 31:37:57.9 & 2.2144 & 1.86 & $\pm$ & 0.46 & 0.87 & $\pm$ & 0.16 & <0.22 & $\pm$ & 0.69 & 0.63 & $\pm$ & 0.13 & <0.08 & $\pm$ & 0.33 & <0.05 & $\pm$ & 0.21 & <0.12 & $\pm$ & 0.27 & <0.15 & $\pm$ & 0.33 \\
HerBS-169 & 08:38:59.47 & 02:13:27.4 & 2.6977 &  & {---} &  & 1.1 & $\pm$ & 0.17 &  & {---} &  & 0.64 & $\pm$ & 0.13 &  & {---} &  & 0.27 & $\pm$ & 0.07 &  & {---} &  & 0.17 & $\pm$ & 0.05 \\
HerBS-171 & 08:39:45.21 & 02:10:18.1 & 2.4793 &  & {---} &  & 0.51 & $\pm$ & 0.16 &  & {---} &  & 0.56 & $\pm$ & 0.15 & <0.16 & $\pm$ & 0.27 & 0.41 & $\pm$ & 0.11 & 0.32 & $\pm$ & 0.08 & <0.15 & $\pm$ & 0.21 \\
HerBS-172 & 14:50:40.47 & 00:33:35.8 & 2.9246 &  & {---} &  & 0.44 & $\pm$ & 0.06 &  & {---} &  & 0.2 & $\pm$ & 0.04 & <0.2 & $\pm$ & 0.21 & <0.09 & $\pm$ & 0.12 & <0.06 & $\pm$ & 0.21 & <-0.01 & $\pm$ & 0.15 \\
HerBS-175 & 12:19:00.83 & 00:33:26.9 & 3.1575 & 1.6 & $\pm$ & 0.21 &  & {---} &  & 1.32 & $\pm$ & 0.18 &  & {---} &  &  & {---} &  & 0.4 & $\pm$ & 0.1 &  & {---} &  & <0.2 & $\pm$ & 0.3 \\
HerBS-176 & 13:12:22.15 & 27:02:17.8 & 2.9805 &  & {---} &  & 0.96 & $\pm$ & 0.16 &  & {---} &  & 0.98 & $\pm$ & 0.14 & 0.41 & $\pm$ & 0.07 & <0.18 & $\pm$ & 0.21 & 0.2 & $\pm$ & 0.06 & 0.24 & $\pm$ & 0.08 \\
HerBS-177 & 11:54:33.72 & 00:50:41.8 & 3.9625 & 4.85 & $\pm$ & 0.36 & 3.36 & $\pm$ & 0.18 & 2.98 & $\pm$ & 0.25 & 2.14 & $\pm$ & 0.16 &  & {---} &  & 0.58 & $\pm$ & 0.11 &  & {---} &  & <0.2 & $\pm$ & 0.21 \\
HerBS-179 & 11:55:20.97 & -02:13:29.4 & 3.9423 & 2.92 & $\pm$ & 0.26 & 2.21 & $\pm$ & 0.18 & 1.78 & $\pm$ & 0.2 & 1.25 & $\pm$ & 0.13 &  & {---} &  & 0.26 & $\pm$ & 0.08 &  & {---} &  & 0.26 & $\pm$ & 0.08 \\
HerBS-180 N \tablefootmark{\ddag}& 13:15:39.73 & 29:22:20.6 &   & 1.3 & $\pm$ & 0.31 & 0.64 & $\pm$ & 0.1 & 0.61 & $\pm$ & 0.2 & <0.24 & $\pm$ & 0.27 &  & {---} &  & <-0.01 & $\pm$ & 0.18 &  & {---} &  & <-0.06 & $\pm$ & 0.18 \\
HerBS-180 NE \tablefootmark{\ddag}& 13:15:39.32 & 29:22:22.4 &   & <0.43 & $\pm$ & 0.93 & 0.45 & $\pm$ & 0.1 & <0.31 & $\pm$ & 0.69 & 0.32 & $\pm$ & 0.09 &  & {---} &  & <0.18 & $\pm$ & 0.21 &  & {---} &  &  & {---} & \\
HerBS-183 & 09:04:53.06 & 02:20:16.9 & 1.891 & 1.36 & $\pm$ & 0.15 & 0.88 & $\pm$ & 0.17 & 0.92 & $\pm$ & 0.1 & 0.7 & $\pm$ & 0.14 & <0.2 & $\pm$ & 0.21 & 0.37 & $\pm$ & 0.06 & 0.22 & $\pm$ & 0.06 & 0.2 & $\pm$ & 0.05 \\
HerBS-185 & 09:24:08.92 & -00:50:18.1 & 4.3238 & 3.94 & $\pm$ & 0.17 &  & {---} &  & 2.85 & $\pm$ & 0.14 &  & {---} &  &  & {---} &  & 0.5 & $\pm$ & 0.06 &  & {---} &  & 0.32 & $\pm$ & 0.06 \\
HerBS-187 \tablefootmark{\dag} & \multicolumn{1}{c}{----} & \multicolumn{1}{c}{----} & 1.828 & 0.62 & $\pm$ & 0.13 & 0.52 & $\pm$ & 0.09 & 0.46 & $\pm$ & 0.12 & 0.42 & $\pm$ & 0.09 &  & {---} &  & <0.08 & $\pm$ & 0.18 &  & {---} &  & <0.13 & $\pm$ & 0.21 \\
HerBS-187 E & 08:37:05.58 & 02:00:31.5 & 1.8285 & <0.13 & $\pm$ & 0.18 & <0.08 & $\pm$ & 0.12 & <0.17 & $\pm$ & 0.21 & <0.15 & $\pm$ & 0.15 &  & {---} &  & <-0.01 & $\pm$ & 0.09 &  & {---} &  & <0.03 & $\pm$ & 0.12 \\
HerBS-187 W & 08:37:05.35 & 02:00:32.7 & 1.8274 & 0.35 & $\pm$ & 0.09 & 0.25 & $\pm$ & 0.06 & 0.26 & $\pm$ & 0.08 & 0.25 & $\pm$ & 0.06 &  & {---} &  & <0.08 & $\pm$ & 0.12 &  & {---} &  & <0.12 & $\pm$ & 0.15 \\
HerBS-187 S & 08:37:05.40 & 02:00:27.9 &   & <0.14 & $\pm$ & 0.21 & 0.19 & $\pm$ & 0.05 & <0.03 & $\pm$ & 0.18 & <0.02 & $\pm$ & 0.12 &  & {---} &  & <0.01 & $\pm$ & 0.09 &  & {---} &  & <-0.02 & $\pm$ & 0.09 \\
HerBS-188 & 08:43:59.98 & 02:50:01.7 & 2.7675 & 1.6 & $\pm$ & 0.17 &  & {---} &  & 1.04 & $\pm$ & 0.11 &  & {---} &  &  & {---} &  & 0.28 & $\pm$ & 0.06 &  & {---} &  & <0.1 & $\pm$ & 0.15 \\
HerBS-190 & 09:04:05.28 & -00:33:33.5 & 2.589 &  & {---} &  & 1.24 & $\pm$ & 0.19 &  & {---} &  & 1.13 & $\pm$ & 0.19 &  & {---} &  & <0.13 & $\pm$ & 0.18 &  & {---} &  & <0.11 & $\pm$ & 0.18 \\
HerBS-191 & 12:47:53.28 & 32:24:45.8 & 3.4413 &  & {---} &  & 1.5 & $\pm$ & 0.19 &  & {---} &  & 1.44 & $\pm$ & 0.19 & 0.72 & $\pm$ & 0.18 &  & {---} &  & <0.24 & $\pm$ & 0.36 &  & {---} & \\
HerBS-193 & 08:53:51.96 & -00:08:05.4 & 3.6951 & 1.57 & $\pm$ & 0.18 & 1.11 & $\pm$ & 0.11 & 1.04 & $\pm$ & 0.19 & 0.88 & $\pm$ & 0.1 &  & {---} &  & 0.21 & $\pm$ & 0.06 &  & {---} &  & <0.19 & $\pm$ & 0.21 \\
HerBS-194 \tablefootmark{\dag} & \multicolumn{1}{c}{----} & \multicolumn{1}{c}{----} &  & 0.99 & $\pm$ & 0.17 &  & {---} &  & 0.82 & $\pm$ & 0.16 &  & {---} &  & 0.33 & $\pm$ & 0.09 & 0.19 & $\pm$ & 0.06 & 0.37 & $\pm$ & 0.08 & <0.25 & $\pm$ & 0.27 \\
HerBS-194 N & 08:55:21.30 & -00:35:56.6 & 2.3335 & <0.32 & $\pm$ & 0.33 &  & {---} &  & 0.39 & $\pm$ & 0.11 &  & {---} &  & <0.19 & $\pm$ & 0.21 & <0.12 & $\pm$ & 0.15 & <0.15 & $\pm$ & 0.15 & <0.09 & $\pm$ & 0.18 \\
HerBS-194 S & 08:55:21.04 & -00:36:11.9 & 2.3316 & 0.67 & $\pm$ & 0.13 &  & {---} &  & 0.43 & $\pm$ & 0.12 &  & {---} &  & <0.14 & $\pm$ & 0.18 & <0.07 & $\pm$ & 0.12 & 0.22 & $\pm$ & 0.06 & <0.16 & $\pm$ & 0.21 \\
HerBS-197 & 12:20:34.05 & -00:38:04.4 & 2.417 &  & {---} &  & 0.66 & $\pm$ & 0.22 &  & {---} &  & <0.52 & $\pm$ & 0.54 & <0.42 & $\pm$ & 0.63 & <0.01 & $\pm$ & 0.21 & <0.05 & $\pm$ & 0.24 & <0.15 & $\pm$ & 0.24 \\
HerBS-199 \tablefootmark{\dag} & \multicolumn{1}{c}{----} & \multicolumn{1}{c}{----} &  & <0.76 & $\pm$ & 0.84 & 1.2 & $\pm$ & 0.16 & <0.32 & $\pm$ & 0.51 & <0.24 & $\pm$ & 0.39 & 0.46 & $\pm$ & 0.09 & <-0.1 & $\pm$ & 0.6 & <0.11 & $\pm$ & 0.21 & 0.79 & $\pm$ & 0.23 \\
HerBS-199 E & 13:33:52.30 & 33:49:13.5 & 1.9248 & <0.34 & $\pm$ & 0.57 & 0.56 & $\pm$ & 0.12 & <0.04 & $\pm$ & 0.36 & <0.14 & $\pm$ & 0.3 & <0.09 & $\pm$ & 0.15 & <0.06 & $\pm$ & 0.42 & <0.08 & $\pm$ & 0.15 & <0.07 & $\pm$ & 0.33 \\
HerBS-199 W & 13:33:51.45 & 33:49:18.9 & 1.9197 & <0.42 & $\pm$ & 0.63 & 0.64 & $\pm$ & 0.11 & <0.28 & $\pm$ & 0.36 & <0.1 & $\pm$ & 0.24 & 0.37 & $\pm$ & 0.08 & <-0.16 & $\pm$ & 0.42 & <0.03 & $\pm$ & 0.15 & 0.72 & $\pm$ & 0.2 \\
HerBS-201 & 14:11:18.06 & -01:06:52.6 & 4.1408 & 1.84 & $\pm$ & 0.15 &  & {---} &  & 1.1 & $\pm$ & 0.1 &  & {---} &  & 0.24 & $\pm$ & 0.07 & 0.2 & $\pm$ & 0.06 & 0.2 & $\pm$ & 0.06 & <0.06 & $\pm$ & 0.27 \\
HerBS-202 & 14:33:28.37 & 02:08:09.8 & 2.0224 & 0.63 & $\pm$ & 0.11 &  & {---} &  & 0.44 & $\pm$ & 0.08 &  & {---} &  & <0.13 & $\pm$ & 0.18 & <0.09 & $\pm$ & 0.18 & <0.07 & $\pm$ & 0.15 & <0.06 & $\pm$ & 0.18 \\
HerBS-204 \tablefootmark{\dag} & \multicolumn{1}{c}{----} & \multicolumn{1}{c}{----} &  &  & {---} &  & 0.87 & $\pm$ & 0.13 &  & {---} &  & 0.45 & $\pm$ & 0.12 &  & {---} &  &  & {---} &  &  & {---} &  &  & {---} & \\
HerBS-204 W & 13:29:09.21 & 30:09:59.3 & 3.4937 &  & {---} &  & 0.56 & $\pm$ & 0.1 &  & {---} &  & <0.2 & $\pm$ & 0.24 &  & {---} &  &  & {---} &  &  & {---} &  &  & {---} & \\
HerBS-204 E & 13:29:09.71 & 30:09:57.1 & 3.4933 &  & {---} &  & 0.31 & $\pm$ & 0.09 &  & {---} &  & <0.25 & $\pm$ & 0.27 &  & {---} &  &  & {---} &  &  & {---} &  &  & {---} & \\
HerBS-205 \tablefootmark{\dag} & \multicolumn{1}{c}{----} & \multicolumn{1}{c}{----} &  & 1.74 & $\pm$ & 0.23 & 0.72 & $\pm$ & 0.1 & 1.43 & $\pm$ & 0.19 & 0.5 & $\pm$ & 0.08 & 0.43 & $\pm$ & 0.09 & <0.25 & $\pm$ & 0.36 & 0.27 & $\pm$ & 0.09 & <-0.01 & $\pm$ & 0.36 \\
HerBS-205 NE & 14:51:32.92 & 02:41:02.8 & 2.96 & 0.6 & $\pm$ & 0.13 & 0.31 & $\pm$ & 0.06 & 0.69 & $\pm$ & 0.12 & 0.23 & $\pm$ & 0.05 & 0.16 & $\pm$ & 0.05 & <0.09 & $\pm$ & 0.21 & <0.1 & $\pm$ & 0.15 & <0.1 & $\pm$ & 0.21 \\
HerBS-205 SE & 14:51:32.91 & 02:40:58.9 & 2.9599 & 0.72 & $\pm$ & 0.14 & 0.23 & $\pm$ & 0.06 & 0.51 & $\pm$ & 0.11 & <0.15 & $\pm$ & 0.15 & <0.15 & $\pm$ & 0.15 & <0.13 & $\pm$ & 0.21 & <0.07 & $\pm$ & 0.12 & <-0.08 & $\pm$ & 0.18 \\
HerBS-205 W & 14:51:32.38 & 02:41:00.5 & 2.963 & 0.42 & $\pm$ & 0.12 & 0.18 & $\pm$ & 0.05 & <0.23 & $\pm$ & 0.3 & 0.12 & $\pm$ & 0.04 & <0.12 & $\pm$ & 0.15 & <0.03 & $\pm$ & 0.18 & <0.1 & $\pm$ & 0.18 & <-0.03 & $\pm$ & 0.21 \\
HerBS-206 & 14:04:21.76 & -00:12:17.1 & 2.8122 & 1.03 & $\pm$ & 0.18 &  & {---} &  & 1.06 & $\pm$ & 0.14 &  & {---} &  & <0.15 & $\pm$ & 0.18 & 0.24 & $\pm$ & 0.07 & <0.19 & $\pm$ & 0.21 & <0.07 & $\pm$ & 0.21 \\
\hline
\end{longtable}
\tablefoot{The flux densities listed are the measured values along with $1\sigma$ uncertainties. In the case of nondetections, the flux density is given with a \textquote{<} sign and $3\sigma$ uncertainty.\\
\tablefoottext{\dag}{The fluxes of source multiples in the field of view at similar redshifts are summed up into one component. The redshift used is the average value of the system.}\\
\tablefoottext{\ddag}{Sources with multiples in the field of view at different redshifts (or with no redshift estimates) discarded from the analysis.} \\
\tablefoottext{a}{HerS-19 and HerBS-82 have tentative redshift values, thus discarded from the analysis.} \\
\tablefoottext{b}{HerBS-105 SW is the only component used in the analysis. The NE1 and NE2 components are very weak with only upper limits the do not contribute to the total flux.} \\
\tablefoottext{c}{HerS-9 is a foreground galaxy that is discarded from the analysis.}
}
\end{landscape}

\begin{longtable}[c]{llcccc}
\caption{Dust continuum properties of the 125 \textit{z}-GAL sources derived using the optically thin modified blackbody in Section \ref{section:zgal_opt_thin}. The best-fit values indicate the 50$^{th}$ percentile and their corresponding uncertainties; i.e., the 16$^{th}$ and 84$^{th}$ percentiles of the posterior likelihood distribution. $\mu$ is the magnification factor for the gravitationally lensed sources.}
\label{tab:zgal-opt-thin-results}\\
\hline \hline
\multirow{2}{*}{Source} & \multirow{2}{*}{$z_{spec}$} & \multirow{2}{*}{$\beta$} & \multirow{2}{*}{$T_{\rm dust}$[K]} & \multirow{2}{*}{$log (\rm \mu M_{\rm dust}/M_\odot)$} & \multirow{2}{*}{$log (\mu L_{\rm FIR}/L_\odot)$} \\ 
& & & & &  \\ \hline
\endfirsthead
\multicolumn{6}{c}%
{{\bfseries Table \thetable\: continued}} \\
\hline \hline
\multirow{2}{*}{Source} & \multirow{2}{*}{$z_{spec}$} & \multirow{2}{*}{$\beta$} & \multirow{2}{*}{$T_{\rm dust}$[K]} & \multirow{2}{*}{$log (\rm \mu M_{\rm dust}/M_\odot)$} & \multirow{2}{*}{$log (\mu L_{\rm FIR}/L_\odot)$} \\ 
& & & & &  \\ \hline
\endhead
\hline
\endfoot
\endlastfoot
\multicolumn{6}{c}{HeLMS Sources} \\ \hline
HeLMS-1 & 1.9047 & $2.54_{-0.20}^{+0.18}$ & $25.29_{-2.10}^{+2.77}$ & $10.16_{-0.06}^{+0.05}$ & $13.52_{-0.04}^{+0.04}$ \\
HeLMS-3 & 1.4202 & $2.55_{-0.22}^{+0.20}$ & $24.63_{-2.31}^{+3.22}$ & $10.15_{-0.08}^{+0.07}$ & $13.44_{-0.05}^{+0.07}$ \\
HeLMS-11 & 2.4829 & $1.87_{-0.16}^{+0.15}$ & $28.80_{-2.18}^{+2.64}$ & $10.13_{-0.06}^{+0.06}$ & $13.34_{-0.03}^{+0.03}$ \\
HeLMS-12 & 2.3699 & $2.20_{-0.22}^{+0.20}$ & $29.36_{-2.53}^{+3.28}$ & $9.88_{-0.07}^{+0.06}$ & $13.38_{-0.03}^{+0.03}$ \\
HeLMS-14 & 1.6169 & $2.65_{-0.26}^{+0.24}$ & $22.69_{-1.95}^{+2.62}$ & $10.00_{-0.09}^{+0.08}$ & $13.14_{-0.04}^{+0.04}$ \\
HeLMS-16 & 2.8187 & $2.41_{-0.23}^{+0.21}$ & $29.81_{-2.32}^{+3.07}$ & $9.78_{-0.08}^{+0.07}$ & $13.48_{-0.03}^{+0.03}$ \\
HeLMS-17 & 2.2978 & $2.32_{-0.24}^{+0.22}$ & $28.97_{-2.65}^{+3.55}$ & $9.81_{-0.07}^{+0.06}$ & $13.36_{-0.03}^{+0.03}$ \\
HeLMS-19 & 4.6879 & $1.76_{-0.32}^{+0.34}$ & $50.08_{-8.88}^{+13.88}$ & $9.47_{-0.07}^{+0.06}$ & $13.76_{-0.04}^{+0.04}$ \\
HeLMS-20 & 2.1947 & $2.24_{-0.19}^{+0.19}$ & $27.72_{-2.28}^{+2.84}$ & $9.89_{-0.06}^{+0.06}$ & $13.27_{-0.04}^{+0.03}$ \\
HeLMS-21 & 2.7710 & $1.88_{-0.19}^{+0.18}$ & $38.77_{-3.73}^{+4.78}$ & $9.65_{-0.07}^{+0.07}$ & $13.54_{-0.03}^{+0.03}$ \\
HeLMS-23 & 1.4888 & $2.49_{-0.27}^{+0.28}$ & $28.05_{-3.36}^{+4.89}$ & $9.69_{-0.10}^{+0.09}$ & $13.30_{-0.07}^{+0.10}$ \\
HeLMS-24 & 4.9841 & $1.26_{-0.28}^{+0.27}$ & $51.66_{-8.59}^{+13.85}$ & $9.78_{-0.11}^{+0.09}$ & $13.73_{-0.03}^{+0.03}$ \\
HeLMS-25 & 2.1404 & $2.35_{-0.21}^{+0.19}$ & $27.50_{-2.25}^{+2.96}$ & $9.83_{-0.08}^{+0.07}$ & $13.27_{-0.04}^{+0.04}$ \\
HeLMS-26 & 2.6887 & $2.31_{-0.62}^{+0.56}$ & $30.22_{-5.91}^{+17.28}$ & $9.56_{-0.19}^{+0.15}$ & $13.26_{-0.08}^{+0.15}$ \\
HeLMS-27 & 3.7620 & $1.37_{-0.16}^{+0.15}$ & $46.31_{-3.94}^{+4.96}$ & $9.79_{-0.05}^{+0.05}$ & $13.62_{-0.03}^{+0.03}$ \\
HeLMS-28 & 2.5322 & $2.05_{-0.20}^{+0.20}$ & $29.01_{-2.31}^{+2.84}$ & $9.92_{-0.07}^{+0.06}$ & $13.28_{-0.03}^{+0.03}$ \\
HeLMS-30 & 1.8197 & $2.63_{-0.26}^{+0.23}$ & $24.30_{-2.17}^{+3.11}$ & $9.80_{-0.08}^{+0.06}$ & $13.12_{-0.05}^{+0.04}$ \\
HeLMS-31 & 1.9494 & $2.62_{-0.25}^{+0.25}$ & $24.77_{-2.18}^{+2.81}$ & $9.81_{-0.09}^{+0.08}$ & $13.17_{-0.04}^{+0.05}$ \\
HeLMS-34 & 2.2714 & $1.96_{-0.19}^{+0.18}$ & $36.42_{-3.56}^{+4.77}$ & $9.63_{-0.07}^{+0.07}$ & $13.46_{-0.04}^{+0.04}$ \\
HeLMS-35 & 1.6684 & $2.67_{-0.36}^{+0.39}$ & $20.54_{-2.04}^{+2.81}$ & $9.98_{-0.15}^{+0.12}$ & $12.86_{-0.04}^{+0.05}$ \\
HeLMS-36 & 3.9802 & $2.03_{-0.19}^{+0.19}$ & $43.13_{-3.76}^{+4.76}$ & $9.34_{-0.05}^{+0.05}$ & $13.58_{-0.03}^{+0.03}$ \\
HeLMS-37 & 2.7576 & $1.93_{-0.23}^{+0.23}$ & $33.85_{-3.41}^{+4.31}$ & $9.71_{-0.07}^{+0.07}$ & $13.34_{-0.03}^{+0.03}$ \\
HeLMS-38 & 2.1898 & $2.36_{-0.23}^{+0.21}$ & $30.56_{-2.99}^{+4.19}$ & $9.58_{-0.09}^{+0.08}$ & $13.30_{-0.04}^{+0.04}$ \\
HeLMS-39 & 2.7659 & $2.58_{-0.28}^{+0.25}$ & $26.91_{-2.32}^{+3.16}$ & $9.68_{-0.08}^{+0.06}$ & $13.24_{-0.03}^{+0.03}$ \\
HeLMS-40 & 3.1420 & $1.77_{-0.19}^{+0.19}$ & $36.01_{-3.06}^{+3.70}$ & $9.77_{-0.06}^{+0.06}$ & $13.43_{-0.03}^{+0.03}$ \\
HeLMS-41 & 2.3353 & $1.94_{-0.30}^{+0.24}$ & $31.96_{-3.42}^{+5.76}$ & $9.71_{-0.09}^{+0.08}$ & $13.23_{-0.04}^{+0.04}$ \\
HeLMS-42 & 1.9558 & $2.67_{-0.29}^{+0.27}$ & $24.71_{-2.39}^{+3.11}$ & $9.74_{-0.10}^{+0.09}$ & $13.12_{-0.04}^{+0.04}$ \\
HeLMS-44 & 1.3700 & $2.73_{-0.46}^{+0.53}$ & $23.35_{-3.72}^{+6.72}$ & $9.62_{-0.22}^{+0.16}$ & $12.89_{-0.10}^{+0.11}$ \\
HeLMS-45 & 5.3994 & $1.51_{-0.24}^{+0.23}$ & $54.79_{-7.78}^{+11.51}$ & $9.52_{-0.07}^{+0.06}$ & $13.77_{-0.03}^{+0.03}$ \\
HeLMS-46 & 2.5765 & $2.09_{-0.22}^{+0.21}$ & $27.16_{-2.38}^{+3.03}$ & $9.96_{-0.07}^{+0.07}$ & $13.19_{-0.04}^{+0.03}$ \\
HeLMS-47 & 2.2232 & $2.08_{-0.20}^{+0.20}$ & $32.93_{-3.29}^{+4.34}$ & $9.63_{-0.07}^{+0.06}$ & $13.31_{-0.04}^{+0.04}$ \\
HeLMS-48 & 3.3514 & $1.72_{-0.18}^{+0.17}$ & $31.50_{-2.30}^{+2.78}$ & $10.05_{-0.06}^{+0.05}$ & $13.37_{-0.03}^{+0.03}$ \\
HeLMS-49 & 2.2154 & $2.14_{-0.27}^{+0.25}$ & $32.16_{-3.46}^{+5.08}$ & $9.61_{-0.09}^{+0.07}$ & $13.29_{-0.05}^{+0.04}$ \\
HeLMS-50 & 2.0531 & $2.46_{-0.25}^{+0.24}$ & $24.60_{-2.14}^{+2.88}$ & $9.81_{-0.08}^{+0.07}$ & $13.04_{-0.03}^{+0.04}$ \\
HeLMS-51 & 2.1567 & $2.79_{-0.32}^{+0.29}$ & $22.52_{-2.02}^{+3.00}$ & $9.79_{-0.09}^{+0.07}$ & $13.01_{-0.03}^{+0.04}$ \\
HeLMS-52 & 2.2092 & $2.42_{-0.25}^{+0.24}$ & $30.03_{-2.98}^{+4.14}$ & $9.57_{-0.10}^{+0.08}$ & $13.29_{-0.05}^{+0.04}$ \\
HeLMS-54 & 2.7070 & $2.87_{-0.66}^{+0.41}$ & $25.56_{-3.16}^{+9.53}$ & $9.50_{-0.15}^{+0.12}$ & $13.14_{-0.06}^{+0.07}$ \\
HeLMS-55 & 2.2834 & $2.18_{-0.25}^{+0.23}$ & $26.66_{-2.37}^{+3.19}$ & $9.86_{-0.08}^{+0.07}$ & $13.11_{-0.03}^{+0.03}$ \\
HeLMS-56 & 3.3896 & $1.68_{-0.19}^{+0.18}$ & $39.03_{-3.61}^{+4.74}$ & $9.65_{-0.07}^{+0.07}$ & $13.41_{-0.03}^{+0.03}$ \\
HeLMS-57 & 1.9817 & $3.22_{-0.33}^{+0.29}$ & $22.17_{-1.96}^{+2.77}$ & $9.59_{-0.11}^{+0.09}$ & $13.05_{-0.04}^{+0.04}$ \\
\hline \multicolumn{6}{c}{HerS Sources} \\ \hline
HerS-2 & 2.0151 & $2.18_{-0.21}^{+0.18}$ & $26.31_{-2.11}^{+2.82}$ & $10.15_{-0.07}^{+0.06}$ & $13.36_{-0.03}^{+0.04}$ \\
HerS-3 & 3.0608 & $2.26_{-0.23}^{+0.18}$ & $35.56_{-3.09}^{+4.60}$ & $9.72_{-0.07}^{+0.07}$ & $13.72_{-0.04}^{+0.03}$ \\
HerS-5 & 1.4493 & $2.52_{-0.32}^{+0.27}$ & $23.49_{-2.39}^{+3.78}$ & $9.91_{-0.12}^{+0.09}$ & $13.05_{-0.05}^{+0.06}$ \\
HerS-8 & 2.2431 & $2.56_{-0.30}^{+0.30}$ & $26.75_{-2.65}^{+3.65}$ & $9.69_{-0.11}^{+0.09}$ & $13.23_{-0.03}^{+0.04}$ \\
HerS-10 & 2.4688 & $2.32_{-0.23}^{+0.21}$ & $27.61_{-2.30}^{+3.11}$ & $9.79_{-0.08}^{+0.07}$ & $13.22_{-0.04}^{+0.03}$ \\
HerS-11 & 4.6618 & $2.12_{-0.30}^{+0.31}$ & $38.59_{-5.19}^{+6.59}$ & $9.59_{-0.06}^{+0.06}$ & $13.67_{-0.04}^{+0.03}$ \\
HerS-12 & 2.2706 & $1.89_{-0.21}^{+0.20}$ & $32.07_{-3.28}^{+4.70}$ & $9.76_{-0.09}^{+0.08}$ & $13.25_{-0.04}^{+0.04}$ \\
HerS-13 & 2.4759 & $1.67_{-0.23}^{+0.21}$ & $39.35_{-4.91}^{+7.69}$ & $9.61_{-0.09}^{+0.08}$ & $13.38_{-0.04}^{+0.04}$ \\
HerS-14 & 3.3441 & $1.69_{-0.17}^{+0.17}$ & $41.53_{-3.69}^{+4.46}$ & $9.66_{-0.05}^{+0.05}$ & $13.54_{-0.03}^{+0.03}$ \\
HerS-15 & 2.3019 & $2.30_{-0.21}^{+0.19}$ & $25.10_{-1.86}^{+2.31}$ & $9.96_{-0.08}^{+0.07}$ & $13.14_{-0.03}^{+0.03}$ \\
HerS-16 & 2.1981 & $2.11_{-0.20}^{+0.19}$ & $25.92_{-2.14}^{+2.73}$ & $10.01_{-0.06}^{+0.06}$ & $13.13_{-0.03}^{+0.03}$ \\
HerS-17 & 3.0182 & $2.11_{-0.19}^{+0.18}$ & $32.28_{-2.71}^{+3.28}$ & $9.74_{-0.06}^{+0.06}$ & $13.41_{-0.03}^{+0.03}$ \\
HerS-18 & 1.6926 & $2.85_{-0.27}^{+0.24}$ & $22.52_{-1.98}^{+2.75}$ & $9.77_{-0.10}^{+0.08}$ & $13.02_{-0.04}^{+0.05}$ \\
HerS-20 & 2.0792 & $3.00_{-0.26}^{+0.26}$ & $21.72_{-1.79}^{+2.21}$ & $9.79_{-0.10}^{+0.08}$ & $13.04_{-0.04}^{+0.04}$ \\
\hline \multicolumn{6}{c}{HerBS Sources} \\ \hline
HerBS-46 & 1.8349 & $3.11_{-0.30}^{+0.26}$ & $20.81_{-1.81}^{+2.56}$ & $9.77_{-0.08}^{+0.07}$ & $12.97_{-0.04}^{+0.04}$ \\
HerBS-48 & 3.1438 & $2.28_{-0.21}^{+0.21}$ & $35.65_{-2.85}^{+3.47}$ & $9.45_{-0.08}^{+0.07}$ & $13.48_{-0.03}^{+0.03}$ \\
HerBS-50 & 2.9283 & $2.31_{-0.16}^{+0.17}$ & $29.77_{-1.96}^{+2.21}$ & $9.79_{-0.06}^{+0.05}$ & $13.41_{-0.03}^{+0.03}$ \\
HerBS-51 & 2.1827 & $2.62_{-0.21}^{+0.19}$ & $26.10_{-2.02}^{+2.58}$ & $9.69_{-0.06}^{+0.05}$ & $13.20_{-0.04}^{+0.03}$ \\
HerBS-53 & 1.4227 & $2.60_{-0.31}^{+0.28}$ & $20.65_{-1.96}^{+2.94}$ & $9.92_{-0.10}^{+0.08}$ & $12.76_{-0.04}^{+0.06}$ \\
HerBS-61 & 3.7293 & $1.89_{-0.15}^{+0.14}$ & $35.68_{-2.08}^{+2.39}$ & $9.77_{-0.05}^{+0.05}$ & $13.49_{-0.03}^{+0.03}$ \\
HerBS-62 & 2.5738 & $2.12_{-0.19}^{+0.19}$ & $30.00_{-2.32}^{+2.73}$ & $9.78_{-0.06}^{+0.06}$ & $13.28_{-0.03}^{+0.03}$ \\
HerBS-65 & 2.6858 & $1.82_{-0.25}^{+0.25}$ & $34.51_{-3.87}^{+4.90}$ & $9.66_{-0.08}^{+0.07}$ & $13.26_{-0.04}^{+0.04}$ \\
HerBS-72 & 3.6380 & $1.98_{-0.30}^{+0.29}$ & $37.98_{-4.89}^{+7.25}$ & $9.52_{-0.09}^{+0.08}$ & $13.44_{-0.07}^{+0.07}$ \\
HerBS-74 & 2.5596 & $2.53_{-0.30}^{+0.26}$ & $28.03_{-2.51}^{+3.91}$ & $9.53_{-0.09}^{+0.08}$ & $13.17_{-0.04}^{+0.04}$ \\
HerBS-76 & 2.3310 & $2.16_{-0.21}^{+0.19}$ & $28.18_{-2.19}^{+2.77}$ & $9.79_{-0.07}^{+0.06}$ & $13.16_{-0.04}^{+0.03}$ \\
HerBS-78 & 3.7344 & $1.63_{-0.26}^{+0.25}$ & $42.98_{-5.31}^{+8.02}$ & $9.57_{-0.08}^{+0.07}$ & $13.48_{-0.06}^{+0.06}$ \\
HerBS-83 & 3.9438 & $2.16_{-0.19}^{+0.18}$ & $32.70_{-2.28}^{+2.56}$ & $9.72_{-0.05}^{+0.04}$ & $13.45_{-0.03}^{+0.03}$ \\
HerBS-85 & 2.8169 & $3.12_{-0.34}^{+0.28}$ & $27.05_{-2.22}^{+3.55}$ & $9.29_{-0.11}^{+0.08}$ & $13.27_{-0.03}^{+0.03}$ \\
HerBS-91 & 2.4047 & $2.02_{-0.24}^{+0.24}$ & $34.30_{-3.78}^{+5.34}$ & $9.53_{-0.07}^{+0.06}$ & $13.27_{-0.04}^{+0.04}$ \\
HerBS-92 & 3.2644 & $1.75_{-0.48}^{+0.32}$ & $33.69_{-4.22}^{+11.06}$ & $9.70_{-0.10}^{+0.09}$ & $13.22_{-0.06}^{+0.05}$ \\
HerBS-105 & 2.6684 & $2.64_{-0.31}^{+0.29}$ & $26.84_{-2.54}^{+3.43}$ & $9.54_{-0.09}^{+0.08}$ & $13.14_{-0.04}^{+0.04}$ \\
HerBS-108 & 3.7168 & $2.02_{-0.16}^{+0.16}$ & $35.74_{-2.53}^{+2.98}$ & $9.67_{-0.05}^{+0.04}$ & $13.50_{-0.03}^{+0.04}$ \\
HerBS-110 & 2.6810 & $2.03_{-0.20}^{+0.18}$ & $28.11_{-1.88}^{+2.37}$ & $9.90_{-0.05}^{+0.05}$ & $13.17_{-0.03}^{+0.03}$ \\
HerBS-115 & 2.3706 & $2.46_{-0.19}^{+0.18}$ & $27.42_{-2.05}^{+2.48}$ & $9.69_{-0.05}^{+0.05}$ & $13.19_{-0.03}^{+0.04}$ \\
HerBS-116 & 3.1547 & $2.04_{-0.19}^{+0.18}$ & $31.26_{-2.15}^{+2.75}$ & $9.77_{-0.06}^{+0.06}$ & $13.31_{-0.03}^{+0.03}$ \\
HerBS-124 & 2.2776 & $2.07_{-0.26}^{+0.25}$ & $31.65_{-3.23}^{+4.55}$ & $9.59_{-0.10}^{+0.08}$ & $13.18_{-0.04}^{+0.04}$ \\
HerBS-125 & 2.5739 & $2.22_{-0.22}^{+0.20}$ & $27.80_{-2.04}^{+2.87}$ & $9.76_{-0.07}^{+0.06}$ & $13.15_{-0.03}^{+0.03}$ \\
HerBS-126 & 2.5875 & $2.53_{-0.29}^{+0.25}$ & $27.40_{-2.50}^{+3.66}$ & $9.57_{-0.08}^{+0.06}$ & $13.15_{-0.04}^{+0.03}$ \\
HerBS-127 & 3.1958 & $2.67_{-0.24}^{+0.23}$ & $32.80_{-2.66}^{+3.41}$ & $9.28_{-0.08}^{+0.06}$ & $13.42_{-0.03}^{+0.03}$ \\
HerBS-128 & 2.0681 & $2.01_{-0.26}^{+0.21}$ & $28.63_{-2.75}^{+4.34}$ & $9.73_{-0.08}^{+0.06}$ & $13.03_{-0.04}^{+0.04}$ \\
HerBS-129 & 3.3074 & $2.08_{-0.17}^{+0.15}$ & $30.75_{-1.94}^{+2.47}$ & $9.79_{-0.06}^{+0.05}$ & $13.31_{-0.03}^{+0.02}$ \\
HerBS-134 & 3.1725 & $1.85_{-0.18}^{+0.16}$ & $34.69_{-2.59}^{+3.30}$ & $9.67_{-0.06}^{+0.05}$ & $13.30_{-0.03}^{+0.03}$ \\
HerBS-136 & 3.2884 & $2.42_{-0.27}^{+0.24}$ & $31.38_{-2.73}^{+3.58}$ & $9.50_{-0.06}^{+0.06}$ & $13.32_{-0.04}^{+0.04}$ \\
HerBS-137 & 3.0408 & $2.48_{-0.27}^{+0.25}$ & $31.70_{-2.96}^{+3.88}$ & $9.41_{-0.07}^{+0.07}$ & $13.31_{-0.03}^{+0.04}$ \\
HerBS-140 & 2.7799 & $1.96_{-0.23}^{+0.23}$ & $34.87_{-3.33}^{+4.65}$ & $9.52_{-0.07}^{+0.06}$ & $13.25_{-0.03}^{+0.03}$ \\
HerBS-143 & 2.2406 & $2.33_{-0.24}^{+0.23}$ & $26.06_{-2.10}^{+2.69}$ & $9.71_{-0.07}^{+0.06}$ & $13.00_{-0.03}^{+0.03}$ \\
HerBS-147 & 3.1150 & $2.12_{-0.21}^{+0.20}$ & $36.18_{-3.32}^{+4.20}$ & $9.43_{-0.06}^{+0.06}$ & $13.36_{-0.04}^{+0.03}$ \\
HerBS-149 & 2.6650 & $1.87_{-0.22}^{+0.22}$ & $34.80_{-3.36}^{+4.23}$ & $9.58_{-0.08}^{+0.07}$ & $13.24_{-0.04}^{+0.03}$ \\
HerBS-150 & 3.6734 & $2.01_{-0.15}^{+0.15}$ & $32.77_{-2.00}^{+2.36}$ & $9.75_{-0.04}^{+0.04}$ & $13.37_{-0.03}^{+0.03}$ \\
HerBS-153 & 3.1501 & $1.81_{-0.18}^{+0.17}$ & $39.66_{-3.41}^{+4.35}$ & $9.56_{-0.06}^{+0.05}$ & $13.45_{-0.03}^{+0.03}$ \\
HerBS-157 & 1.8964 & $2.26_{-0.24}^{+0.22}$ & $22.37_{-1.95}^{+2.53}$ & $10.01_{-0.07}^{+0.06}$ & $12.85_{-0.04}^{+0.03}$ \\
HerBS-162 & 2.4738 & $3.33_{-0.32}^{+0.28}$ & $22.24_{-1.80}^{+2.50}$ & $9.52_{-0.10}^{+0.08}$ & $13.07_{-0.04}^{+0.04}$ \\
HerBS-164 & 2.0126 & $2.92_{-0.28}^{+0.27}$ & $21.78_{-1.76}^{+2.41}$ & $9.69_{-0.10}^{+0.08}$ & $12.89_{-0.03}^{+0.04}$ \\
HerBS-165 & 2.2251 & $2.45_{-0.28}^{+0.26}$ & $25.24_{-2.41}^{+3.26}$ & $9.65_{-0.10}^{+0.09}$ & $12.94_{-0.04}^{+0.03}$ \\
HerBS-167 & 2.2144 & $2.94_{-0.26}^{+0.24}$ & $20.79_{-1.59}^{+2.05}$ & $9.80_{-0.10}^{+0.08}$ & $12.88_{-0.03}^{+0.03}$ \\
HerBS-169 & 2.6977 & $2.17_{-0.23}^{+0.23}$ & $31.02_{-2.88}^{+3.68}$ & $9.62_{-0.07}^{+0.07}$ & $13.23_{-0.04}^{+0.03}$ \\
HerBS-171 & 2.4793 & $2.37_{-0.53}^{+0.32}$ & $27.58_{-2.91}^{+6.75}$ & $9.60_{-0.11}^{+0.09}$ & $13.08_{-0.04}^{+0.04}$ \\
HerBS-172 & 2.9246 & $2.90_{-0.31}^{+0.28}$ & $28.80_{-2.78}^{+3.92}$ & $9.22_{-0.08}^{+0.07}$ & $13.20_{-0.03}^{+0.03}$ \\
HerBS-175 & 3.1575 & $1.93_{-0.23}^{+0.22}$ & $32.91_{-2.80}^{+3.47}$ & $9.64_{-0.06}^{+0.06}$ & $13.21_{-0.03}^{+0.03}$ \\
HerBS-176 & 2.9805 & $2.16_{-0.21}^{+0.19}$ & $32.30_{-2.69}^{+3.58}$ & $9.55_{-0.07}^{+0.06}$ & $13.25_{-0.03}^{+0.03}$ \\
HerBS-177 & 3.9625 & $1.88_{-0.14}^{+0.15}$ & $34.77_{-2.12}^{+2.34}$ & $9.84_{-0.04}^{+0.04}$ & $13.50_{-0.03}^{+0.03}$ \\
HerBS-179 & 3.9423 & $1.88_{-0.18}^{+0.18}$ & $38.24_{-2.95}^{+3.54}$ & $9.57_{-0.05}^{+0.04}$ & $13.43_{-0.03}^{+0.03}$ \\
HerBS-183 & 1.8910 & $2.15_{-0.27}^{+0.23}$ & $23.54_{-2.23}^{+3.30}$ & $9.96_{-0.08}^{+0.06}$ & $12.87_{-0.04}^{+0.04}$ \\
HerBS-185 & 4.3238 & $1.89_{-0.21}^{+0.20}$ & $36.16_{-3.60}^{+4.75}$ & $9.74_{-0.06}^{+0.06}$ & $13.49_{-0.04}^{+0.04}$ \\
HerBS-187 & 1.8279 & $2.46_{-0.26}^{+0.25}$ & $24.59_{-2.54}^{+3.39}$ & $9.66_{-0.08}^{+0.07}$ & $12.89_{-0.05}^{+0.04}$ \\
HerBS-188 & 2.7675 & $1.89_{-0.21}^{+0.20}$ & $34.37_{-3.09}^{+4.02}$ & $9.57_{-0.06}^{+0.06}$ & $13.21_{-0.04}^{+0.04}$ \\
HerBS-190 & 2.5890 & $2.14_{-0.24}^{+0.23}$ & $29.17_{-2.71}^{+3.51}$ & $9.70_{-0.09}^{+0.07}$ & $13.14_{-0.04}^{+0.04}$ \\
HerBS-191 & 3.4428 & $1.81_{-0.23}^{+0.22}$ & $36.07_{-3.25}^{+4.12}$ & $9.62_{-0.06}^{+0.06}$ & $13.30_{-0.03}^{+0.03}$ \\
HerBS-193 & 3.6951 & $2.17_{-0.20}^{+0.20}$ & $38.29_{-3.34}^{+3.97}$ & $9.36_{-0.05}^{+0.05}$ & $13.46_{-0.04}^{+0.03}$ \\
HerBS-194 & 2.3325 & $2.12_{-0.25}^{+0.24}$ & $29.59_{-3.09}^{+4.30}$ & $9.63_{-0.08}^{+0.07}$ & $13.10_{-0.05}^{+0.04}$ \\
HerBS-197 & 2.4170 & $2.58_{-0.34}^{+0.37}$ & $26.90_{-2.70}^{+3.62}$ & $9.50_{-0.16}^{+0.12}$ & $13.07_{-0.04}^{+0.03}$ \\
HerBS-199 & 1.9223 & $2.07_{-0.66}^{+0.36}$ & $28.88_{-4.44}^{+22.18}$ & $9.60_{-0.24}^{+0.11}$ & $12.98_{-0.06}^{+0.13}$ \\
HerBS-201 & 4.1408 & $2.41_{-0.26}^{+0.24}$ & $33.57_{-3.45}^{+4.64}$ & $9.42_{-0.06}^{+0.06}$ & $13.41_{-0.04}^{+0.04}$ \\
HerBS-202 & 2.0224 & $2.53_{-0.28}^{+0.26}$ & $26.18_{-2.73}^{+3.72}$ & $9.56_{-0.09}^{+0.08}$ & $13.01_{-0.05}^{+0.04}$ \\
HerBS-204 & 3.4935 & $1.88_{-0.28}^{+0.30}$ & $47.78_{-7.55}^{+10.50}$ & $9.15_{-0.09}^{+0.10}$ & $13.46_{-0.04}^{+0.05}$ \\
HerBS-205 & 2.9610 & $1.94_{-0.40}^{+0.36}$ & $35.84_{-5.37}^{+10.62}$ & $9.49_{-0.12}^{+0.11}$ & $13.29_{-0.07}^{+0.08}$ \\
HerBS-206 & 2.8122 & $2.14_{-0.22}^{+0.20}$ & $32.90_{-2.75}^{+3.73}$ & $9.48_{-0.07}^{+0.06}$ & $13.21_{-0.03}^{+0.03}$ \\
\hline \multicolumn{6}{c}{Pilot Program Sources} \\ \hline 
HerBS-34 & 2.6637 & $1.82_{-0.14}^{+0.13}$ & $32.69_{-2.18}^{+2.69}$ & $9.92_{-0.05}^{+0.04}$ & $13.39_{-0.03}^{+0.03}$ \\
HerBS-44 & 2.9268 & $2.00_{-0.16}^{+0.16}$ & $37.78_{-2.85}^{+3.53}$ & $9.58_{-0.06}^{+0.05}$ & $13.52_{-0.03}^{+0.03}$ \\
HerBS-54 & 2.4417 & $2.17_{-0.18}^{+0.17}$ & $26.14_{-1.84}^{+2.24}$ & $10.00_{-0.06}^{+0.06}$ & $13.18_{-0.03}^{+0.03}$ \\
HerBS-58 & 2.0842 & $2.36_{-0.21}^{+0.20}$ & $23.95_{-1.87}^{+2.38}$ & $9.93_{-0.06}^{+0.05}$ & $13.02_{-0.04}^{+0.03}$ \\
HerBS-79 & 2.0782 & $2.48_{-0.21}^{+0.22}$ & $24.14_{-1.91}^{+2.28}$ & $9.83_{-0.08}^{+0.07}$ & $13.02_{-0.03}^{+0.03}$ \\
HerBS-89a & 2.9497 & $1.90_{-0.13}^{+0.13}$ & $28.63_{-1.67}^{+1.94}$ & $10.07_{-0.04}^{+0.04}$ & $13.30_{-0.03}^{+0.03}$ \\
HerBS-113 & 2.7870 & $1.80_{-0.20}^{+0.18}$ & $33.24_{-2.81}^{+3.96}$ & $9.74_{-0.07}^{+0.05}$ & $13.23_{-0.04}^{+0.03}$ \\
HerBS-154 & 3.7070 & $2.30_{-0.19}^{+0.18}$ & $34.29_{-2.55}^{+3.03}$ & $9.49_{-0.04}^{+0.04}$ & $13.44_{-0.03}^{+0.03}$ \\
\hline 
\end{longtable}

\begin{appendix}
\section{\textit{z}-GAL sources spectral energy distribution}
\begin{figure}[h!]
    \centering
    \includegraphics[width=\textwidth]{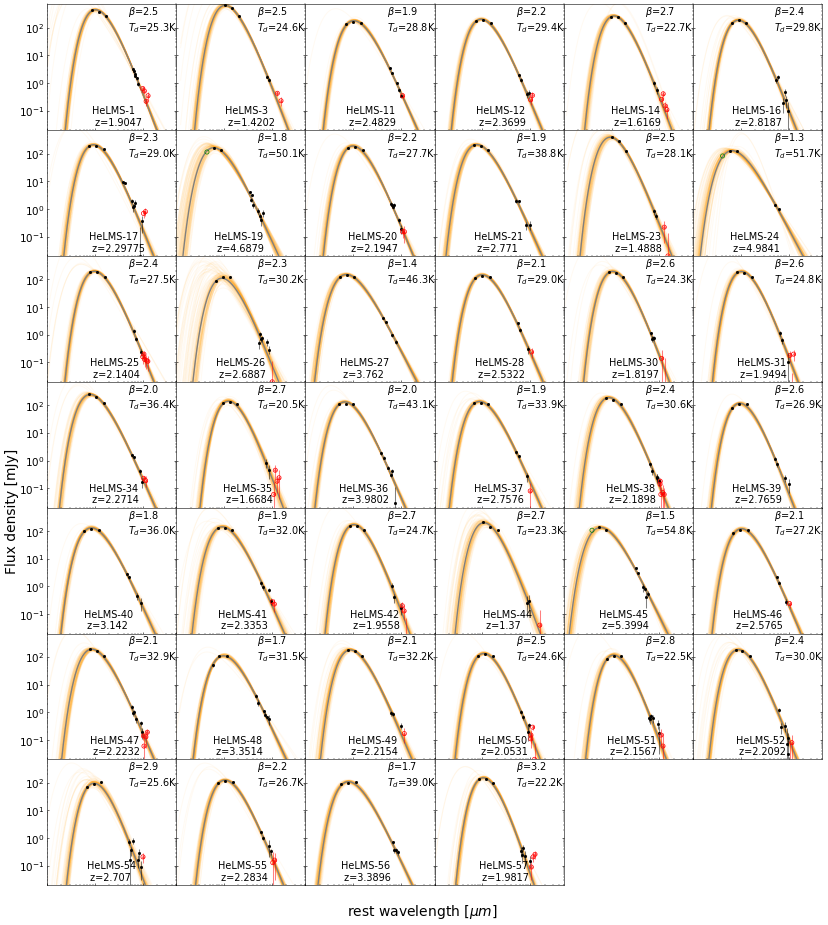}
    \caption{Spectral energy distributions of the HeLMS sources. The observed flux densities that are used in the computation of the dust properties are shown as black dots and the error bars correspond to their 1$\sigma$ uncertainties. We also plot the millimeter flux densities at which the wavelength is above 1000 $\rm \mu m$ in rest-frame as open red circles and the flux densities at which the rest-frame wavelength is below 50 $\rm \mu m$ as open green circles. The solid gray line is the MBB best fit and the orange lines are the output sampling of 100 random walks from the EMCEE output. The names of the sources and their spectroscopic redshifts are given in the lower right corner of each panel.}
    \label{fig:SED-all-sources-HeLMS}
\end{figure}

\begin{figure}[h!]
    \ContinuedFloat
    \centering
    \vspace*{4cm}
    \includegraphics[width=\textwidth]{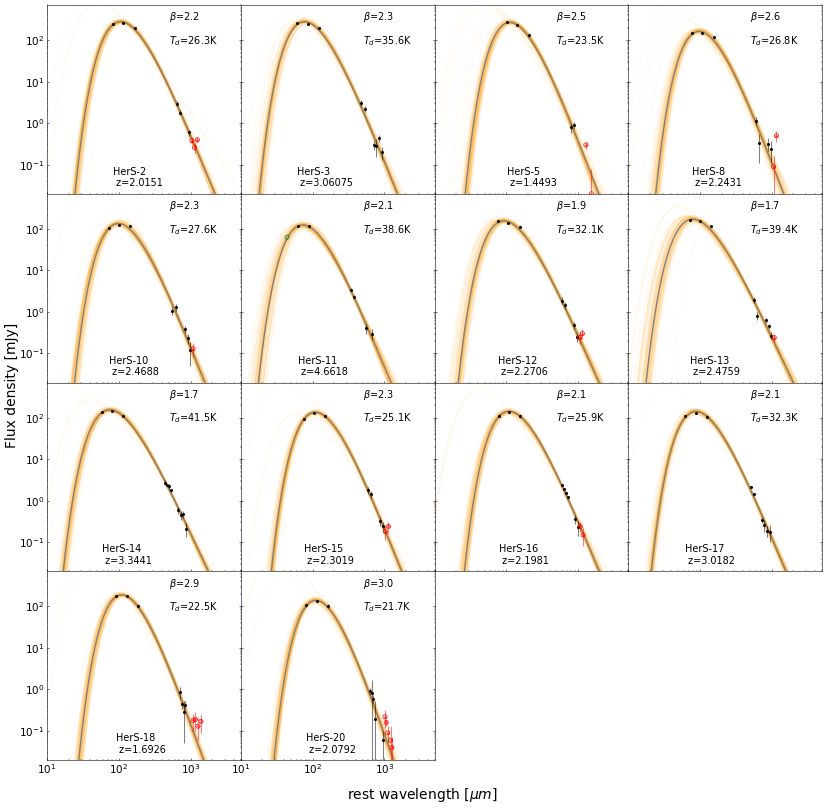}
    \caption{continued, for the HerS sources.}
    \label{fig:SED-all-sources-HerS}
\end{figure}

\begin{figure}[h!]
    \ContinuedFloat
    \centering
    \includegraphics[width=\textwidth]{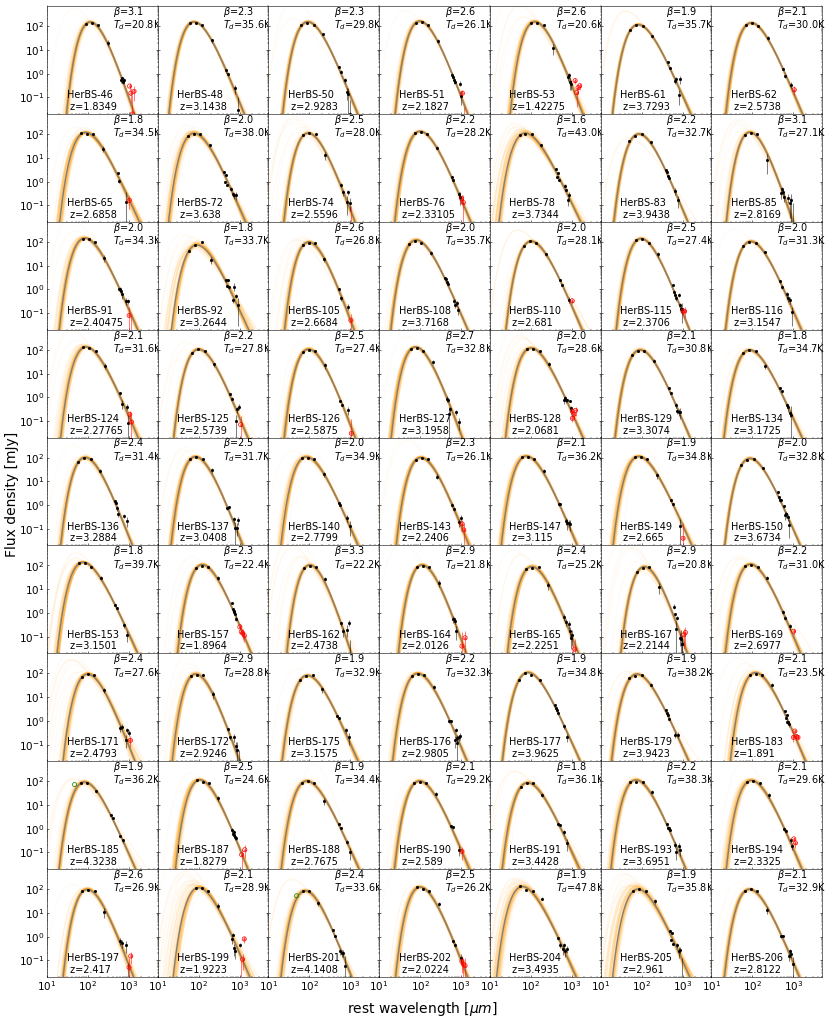}
    \caption{continued, for the HerBS sources.}
    \label{fig:SED-all-sources-HerBS}
\end{figure}

\begin{figure}[h!]
    \ContinuedFloat
    \centering
    \vspace*{7cm}
    \includegraphics[width=\textwidth]{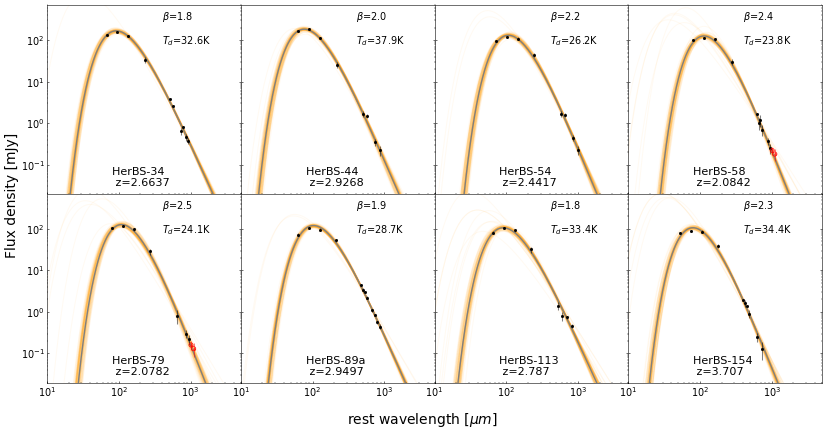}
    \caption{continued, for the Pilot Program sources.}
    \label{fig:SED-pilot-sources}
\end{figure}

\FloatBarrier
\section{MBB redshift-binned results}
\begin{figure}[h!]
    \centering
    \includegraphics[width=0.9\textwidth]{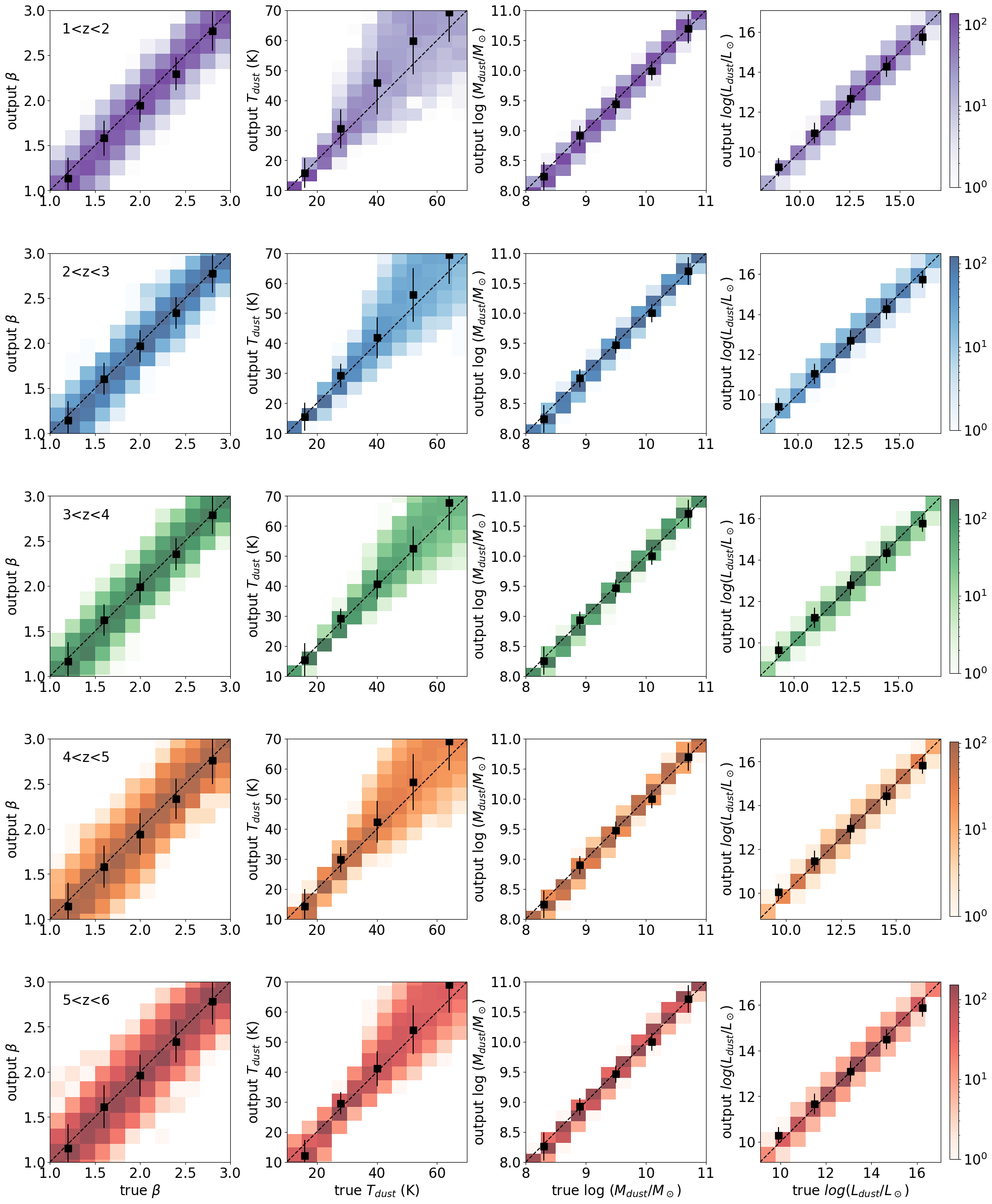}
    \caption{Same as Fig. \ref{fig:mock-thin-hist2d} but redshift-binned results of the mock catalog.}
    \label{fig:mock-thin-hist2d-redshift-binned}
\end{figure}

\FloatBarrier
\section{GMBB details}\label{Appendix:forced-thick-thin}
\begin{figure}[h!]
    \centering
    \vspace*{2cm}
    \includegraphics[width=\textwidth]{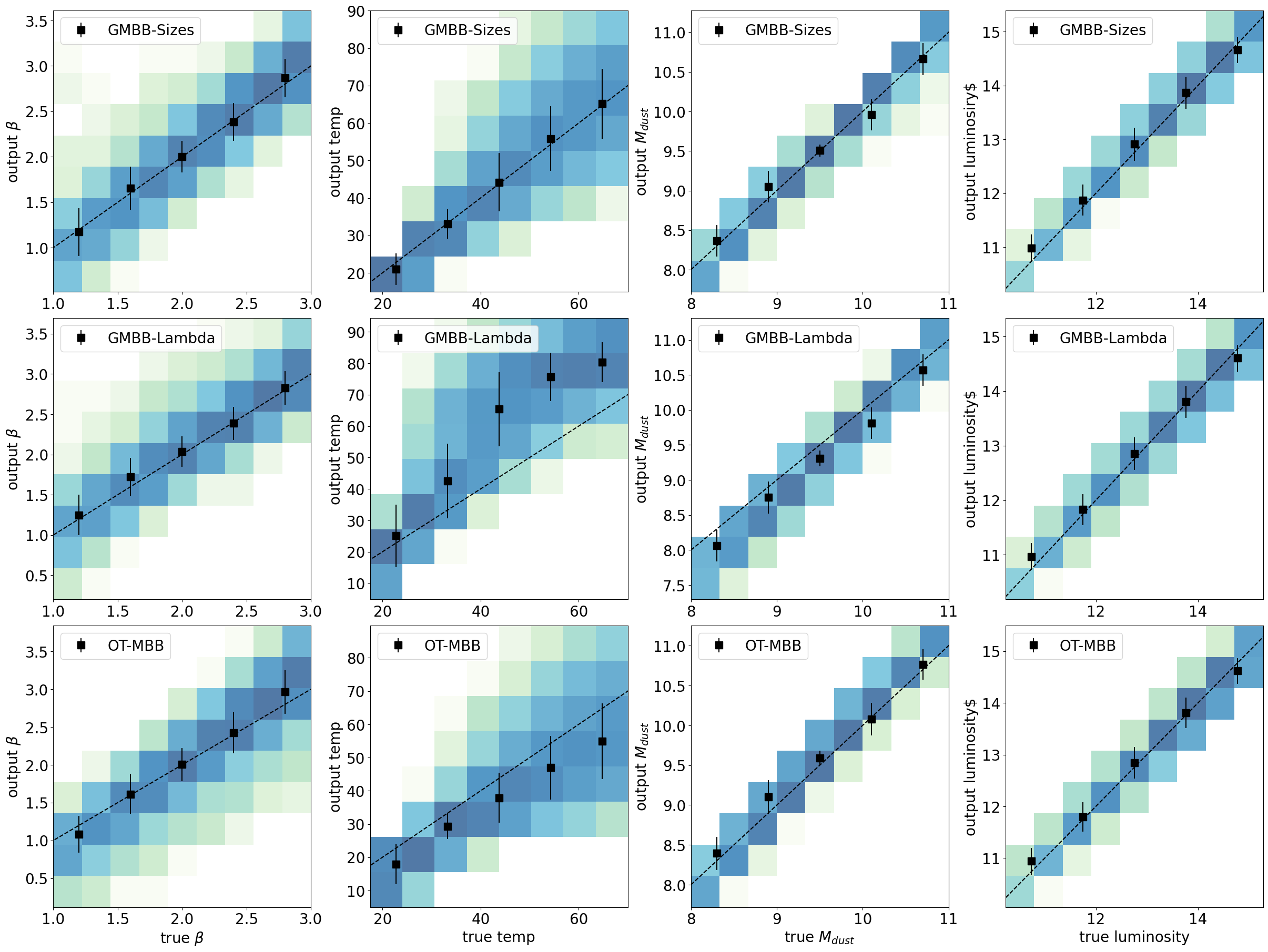}
    \caption{Same as Fig. \ref{fig:density_hist2d_gmbb} but binned to where the medium becomes optically thin at lower wavelengths ($\rm \lambda_{thick}<100 \mu m$). This shows that GMBB-Sizes (first row) estimates the output parameters with good accuracy, as does the MBB (third row), but that GMBB-Lambda (second row) will always overestimate the dust temperatures within this range of $\rm \lambda_{thick}$.}
    \label{fig:mock-thick-hist2d-forced-thin}
\end{figure}

\begin{figure}[h!]
    \centering
    \vspace*{4cm}
    \includegraphics[width=\textwidth]{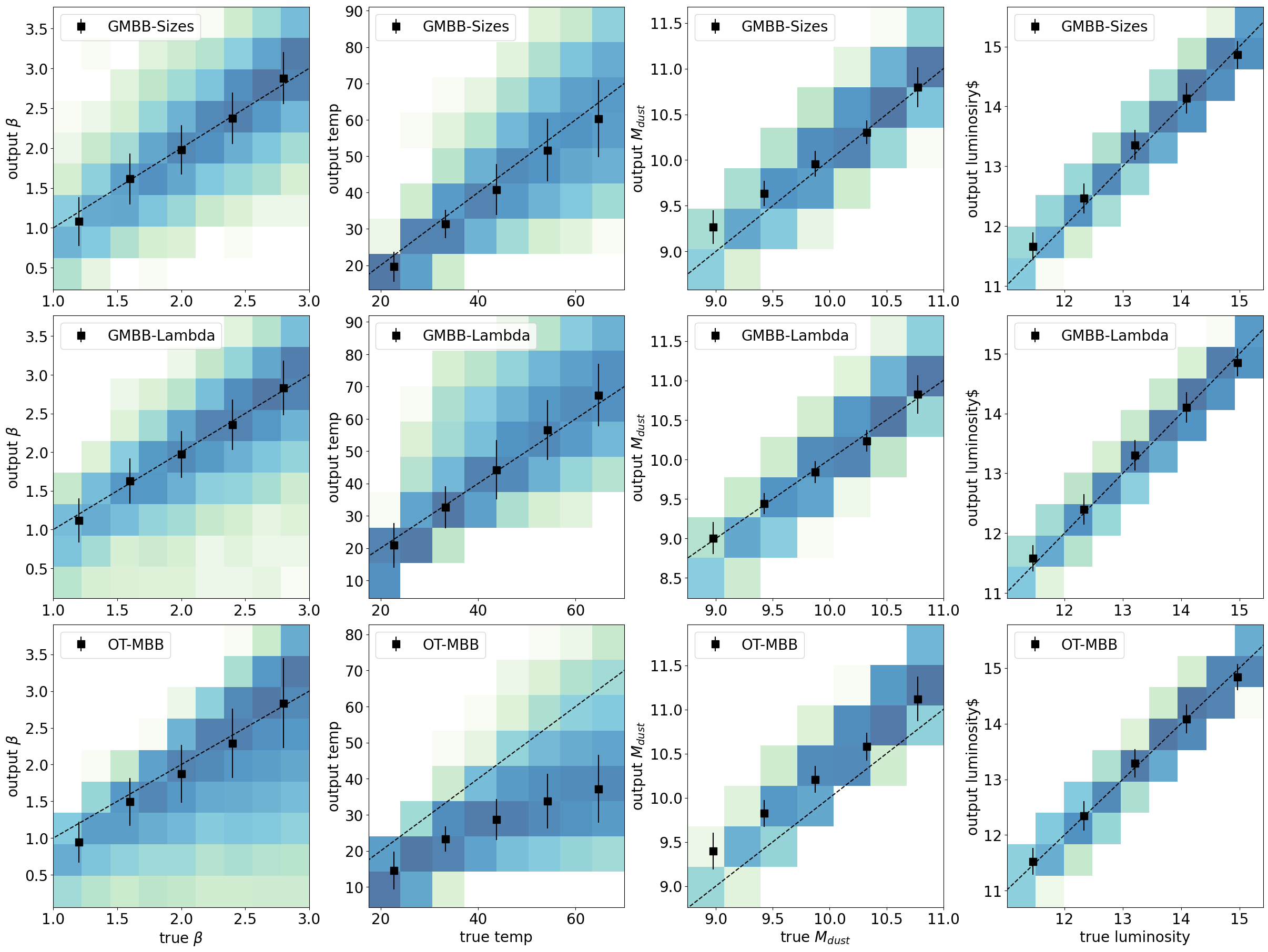}
    \caption{Same as Fig. \ref{fig:density_hist2d_gmbb} but binned to where the medium becomes optically thick at higher wavelengths($\rm \lambda_{thick} > 100 \rm \mu m$). This shows us that the optically thin method will always underestimate the dust temperatures and consequently $\rm M_{\rm dust}$. On the other hand, both versions of the GMBB are able to retrieve (on average) the true dust parameters within this range of $\rm \lambda_{thick}$.}
    \label{fig:mock-thick-hist2d-forced-thick}
\end{figure} 

\FloatBarrier
\section{\textit{z}-GAL sources posterior distribution}
\begin{figure}[h!]
    \centering
    \includegraphics[width=0.7\textwidth]{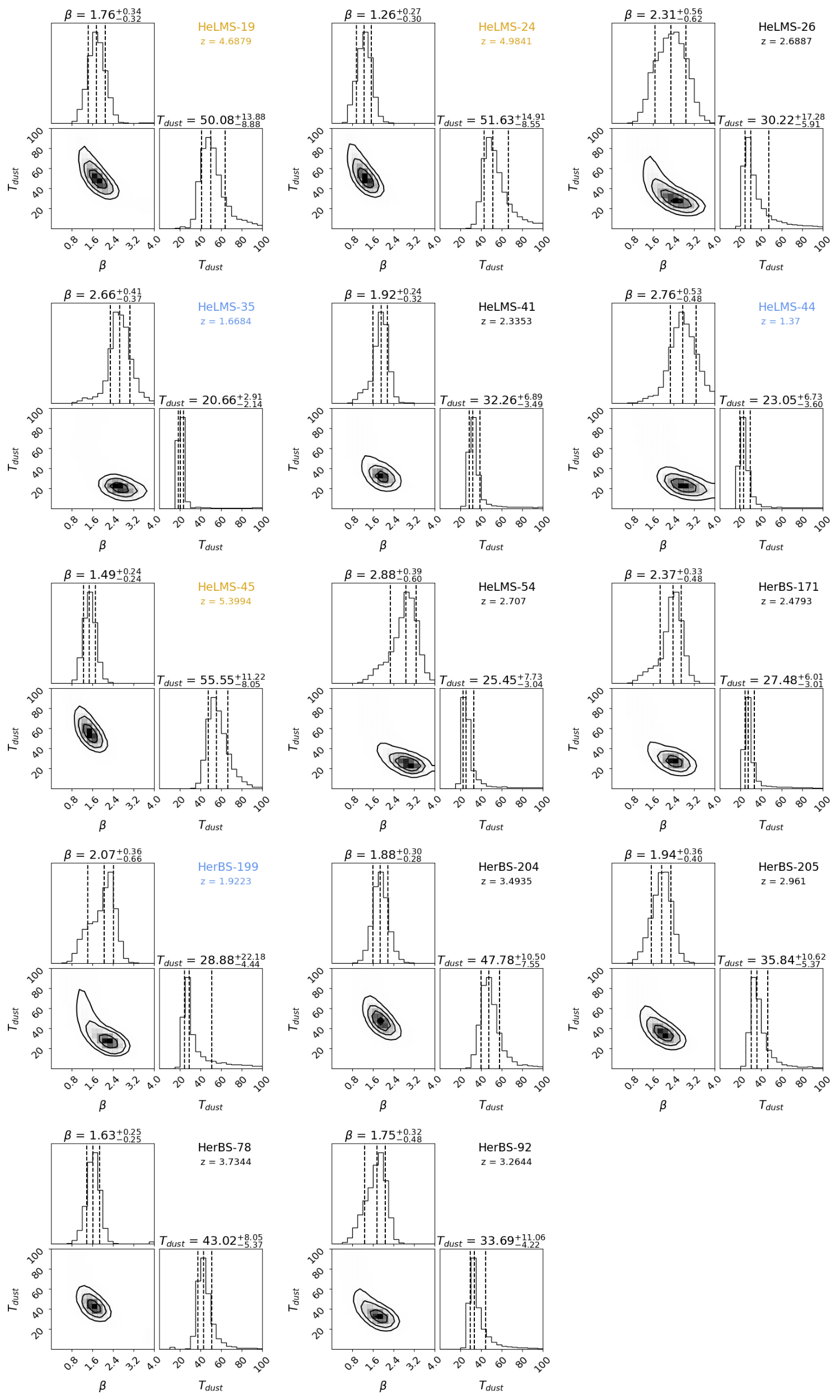}
    \caption{ $\beta - T_{\rm dust}$ posterior distribution plots for sources whose $\delta \beta / \beta > 15\%$. The yellow names represent sources at $\textit{z} > 4$, the blue names are sources at \textit{z} < 2, and black names are sources at $2 \leq \textit{z} \leq 4$.}
    \label{fig:corner-bad-sources}
\end{figure}

\begin{figure}[h!]
    \centering
    \includegraphics[width=0.6\textwidth]{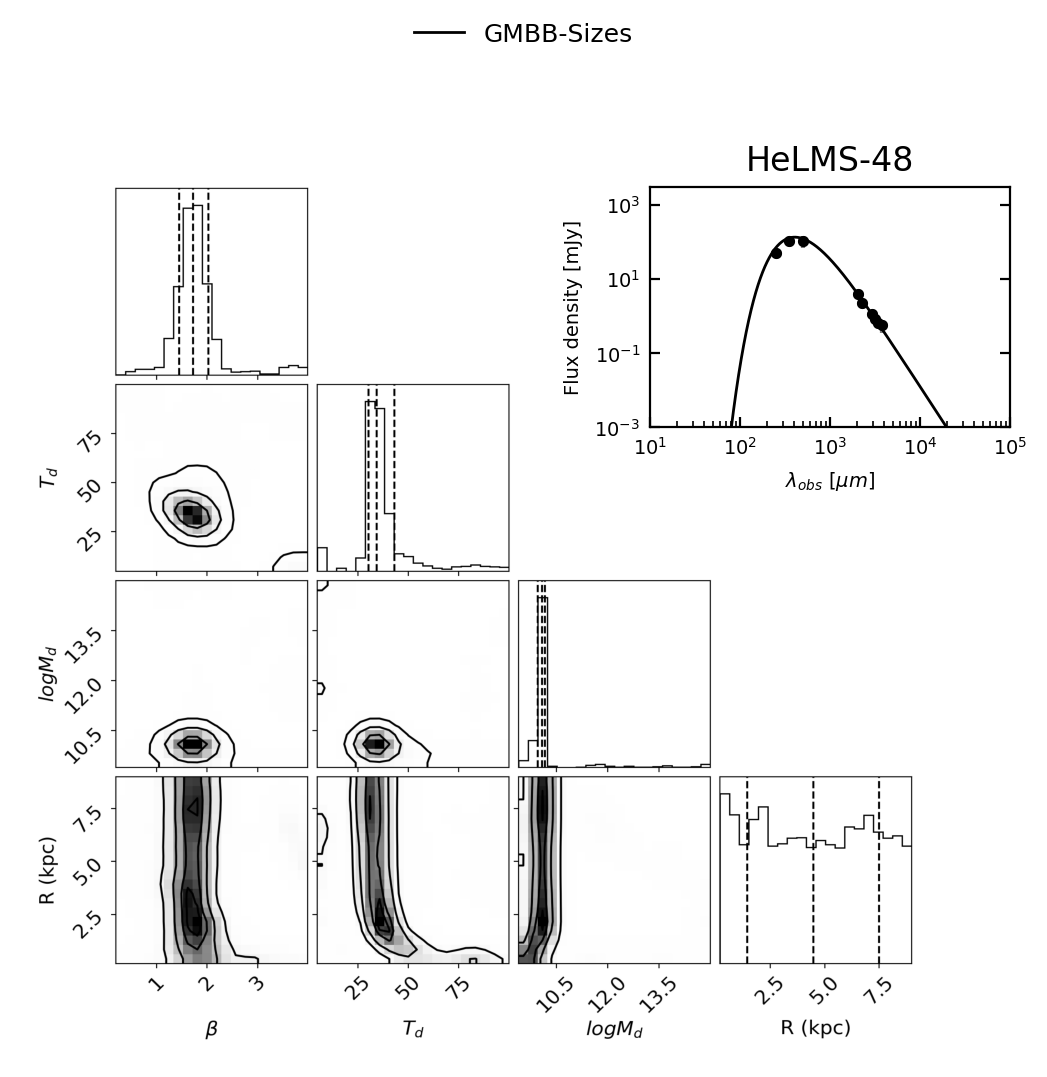}
    \caption{Corner plot of HeLMS-48 demonstrating the posterior likelihood distribution of estimated parameters using GMBB-Sizes and the SED fit. In the corner plot, the vertical lines represent the 16$^{th}$, 50$^{th}$, and 84$^{th}$ percentiles, respectively. The flux densities are shown as black dots in the SED, and the black line is the best fit.}
    \label{fig:helms48-corner-plot}
\end{figure}

\begin{figure}[h!]
    \centering
    \includegraphics[width=0.8\textwidth]{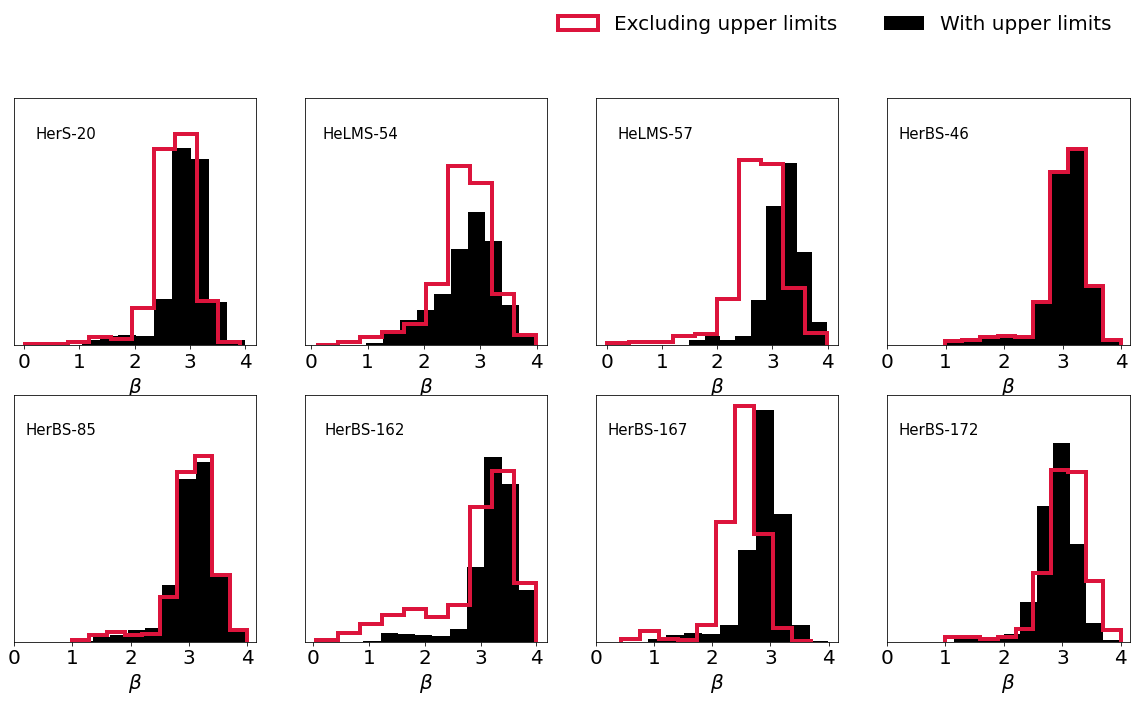}
    \caption{Posterior distribution of $\beta$ as estimated when using the measured flux densities (which include those that are upper limits) along the RJ tail in black and when removing them in red, for eight sources whose $\beta \geq$ 3. This shows that the large $\beta$ estimate is not biased by the use of measured flux densities on nondetections.}
    \label{fig:beta-corner-plots}
\end{figure}

\end{appendix}
\end{document}